\documentclass[12pt]{article}
\usepackage{amsfonts}
\usepackage{amsmath}
\usepackage{amssymb}
\usepackage{color}
\usepackage{graphicx}
\usepackage{mathcomp}
\usepackage[top=1in, bottom=1in, left=1in, right=1in]{geometry}
\usepackage{setspace}

\makeatletter
\@addtoreset{equation}{section} %%% Latex Companion, p.~16
\makeatother
\def\appendix#1
{
\addtocounter{section}{1}\setcounter{equation}{0}%\setcounter{section}{1}
 \renewcommand{\thesection}{\Alph{section}}
 \section*{%Appendix~
 \thesection\protect\indent \parbox[t]{16.715cm}{#1}}
 \addcontentsline{toc}{section}{Appendix \thesection\ \ \ #1}}

\begin{document}

\thispagestyle{empty}
		\begin{centering}
			\textbf{Emergence of General Relativity from Loop Quantum Gravity}\rule{0.0in}{0in}\\
			\vspace{0.5em}
			By \rule{0.0in}{0in}\\
			\vspace{0.5em}
			Chun-Yen Lin\rule{0.0in}{0in}\\
			B.S.  (National Taiwan University) 2003 \rule{0.0in}{0in}\\
			\vspace{0.5em}
			DISSERTATION\rule{0.0in}{0in} \\
			\vspace{0.5em}
			Submitted in partial satisfaction of the requirements for the degree of\rule{0.0in}{0in} \\
			\vspace{0.5em}
			DOCTOR OF PHILOSOPHY \rule{0.0in}{0in}\\
			\vspace{0.5em}
			in \rule{0.0in}{0in}\\
			\vspace{0.5em}
			Physics \rule{0.0in}{0in}\\
			\vspace{0.3em}
			in the \rule{0.0in}{0in}\\
			\vspace{0.3em}
			OFFICE OF GRADUATE STUDIES \rule{0.0in}{0in}\\
			\vspace{0.3em}
			of the \rule{0.0in}{0in}\\
			\vspace{0.3em}
			UNIVERSITY OF CALIFORNIA\rule{0.0in}{0in} \\
			\vspace{0.5em}
			DAVIS\rule{0.0in}{0in} \\
			\vspace{0.5em}
			Approved:\rule{0.0in}{0in} \\
			\vspace{1em}
			\rule{50mm}{0.2mm} \rule{0.0in}{0in}\\
			\vspace{-0.5em} Steven Carlip, Chair\rule{0.0in}{0in} \\
			\vspace{1em}
			\rule{50mm}{0.2mm}\rule{0.0in}{0in} \\
			\vspace{-0.5em} Andreas Albrecht\rule{0.0in}{0in} \\
			\vspace{1em}
			\rule{50mm}{0.2mm}\rule{0.0in}{0in} \\
			\vspace{-0.5em} John Terning \rule{0.0in}{0in}\\
			\vspace{0.5em}
			Committee in Charge\rule{0.0in}{0in} \\
                                   \vspace{0.5em}
			{2011}\rule{0.0in}{0in}\\
			\vspace{0.5em}
			\text{i}\rule{0.0in}{0in}\\		
                      \end{centering}

\newpage
\pagenumbering{roman}
\setcounter{page}{2}

\section*{Acknowledgments}
This work is only the first monument of a journey, that began five and a half years ago and will continue for as long as it possibly can. Yet I am already in many people's debt.

I would like to express my gratitude to my advisor, Prof. Steven Carlip, who taught me the elements of quantum gravity, through its fundamental essence instead of a specific theory. If there is any insight I possess in this deeply mysterious subject, it is with that scope he provided. As a mentor, he not only offered so much great advice and instruction on my research, but also tirelessly reminded me how important it is to be able to convey and communicate ideas in science. 

I was privileged to learn quantum field theory from Prof. John Gunion, Prof. Joseph Kiskis and Prof. John Terning, and to learn cosmic inflation theory from Prof. Andreas Albrecht. I especially owe Prof. Joseph Kiskis and Prof. Andreas Albrecht gratitude for the inspirations from their styles in theoretical physics and their encouragements.

Also, I am grateful to physics department of U.C. Davis, for providing such a supportive resource and a wonderful company to be with throughout my graduate program. Especially, I would like to thank Laura Peterson, Georgie Tolle, Onelia Yan and Tracy Lade for ensuring my smooth and trouble-free progress. 

Last but not ever the least, I thank my dearest family for their all-present support and, for many times, their wise and practical advices.

\newpage

 \rule{0pt}{180pt}      
    \centerline{\emph{To my father, mother and sister, who plant their love wherever I am.}}

\newpage

\begin{abstract}
		A model is proposed to demonstrate that classical general relativity can emerge from loop quantum gravity, in a relational description of gravitational field in terms of coordinates given by matter. Local Dirac observables and coherent states are defined to explore physical content of the model. Expectation values of commutators between the observables for the coherent states recover the four-dimensional diffeomorphism algebra and the large-scale dynamics of the gravitational field relative to the matter coordinates. Both results conform with general relativity up to calculable corrections near singularities.
	\end{abstract}

\newpage
	
	\tableofcontents
           \listoffigures

\begin{quotation}
\begin{centering}
\newpage
\rule{0pt}{180pt}
\emph{In our dreams (writes Coleridge) images represent the sensations we think they cause ; we do not feel horror because we are threatened by a sphinx ; we dream a sphinx in order to explain the horror we feel. If this is so, how could a mere chronicle of its forms transmit the stupor, the exaltation, the alarm, the menace and the jubilance which made up the fabric of that dream that night ?
}

\rule{0pt}{30pt}
-- J. L. Borges, \emph{Ragnar\"{o}k}
\end{centering}
\end{quotation}

\newpage
\pagestyle{myheadings}
\markright{}

\pagenumbering{arabic}

\section{Introduction}

    Loop quantum gravity \cite{intro1}\cite{intro}\cite{perez} is a candidate quantum theory of gravity. The theory strives to construct the deep quantum structure of spacetime, while holding on to background independence that is the heart of general relativity. Following the approach of Dirac quantization in the Arnowitt-Deser-Misner (ADM) formalism \cite{intro1}\cite{intro}\cite{perez}, the theory's kinematical Hilbert space, called knot space, is rigorously constructed and is shown to be almost unique. Knot space describes spatial geometry in terms of quanta of areas and volumes, realizing the idea that the geometry of space should be discrete with Planck-sized units.

Because of its novelty, the theory also faces new challenges \cite{intro1}\cite{intro}\cite{perez}: based on the Planck-scale discrete structure of the knot states, the recovery of the observed smooth geometry has been a major issue for the theory; further, the background independence demands one to \emph{solve} the Hamiltonian constraint upon knot space to obtain a physical Hilbert space, and also to construct diffeomorphism-invariant local Dirac observables. The difficulty in these tasks makes the semi-classical limit of the theory hard to obtain, and thus it is unknown whether loop quantum gravity describes the gravitational field at observed scales. However, the tasks are completed in a symmetrically reduced setting in loop quantum cosmology \cite{lqc1}\cite{lqc2}\cite{marolf1}\cite{marolf2}, minisuperspace models that carry the main features of loop quantum gravity. Moreover, it has been shown that loop quantum cosmology has correct semi-classical limits, with meaningful corrections in the regions where the classical theory breaks down \cite{lqc1}\cite{lqc2}. Therefore, loop quantum cosmology provides a valuable guidance for the full theory.

Following this guidance, the model proposed in this thesis explores the semi-classical limit of loop quantum gravity. The model shows that the above tasks can be accomplished with a few explicitly specified assumptions. Further, the obtained semi-classical limit is also shown to reproduce general relativity. The system considered by the model contains both gravitational and matter fields, and both sectors are treated with the standard loop quantization procedure to obtain the kinematical Hilbert space for the system. 
To go beyond the kinematical level, three explicit assumptions are made: 1) that a modified Hamiltonian constraint operator is valid; 2) that group averaging method solves the modified Hamiltonian constraint and gives the physical Hilbert space for the model; 3) that the matter back-reactions on the gravitational dynamics can be ignored in this context.

 The matter fields, with their negligible back reactions, provide internal spacetime coordinates for the local observables in the model. Using matter coordinates to describe local dynamics is a long existing idea in general relativity \cite{torre1}\cite{torre}\cite{rovelli}, the implementation of which in the quantum theory leads to the local quantum observables \cite{kuchar}\cite{kuchartorre}\cite{torre}\cite{rovelli}. The internal coordinates are provided by composite matter fields, and are also quantized.  Since the coordinate fields are part of the fully coupled system, the observables are thus defined in a background independent manner and are diffeomorphism invariant. 

Subsequently, the appropriate coherent states in the model are defined to minimize the uncertainty of the gravitational local observables. The large scale physics is then obtained by explicit calculations based on the observables and the coherent states.  First of all, the matter coordinates provide a natural correspondence between the expectation values of the observables and a classical emergent gravitational fields. In terms of the emergent gravitational field, the quantum constraint algebra of the model reproduces the classical diffeomorphism algebra up to quantum corrections. Further, the emergent gravitational field satisfies classical constraint equations up to quantum corrections. Moreover, the clock time derivatives of the emergent fields are shown to be given by the temporal translations generated by the quantum algebra. Finally, the equations of motion recover the classical gravitational dynamics of general relativity up to quantum corrections. All of the correction terms for the semi-classical limits are small in large scales and nonsingular regions of spacetime. Therefore, the model's semi-classical limit reproduces general relativity, up to significant and calculable corrections where the classical theory breaks down.

\section{Loop Quantum Gravity}

In the first part of the thesis, I briefly introduce the basics of the loop quantum gravity, and describe how it couples to the matter fields. The purpose is mainly to explain the kinematics of the theory, which captures quantum geometry of space background independently, and provides a foundation for the dynamics.

\subsection{Classical Hamiltonian Formalism }
Loop quantum gravity is the result of quantizing general relativity in a special Hamiltonian formulation, which uses triad and connection fields to describe gravity \cite{form1}\cite{form2}\cite{form3}. 

The most common form of Hamiltonian general relativity uses metric and extrinsic curvature of space as phase space variables \cite{form1}\cite{form2}\cite{form3}. Instead of using the spatial metric $q_{ab} (\text{x})$, we can introduce triad fields $\{e^a_i (\text{x})\} (\text{x}\equiv (x,y,z); a=x,y,z;  i=1,2,3)$ consisting of three vector fields that are orthonormal according to the metric. In this correspondence, each set of triad fields specifies a spatial metric by $q_{ab} (\text{x})=\delta_{ij}e^i_a (\text{x})e^j_b(\text{x})$ uniquely $(e^i_a (\text{x})e^a_j(\text{x})=\delta^i_j)$. On the other hand, setting $U^i_j(\text{x})$ to be a local $SO(3)$ transformation, $e^a_i(\text{x})$ and $e'^a_i(\text{x})=U^j_i(\text{x})e^a_j(\text{x})$ would determine the same $q_{ab} (\text{x})$. Thus the use of the triad fields in place of the metric introduces an additional local $SO(3)$ symmetry.  

The triad fields also serve as a spatial frame in space, such that we can write every spatial tensor using the triad fields as the basis. For instance, a vector field will be written as $V^i(\text{x})$. In this frame, the spatial Levi-Civita connection will be written in the form:
\begin{equation}
\begin{split}
\Gamma^i_a(\text{x})=-\frac{1}{2}{\epsilon^{ij}}_k e^b_j\left(\partial_{[a}e^k_{b]}+\delta^{kl}\delta_{ms}e^c_l e^m_a \partial_b e^s_c\right)(\text{x})
\nonumber
\end{split}
\end{equation}
Moreover, the canonical conjugate momenta of $e^a_i (\text{x})$ is the spatial extrinsic curvature field $K^i_a(\text{x})$, whose explicit form is given by:
\begin{equation}
\begin{split}
K^i_a(\text{x})\equiv\delta^{ij} K_{ab}e^b_j (\text{x})\rule{2pt}{0pt} ;\rule{10pt}{0pt} K_{ab}(\text{x})\equiv \mathcal{L}_n q_{ab}(\text{x})
\nonumber
\end{split}
\end{equation}
where $\mathcal{L}_n$ denotes the Lie derivative with respect to the unit normal field $n$ perpendicular to the spatial slices (according to the spacetime metric).
There's no surprise that general relativity can be written in terms of the triad fields serving as an alternative basis. However, this new point of view leads to the discovery of Ashtekar formalism \cite{form2}\cite{form3}, which reveals the striking similarities between general relativity and Yang-Mills gauge theory. 

The fundamental variables in Ashtekar formalism are related to the triad and spatial extrinsic curvature fields, through a canonical transformation
\begin{equation}
\begin{split}
E^a_i (\text{x})\equiv {\det (e)} e^a_i(\text{x})\rule{10pt}{0pt}
\\
A^i_a(\text{x}) \equiv \Gamma ^i_a (\text{x})+ \gamma K^i_a(\text{x})
\end{split}
\end{equation}
where the real number $\gamma$ is called Immirzi parameter, and classically different values for the parameter correspond to different canonical transformations in $(2.1)$. By construction, the fields $E^a_i(\text{x})$ are densitized triad fields, and the field $A^i_a(\text{x})$ are $SO(3)$ gauge fields. The variables also satisfy the simple Poisson algebra, with the only non-vanishing bracket
\begin{equation}
\begin{split}
\{A^i_a(\text{x}), E^b_j(\text{y})\}= 8\pi( G/c^3 )\gamma \delta^b_a \delta^i_j \delta (\text{x},\text{y})
\end{split}
\end{equation}
where $G$ is the Newton's constant. Using $(2.2)$, the Hamiltonian general relativity can be reformulated in terms of the new canonnical conjugate variables $A^i_a(\text{x})$ and $E^a_i(\text{x})$. Details of the reformulation can be found in chapter 2 of \cite{perez} and chapter 3 of \cite{outsideview}. The resulting theory has the Hamiltonian
\begin{equation}
\begin{split}
\text{H}_g(\bar N, \bar{\Lambda}, \bar V)= H_g (\bar N)+G_g(\bar{\Lambda})+M_g(\bar V)
\end{split}
\end{equation}
where each of the three terms comes from smearing  a constraint functional with an arbitrary Lagrangian multiplier (with the bars to emphasize the non-dynamical nature) over the spatial manifold $M$
\begin{equation}
\begin{split}
\\
H_g (\bar N)\equiv\int_M d^3 \text{x} \bar N(\text{x}) H_g (\text{x}) \equiv  \int_M d^3 \text{x}\bar  N(\text{x}) \frac{E^a_i E^b_j}{\sqrt{\det E}} \left[ {\epsilon^{ij}}_k F^k_{ab} +2(1-\gamma^2) K^i_{[a} K^j_{b]}\right](\text{x})
\\\\
G_g(\bar{\Lambda})\equiv \int_M d^3 \text{x} \bar{\Lambda}^i (\text{x})G_{g,i}(\text{x})\equiv \int_M d^3 \text{x} \bar{\Lambda}^i \left( \partial_a E^a_i +{\epsilon_{ij}}^k A_a^j E^a_k \right)(\text{x})\rule{56pt}{0pt}
\\\\
M_g(\bar V)\equiv \int_M d^3 \text{x} \bar V^a(\text{x}) M_{g,a}(\text{x})\equiv \int_M d^3 \text{x} \bar V^a(\text{x}) \left( E^b_i F^i_{ab} -\frac{1-\gamma^2}{ \gamma} K^i_a G_{g,i}\right)(\text{x})
\\
\end{split}
\end{equation}
where $K^i_a$ is now considered a function of the canonical variables $A$ and $E$, as determined from $(2.2)$ and $(2.1)$.
It is known that general relativity is a theory of pure constraints, and in this formalism the solutions to the theory are given by the constraint equations
\begin{equation}
\begin{split}
 H_g (\text{x})= M_{g,a}(\text{x})=G_{g,i}(\text{x})=0
\end{split}
\end{equation}
The smeared Hamiltonian constraint $H_g(\bar N)$, with any lapse function $\bar N(\text{x})$, generates a diffeomorphism on the fields that satisfy the on-shell condition $(2.5)$ in the directions perpendicular to spatial slices. The smeared momentum constraint $M_g(\bar V)$, with any shift function $\bar V^a(\text{x})$, generates a spatial diffeomorphism. Together, the smeared Hamiltonian and momentum constraints generate spacetime diffeomorphisms. Additionally, the smeared Gauss constraint $G_g(\bar{\Lambda})$, with any $ \bar{\Lambda}^i $, generates a local rotation of the triads. Clearly, the presence of Gauss constraint is due to the additional local $SO(3)$ symmetry introduced along with the triad fields \cite{intro}\cite{intro1}\cite{perez}.

From here on, we replace the $SO(3)$ symmetry group with $SU(2)$ group, exploiting the fact that $SU(2)$ group is the universal covering of $SO(3)$ group and thus has the same Lie algebra. The constraints $(2.4)$ and Poisson bracket $(2.2)$ remain exactly the same under the replacement, so the physical content of the theory is unchanged. Under such illumination, the similarities between general relativity and $SU(2)$ Yang-Mills theory are revealed. The two theories share the same preliminary phase space parameterized by the fields $E^a_i (\text{x})$ and $A^i_a (\text{x})$, which represent (generalized) electric fields and magnetic potential fields in $SU(2)$ Yang-Mills theory. Further, they are both $SU(2)$ symmetric and they share exactly the same Gauss constraint. These similarities inspire the introduction of the loop representation from gauge theories into quantum gravity, and result in significant parallelism of the two in their quantization procedures. However, there is a major distinction between the theories that causes drastic departures of their physical Hilbert spaces.

The difference lies in the background independence of general relativity. Treating spacetime as a dynamical entity instead of a fixed physical background, general relativity uniquely possesses diffeomorphism symmetry. As a result, in contrast to the case of $SU(2)$ Yang-Mills theory, the value of an $SU(2)$ invariant field at a spacetime coordinate point $(\text{x},t)$ is not an observable, since it is not invariant under any diffeomorphism that moves the coordinate point. Further, the Hamiltonian of the theory is composed of the constraints. That means the evolution along a Hamiltonian flow is merely a gauge transformation, and any observable must be constant along the flow. To incorporate these features, the central theme of the quantization procedure toward loop quantum gravity is about achieving a quantum theory that is devoid of any fixed physical background.

As a theory of pure symmetry, general relativity is governed by the algebra of diffeomorphism and local $SU(2)$ group. Denote $[ \bar\Lambda, \bar\Lambda']$ to be the $SU(2)$ commutator, and set $[ \bar V, \bar V']^a\equiv \bar V^b \partial_b  \bar V'^a- \bar V'^b \partial_b  \bar V^a$. In terms of the generators $H_g(\bar N)$, $M_g(\bar V)$ and $G_g(\bar{\Lambda})$ the algebra is 
\begin{equation}
\begin{split}
\\
\{G_g(\bar{\Lambda}), G_g(\bar{\Lambda}')\}= 8\pi( G/c^3 ) \gamma G_g([\bar{\Lambda},\bar{\Lambda}']);\rule{20pt}{0pt}  \{G_g(\bar{\Lambda}), M_g(\bar V)\}= 8\pi( G/c^3 ) \gamma G_g(\mathcal{L}_{\bar V} \bar{\Lambda})\rule{90pt}{0pt}
\\\\
\{M_g(\bar V), M_g(\bar V')\}=  8\pi( G/c^3 ) \gamma M_g([\bar V,\bar V'])\rule{308pt}{0pt}
\\\\
\{G_g(\bar{\Lambda}), H_g(\bar N)\}= 0;\rule{20pt}{0pt} \{M_g(\bar V), H_g(\bar N)\}=  8\pi( G/c^3 ) \gamma H_g(\mathcal{L}_{\bar V}\bar N)\rule{190pt}{0pt}
\\\\
\{H_g(\bar N), H_g(\bar N')\}= 8\pi( G/c^3 )\gamma\left( M_g(\bar S)+G_g(\bar S^a A_a)\right)+ \frac{1-\gamma^2}{8\pi( G/c^3 )}\gamma G_g\left(\frac{[E^a\partial_a \bar N,E^b\partial_b \bar N']}{|\det E|}\right)\rule{65pt}{0pt}
\\
\\
\end{split}
\end{equation}
where $\mathcal{L}_{\bar V}$ denotes a Lie derivative and
\begin{equation}
\begin{split}
\bar S^a=( \bar N\partial_b \bar N'-\bar N'\partial_b \bar N) \frac{E^b_i E^{ai}}{|\det E|}\rule{80pt}{0pt}\\
\end{split}
\end{equation}
Note that instead of structure constants, the algebra has structure functionals depending on the phase space variables.

\subsection{Quantization in Loop Representation}

Loop quantum gravity results from Dirac quantization of general relativity, in the Ashtekar formalism based on loop variables \cite{intro}\cite{intro1}\cite{perez}. Explicitly background independent, the theory is non-perturbative, in contrast to the usual forms of particle theories. Originating in non-perturbative quantization of Yang-Mills theory, the loop representation is introduced to quantize gravitational fields in the theory. As a kinematical quantum space, knot space is rigorously constructed to solve Gauss and momentum constraints. The knot states in the space describe spatial quantum geometry. The construction of the physical Hilbert space is an ongoing project involving quantizing and solving Hamiltonian constraint in (dual) knot space.

\subsubsection{Cylindrical Functions}

Based on the Ashtekar formalism given in the previous chapter, loop quantum gravity uses the connection fields $A^i_a(\text{x}) $ as configuration variables to construct quantum states. To employ the loop representation we introduce a special basis, called cylindrical functions, for general wave functionals of the connection fields \cite{intro}\cite{intro1}\cite{perez}. 

 Cylindrical functions are wave functionals that depend on the connection fields specifically through holonomies. A holonomy is a group-valued functional of connection fields defined with an oriented path $\bar{e}\in M$ in the spatial manifold $M$. The bar indicates that $\bar{e}$ is  embedded in $M$. Here, each holonomy will be $SU(2)$ valued, and has an explicit spin $j$ matrix representation as
\begin{equation}
 {h}^{(j)}(\bar{e})^{\bar{k}}_{\bar{l}}[A]\equiv [\mathcal{P} \exp \int_{\bar{e}}d\bar{e}^{b} A^{i}_{b}(\text{x})\tau_{i}^{(j)}]^{\bar{k}}_{\bar{l}}  
\end{equation}
where $\bar{k}$ denotes $SU(2)$ indices in the spin $j$ representation, and $\mathcal{P}$ denotes path ordering. Under the transformations of the connection fields, a holonomy behaves simply under local $SU(2)$ transformations $U^{\bar{k}}_{\bar{l}}(\text{x})$: 
\begin{equation}
 {h'}^{(j)}(\bar{e})^{\bar{k}}_{\bar{l}}=U^{\bar{k}}_{\bar{i}}(\bar{e}_{(1)}) {h}^{(j)}(\bar{e})^{\bar{i}}_{\bar{j}} U^{\bar{j}}_{\bar{l}}(\bar{e}_{(0)}) \equiv U{h}^{(j)}(\bar{e})^{\bar{k}}_{\bar{l}}
\end{equation} 
where $\bar{e}_{(0)}$ and $\bar{e}_{(1)}$ denote the beginning and end point of $\bar{e}$.
Given a specific connection fields $A^{i}_{b}(\text{x})$, a holonomy specifies the parallel transportation of a spinor across the path $\bar{e}$ through the linear map $(2.8)$ . Moreover, if we consistently assign group values to all holonomies along all paths in $M$, we uniquely specify a distributional connection field.\footnote{Every smooth connection field on $M$ gives a consistent assignment to the values for all holonomies. However, all possible consistent assignments of holonomy values, in general, correspond to non-smooth connection fields that are distributional.}  Relying on this fact, loop representation uses holonomies to build cylindrical functions for the basis.

 Define a graph $\bar{\gamma}$ in $M$ to be a set $\{\bar{e}_i\}$ containing $N_e$ oriented paths, called edges, meeting at most at their end points. Again, the bars indicate that $\bar{\gamma}$ is embedded in $M$.  A cylindrical function $\psi_{\bar{\gamma},f}$ is specified by a graph $\bar{\gamma}$ and a complex function $f: SU(2)^{N_e}\to \mathbb{C}$. Also, each cylindrical function is a wave functional of the connection fields through the relation
\begin{equation}
\psi_{\bar{\gamma},f}[A]\equiv f({h}(\bar{e}_1),{h}(\bar{e}_2),{h}(\bar{e}_3),....,{h}(\bar{e}_{N_e}))[A]
\end{equation}
The vector space spanned by cylindrical functions is denoted by $Cyl$, and carries an (over) complete representation of the wave functions. The space is endowed with  inner products given by
\begin{equation}
\begin{split}
\langle\psi_{\bar{\gamma},f}|\psi'_{\bar{\gamma}',f'}\rangle\rule{340pt}{0pt}\\\equiv \int \prod_{\bar{e}\in \bar{\gamma} \bigcup \bar{\gamma}'} dh_{(\bar{e})}   f^*({h}(\bar{e}_1),{h}(\bar{e}_2),{h}(\bar{e}_3),....,{h}(e_{N_e}))  f'({h}(\bar{e}'_1),{h}(\bar{e}'_2),{h}(\bar{e}'_3),....,{h}(\bar{e}'_{N'_e}))
\end{split}
\end{equation}
where $dh$ is the normalized Haar measure of $SU(2)$. The space $Cyl$ with its inner product then serves as the primary space for the quantization to proceed.

Recall that general relativity is given by the three constraints $(2.4)$. Following the Dirac quantization approach, loop quantum gravity imposes the constraint equations $(2.5)$ by demanding the physical states to be annihilated by the quantized constraints. We denote $S$ the space obtained from imposing Gauss constraint on $Cyl$, and $K$ the space obtained from imposing  momentum constraint on $S$. The currently unfinished step is the construction of the physical Hilbert space $H$ by imposing Hamiltonian constraint on $K$. In the following we will briefly describe the well-established part of the procedure that successfully leads to $S$ and $K$.

\subsubsection{Space $\mathcal{S}$ of Spin Network States}

The first constraint to be imposed is the Gauss constraint, which generates local $SU(2)$ transformations in the classical theory. The action of an arbitrary local $SU(2)$ transformation $U (\text{x})$ on $Cyl$ is directly given by $\hat{\mathcal{U}}_U$
\begin{equation}
\begin{split}
\hat{\mathcal{U}}_U \psi_{\bar{\gamma},f} = \psi_{\bar{\gamma},f_U}\rule{265pt}{0pt}\\\\
f_U({h}(\bar{e}_1),{h}(\bar{e}_2),{h}(\bar{e}_3),....,{h}(\bar{e}_{N_e}))\equiv f(U{h}(\bar{e}_1),U{h}(\bar{e}_2),U{h}(\bar{e}_3),....,U{h}(\bar{e}_{N_e}))\\
\end{split}
\end{equation}
The quantized Gauss constraints are the generators of the finite transformations $(2.12)$, thus the states satisfying the constraints should be invariant under the transformations.

It turns out that the $SU(2)$ invariant subspace of $Cyl$ is spanned by the cylindrical functions called $spin$ $network$ $states$ \cite{intro}\cite{intro1}\cite{perez}. Each spin network state is characterized by the following information (fig.1): 1) a closed embedded graph $\bar\gamma$ in $M$, with its $N_e$ oriented edges $\{\bar{e}_i\}$ connecting to its $N_v$ nodes $\{\bar{v}_n\}$ without any free end;  2) a non-trivial $SU(2)$ spin representation $j_{i}$ assigned to each edge;  3) a generalized Clebsch-Gordan coefficient (intertwiner) $i_{n}$ assigned to each node, with the exactly right indices to contract invariantly with the matrix indices of the holonomies along the edges meeting at $\bar{v}_n$. The explicit form of a spin network state $S_{\bar{\gamma}, j_i, i_n}$ as a cylindrical function is
\begin{equation}
\begin{split}
S_{\bar{\gamma}, j_i, i_n}\equiv \psi_{\bar{\gamma},f_{ j_i, i_n}}\equiv Inv\left\{ \bigotimes_n^{N_v}  i_n \bigotimes_i^{N_e} {h}^{(j_i)}(\bar{e}_i)\right\}\\
\end{split}
\end{equation}
where $Inv\{...\}$ denotes the local $SU(2)$ invariant tensorial product. Thus, $(2.13)$ explicitly shows the invariance of $S_{\bar{\gamma}, j_i, i_n}$ under the transformations $(2.12)$. Note that the closure of $\bar{\gamma}$ is essential for the application of $Inv\{...\}$. With the inner products given by $(2.11)$, the set of all spin network states $\{S_{\bar \gamma, j_i, i_n}\}$ provides a complete (but not over complete) orthonormal basis for $\mathcal S$ which solves Gauss constraint \cite{intro}\cite{intro1}\cite{perez}.

Note that the orthonormal basis $\{S_{\bar\gamma, j_i, i_n}\}$ contains spin network states that differ only by arbitrarily small changes of the embedded graphs. That means, in contrast to the case of Fock space, there are orthogonal states in $\mathcal S$ that are arbitrarily close in their wave functional values on continuous $A$ fields. Because of this feature, the space is said to be inseparable. Detailed comparisons have been made between Schroedinger and polymer representations in the context of quantum mechanics \cite{corichi}\cite{corichi1}. It is shown that polymer representation, which is inseparable in the same way as loop representation, leads to a quantum mechanics that is unitarily inequivalent to Schroedinger theory. The cause of the inequivalence is the inseparability of the Hilbert space obtained from the polymer representation. For the same reason, we expect the quantum theory built upon $\mathcal S$ to be unitarily inequivalent to a theory built upon Fock space representation.

\subsubsection{Operators in $\mathcal S$ and Quantum Geometry}

The natural elementary operators in $\mathcal S$ are constructed from the flux and holonomy variables, respectively defined on oriented surfaces and paths embedded in $M$. Let $\bar S$ be an oriented surface in $M$, and let $f^i$ be an $SU(2)$ valued test function on $\bar S$. Then a flux operator is given by \cite{intro}\cite{intro1}\cite{perez}:
\begin{equation}
\begin{split}
 \hat{F}_f(\bar{S})
\equiv\frac{1}{2} \int_{\bar{S}}f^{i}\hat{E}^{a}_{i} \epsilon_{abc}d\bar{S}^{bc} 
\equiv \int_{\bar{S}}f^{i}\hat{E}^{a}_{i}d\bar{S}_{a} 
=-i\hbar 8\pi( G/c^3 )\gamma\int_{\bar{S}}f^{i} \frac{\delta}{\delta A^i_a} d\bar{S}_{a}
\\
\end{split}
\end{equation}
where $d\bar{S}^{bc}$ denotes the oriented differential surface element; the last equality comes from the fact that in the connection representation, $\hat E_i$ can be expressed as a functional derivative (see (2.3)). Let $\bar e$ be an oriented path in $M$, and let $ g^{\bar{l}}_{\bar{k}} $ be an $SU(2)$ test function along $\bar e$. Then a holonomy operator is given by \cite{intro}\cite{intro1}\cite{perez}:
\begin{equation}
\begin{split}
 \hat{h^{(j)}}(\bar{e})_g\equiv g^{\bar{l}}_{\bar{k}}  [\mathcal{P}\hat{ \exp} \int_{\bar{e}}d\bar{e}^{b} A^{i}_{b}(\text{x})\tau_{i}^{(j)}]^{\bar{k}}_{\bar{l}}  
\end{split}
\end{equation}
where $d\bar{e}^{b}$ denotes the oriented differential path element; in the connection representation a holonomy operator is just a multiplicative operator.

Specifically, these elementary operators act on a cylindrical function of a single holonomy in the following way
\begin{equation}
\begin{split}
\widehat {h}^{(j)}(\bar{e}_2)_g\cdot {h}^{(i)}(\bar{e}_1)_{g'}[A]= {h}^{(j)}(\bar{e}_2)_g {h}^{(i)}(\bar{e}_1)_{g'}[A]\rule{60pt}{0pt}\\\\
\widehat{{F}_{f}(\bar{S})}\cdot {h}^{(i)}(\bar{e}_1)_{g}[A] =i\hbar G \text{sgn} (\bar{S},\bar{e}_1)f^{i}(\bar{p})(\tau_{i}^{(i)})^{\bar{m}}_{\bar{n}}{g}^{\bar{l}}_{\bar{k}}({h}^{(i)}(\bar{e}_{1_{\bar{p}}^+})^{\bar{k}}_{\bar{m}}) ({h}^{(i)}(\bar{e}_{1_{\bar{p}}^-})^{\bar{n}}_{\bar{l}})[A]
\end{split}
\end{equation}
where $\text{sgn} (\bar{S},\bar{e}_1) $ is equal to $+1$ if $e_1$ intersect $\bar{S}$ with the same orientation of $\bar{S}$ at a point $\bar p$, $-1$ for the opposite case, and $0$ otherwise. The paths $\bar{e}_{1_{\bar{p}}^+}$ and $\bar{e}_{1_{\bar{p}}^-}$ result from dividing $e_1$ into two new edges from the point $\bar{p}$ (fig.1$(a)$). Recall that spin network states are cylindrical functions built from invariant products of holonomies, so $(2.15)$ also determine the actions of the operators on $\mathcal S$.

The holonomy operator $\hat{h}^{(j)}(\bar{e}_2)_g $ acts on a spin network state  in the straightforward way:
\begin{equation}
\begin{split}
\widehat{{h}^{(j)}}(\bar{e}_2)_g\cdot S_{\bar{\gamma}, j_i, i_n}[A]
= ({h}^{(j)}(\bar{e}_2)_g S_{\bar{\gamma}, j_i, i_n})[A]
\end{split}
\end{equation}
The result in (2.18) is a linear combination of spin network states that base on graphs covering both $\bar{\gamma}$ and $\bar{e}_2$. Thus the action is in general graph-changing by the inclusion of a new edge $\bar{e}_2$. When $\bar{\gamma}$ overlaps with $\bar{e}_2$, the action $(2.18)$ also changes the spin representations by coupling $ \hat{h}^{(j)}(\bar{e}_2)_g $ to the existing holonomies in $S_{\bar{\gamma}, j_i, i_n}$ . 

\begin{figure}
\begin{center}
\includegraphics[angle=0,width=4in,clip=true]{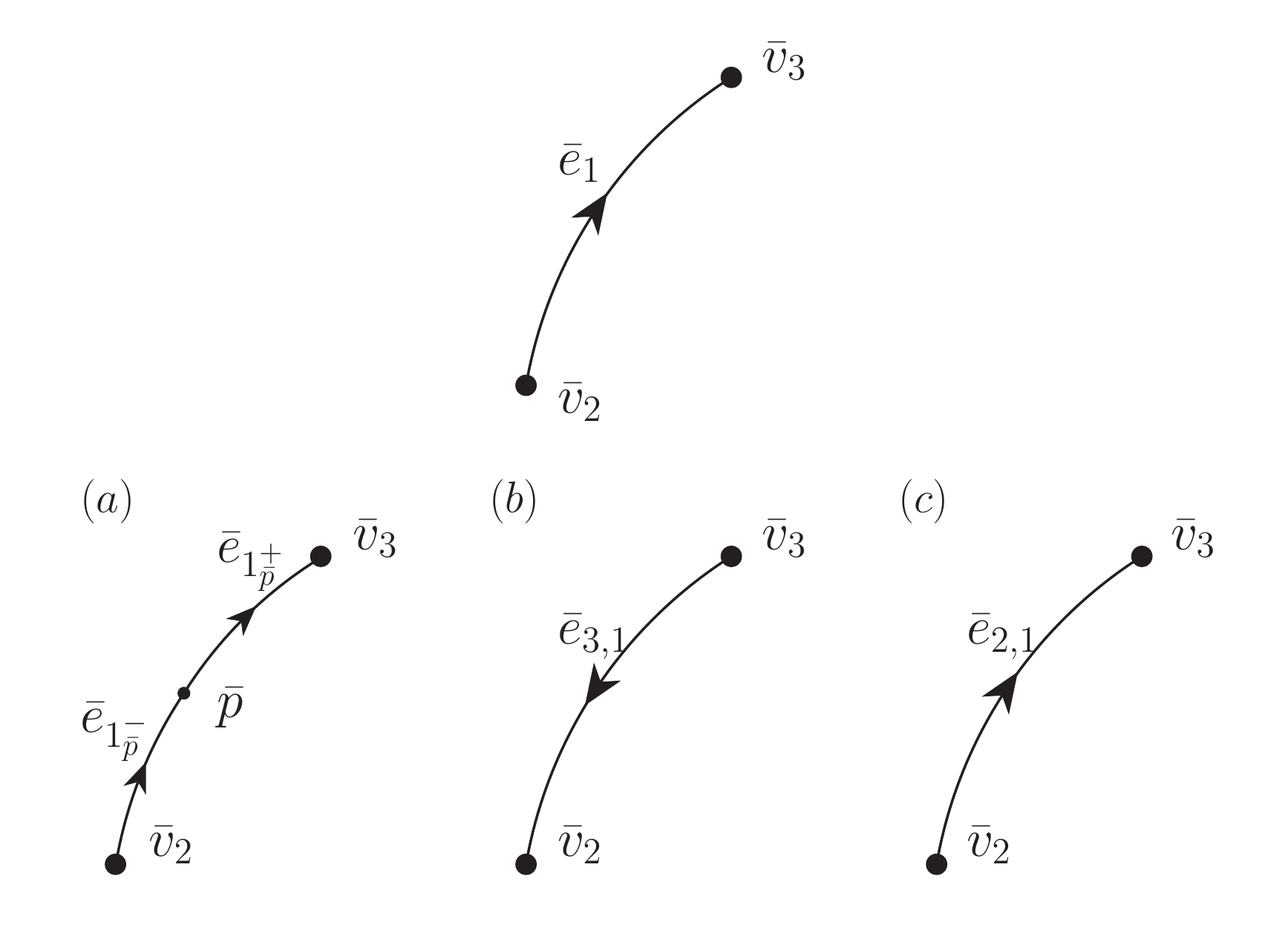}
\begin{spacing}{1}
\caption{The upper figure pictures a generic part of an embedded graph, which is an embedded edge $\bar{e}_1$ connecting the two nodes $\bar{v}_2$ and $\bar{v}_3$. The lower figures demonstrate the corresponding definitions of $\bar{e}_{1_{\bar p}^+}$, $\bar{e}_{1_{\bar p}^-}$, $\bar{e}_{3,1}$ and $\bar{e}_{2,1}$. In this case, $\bar{e}_{2,1}$ is the same as $\bar{e}_1$. }
\end{spacing}
\end{center}
\end{figure}

As a differential operator, $\hat{F}_{f}(\bar{S})$ acts on a spin network state following $(2.15)$ and Leibniz rules. An important example is given by the case when $\bar\gamma$ intersects with $\bar S$ at one and only one node $\bar v_1$. In this case we have
\begin{equation}
\begin{split}
\hat{F}_{f}(\bar{S})\cdot S_{\bar{\gamma}, j_i, i_n}[A]
\equiv\sum_{\bar e_{i'}|_{\bar v_1 \in \bar e_{i'}}} \iota(\bar{S},\bar{e}_{i'})f^{i}(\bar v_1) \hat{J}_{i}(\bar e_{1,i'}) \cdot Inv\left\{ \bigotimes_n^{N_v}  i_n \bigotimes_i^{N_e} {h}^{(j_i)}(\bar{e}_i)\right\}[A]\\
\end{split}
\end{equation}
where $\iota(\bar{S},\bar{e}_{i'})$ is equal to $+1$ or $-1$ when $\bar{e}_{i'}$ is above or below $\bar S$ in an infinitesimal neighborhood of $\bar v_1$. The $\bar e_{n,i'}$ (with $\bar v_n \in \bar e_{i'}$) denotes the oriented path that start from $\bar v_n$ and overlaps completely with the edge $\bar e_{i'}$ (fig.1$(b)(c)$). The action of $\hat{J}_{i}(\bar e)$ with an arbitrary path $\bar e$ on a cylindrical function is given by
\begin{equation}
\begin{split}
\hat{J}_{i}(\bar e)\cdot \psi_{\bar{\gamma},f}[A]= \psi_{\bar{\gamma},f'}[A]\rule{233pt}{0pt} \\\\
f'({h}(\bar{e}_1),{h}(\bar{e}_2),{h}(\bar{e}_3),....,{h}(\bar{e}_{N_e}))[A]\rule{185pt}{0pt}\\
\equiv i \hbar G\sum_k [\delta_{\bar e, \bar{e}_k}\cdot\partial^R_i(\bar{e}_k)-\delta_{\bar{e}, \bar{e}^{-1}_k}\cdot \partial^L_i(\bar{e}_k)]f({h}(\bar{e}_1),{h}(\bar{e}_2),{h}(\bar{e}_3),....,{h}(\bar{e}_{N_e}))[A]\\
\\
\end{split}
\end{equation}
where $\partial^L_i(\bar{e}_k)$ and $\partial^R_i(\bar{e}_k)$ denote the left and right Lie derivatives on $h(\bar{e}_k)$, with the $i$th generator. Applying it to $S_{\bar{\gamma}, j_i, i_n}$ one obtains :
\begin{equation}
\begin{split}\\
f^i(\bar e_{(0)})\hat{J}_{i}(\bar e)S_{\bar{\gamma}, j_i, i_n}[A]\rule{353pt}{0pt}\\
\equiv Inv\left\{ i \hbar G \sum_k f^i(\bar e_{(0)})\left [\delta_{\bar e, \bar{e}_k}\cdot{h}^{(j_k)}(\bar{e}_k)\tau^{(j_k)}_i-\delta_{\bar e, {\bar{e}_k}^{-1}}\cdot\tau^{(j_k)}_i{h}^{(j_k)}(\bar{e}_k)\right] \otimes\bigotimes_{n}^{N_v}  i_n \bigotimes_{i\neq k}^{N_e} {h}^{(j_i)}(\bar{e}_i)\right\}[A]\\\\
\end{split}
\end{equation}
where $\bar e_{(0)}$ and $\bar e_{(1)}$ stand for the origin and the end points of $\bar e_{}$.

Combining flux and holonomy operators, one can construct operators that characterize the geometry of space. In particular, the spatial area and volume operators have been rigorously constructed in loop quantum gravity. Moreover, their spectra have  led to significant understanding of the possible Planck-scale structure of space.

The area operator in loop quantum gravity is derived from regularizing the classical expression in terms of the flux variables. The resulting operator for the area of a surface $\bar{S}$ is \cite{area}
\begin{equation}
\hat{A}_{\bar{S}}|S_{\bar{\gamma}, j_i, i_n}\rangle \equiv \sqrt{\hat{F}_{i}(\bar{S})  \hat{F}^{i}(\bar{S})}|S_{\bar{\gamma}, j_i, i_n}\rangle
\end{equation}
Remarkably, spin network states are eigenstates of $\hat{A}_{\bar{S}}$ for any surface $\bar{S}$. Actually, one can easily check that 
\begin{equation}
\begin{split}
\hat{A}_{\bar{S}}|S_{\bar{\gamma}, j_i, i_n}\rangle=8\pi l^2_p \gamma\sum_{\bar{p}} \sqrt{j_{i_p}(j_{i_p}+1)} |S_{\bar{\gamma}, j_i, i_n}\rangle
\end{split}
\end{equation}
where we introduce Planck length through $l^2_p\equiv\hbar (G/c^3)$. The summation above goes over all points $\bar{p}$ where $\bar{\gamma}$ intersects transversely with $\bar{S}$, and $j_{i_p}$ represents the spin carried by the edge $e_{i_p}$ containing the point $\bar{p}$.
From above, it is clear that the spectrum of area operators in $\mathcal S$ is discretized, built from multiples of small quanta of order of $l^2_p$.

The volume operators in loop quantum gravity also result from the regularization of classical expressions using the flux variables, but with much more complicated technicalities. Also, there are a variety of different regularization approaches that lead to different operator forms for volume \cite{volume}\cite{volume1}. Here we will follow the one introduced in \cite{volume1}. The volume operators given by this approach act on $|S_{\bar{\gamma}, j_i, i_n}\rangle$ with a set of flux operators over the surfaces intersecting $\bar\gamma$ only at the nodes. With the help of $(2.18)$, the explicit action boils down to  \cite{volume1}:
\begin{equation}
\begin{split}
\hat{V}_{\bar{R}} |S_{\bar{\gamma}, j_i, i_n}\rangle\equiv \sum_{\bar{v}_n\in \bar{R}} \sqrt{\big|\hat{q}_{\bar{v}_n} \big|} |S_{\bar{\gamma}, j_i, i_n}\rangle\rule{255pt}{0pt}\\\\
\hat{q}_{\bar{v}_n}  |S_{\bar{\gamma}, j_i, i_n}\rangle\equiv  \frac{1}{48}\sum_{\bar{v}_n\in\bar{e}_i,\bar{v}\in\bar{e}_j,\bar{v}_n\in\bar{e}_k}\text{sgn}(\bar{e}_{n,i},\bar{e}_{n,j},\bar{e}_{n,k}) \epsilon^{pqr}  \hat{J}_{p}(\bar{e}_{n,i})  \hat{J}_{q}(\bar{e}_{n,j}) \hat{J}_{r}(\bar{e}_{n,k}) |S_{\bar{\gamma}, j_i, i_n}\rangle\rule{5pt}{0pt}
\end{split}
\end{equation}
where $\hat J$ was defined in $(2.19)$ and $(2.20)$.
It has been shown that the spectrum for the volume operators are also discretized in $S$, built from multiples of small quanta of orders of $l^3_p$.
By $(2.23)$ the volume operator for a region $\bar{R}$ acts only on the nodes of a spin network state contained in $\bar{R}$. Further, each node-wise operation on the spin network state is given by the sum of the triplets of the action $(2.19)$ on the edges merging at the node. Therefore, the volume operator replaces the intertwiner for each of the nodes in $\bar R$ with a specific linear combination of others, preserving both $\bar{\gamma}$ and $j_i$. Consequently, there is an volume and area eigenbasis for $S$, consisting of states with definite $\bar{\gamma}$ and $j_i$. This diagonalizing basis has the following interpretation:  a certain quantum of volume is assigned to each node of the state, and a certain quantum of area is assigned to each edge of the state.  Through this basis, $S$ provides a concrete description of quantum spatial geometry -- the networks carrying Planck sized units of areas and volumes, associated with their edges and nodes.

Finally, notice that the action of an area operator $\hat{A}_{\bar S}$ on a spin network state $|S_{\bar{\gamma}, j_i, i_n}\rangle$ is solely determined by the edges of $\bar{\gamma}$  intersecting transversely with $\bar{S}$. Similarly, the action of an volume operator $\hat{V}_{\bar{R}}$ on the same state is solely determined by the nodes of $\bar{\gamma}$ contained in $\bar R$.  Such information is insensitive to the details about the embeddings of $\bar{S}$, $\bar{R}$, and $\bar{\gamma}$; only the connectedness and transversality between $\{\bar e_i, \bar v_n, \bar S, \bar R\}$ are relevant in the quantum geometry given by $|S_{\bar{\gamma}, j_i, i_n}\rangle$.

\subsubsection{ Knot space K}

The next crucial step is to solve the momentum constraint. Since $M_g(\bar V)$ generates spatial diffeomorphisms in general relativity \cite{form1}\cite{form2}\cite{form3}, to solve the quantum momentum constraint  is to construct a quantum space invariant under spatial diffeomorphisms. 

Set $\mu$ to be a diffeomorphism on the spatial manifold $M$, which transforms the connection fields $\mu: A\to A'$. The action of $\hat\mu$ on cylindrical functions is induced naturally by the transformation on the connection fields, through the requirement 
\begin{equation}
\begin{split}
\hat{\mu}:\psi_{\bar{\gamma}, f}\to \psi'_{\bar{\gamma}', f'}\\
\psi_{\bar{\gamma}, f}[A] \equiv \psi'_{\bar{\gamma}', f'}[A']
\nonumber
\end{split}
\end{equation}
for all $A$. From now on we introduce the abbreviation $\bar\Gamma\equiv (\bar{\gamma}, j_i, i_n)$. Since $\bar\Gamma$ is effectively a graph $\bar{\gamma}$ carrying the quantum numbers $j_i$ and $i_n$ with its edges and nodes, it is called a colored graph. Accordingly, we will denote a spin network state as $|S_{\bar \Gamma}\rangle$. It follows that the spatial diffeomorphisms act on $\mathcal S$ and $\mathcal{S}^*$ as
\begin{equation}
\begin{split}
\hat{\mu} |S_{\bar{\Gamma}}\rangle = |S_{\mu\bar{\Gamma}}\rangle\rule{2pt}{0pt};\rule{10pt}{0pt}
 \langle S_{\bar{\Gamma}}|\hat{\mu} = \langle S_{\mu^{-1} \bar{\Gamma}}|
\end{split}
\end{equation}
where $\mu \bar{\Gamma}$ denotes the colored graph obtained by applying a spatial diffeomorphism $\mu$ to the colored graph $\bar{\Gamma}$.
With the inner product $(2.11)$, there is no normalizable state in $\mathcal S$ that is invariant under $(2.23)$ \cite{intro}\cite{intro1}\cite{perez}. However, normalizable solutions do exist in the algebraic dual space $\mathcal{S}^*$ \cite{intro}\cite{intro1}\cite{perez}, and they can be constructed by the group averaging procedure described in the following \cite{intro}\cite{intro1}\cite{perez}.

The method of group averaging \cite{ave3}\cite{ave4} is a constructive approach to obtain a physical Hilbert space for a system that is constrained due to its symmetry. The procedure starts from a unconstrained Hilbert space, which is analogous to the off-shell classical phase space. The states in this Hilbert space are in general non-invariant under the transformations of the symmetry group. To construct invariant states, the procedure starts with a state in the Hilbert space and averages over the transformations of that state by the symmetry group. By construction, the results of the averaging will be invariant under the actions of the symmetry group. When this procedure can be properly carried out, the resulting elements form a physical Hilbert space which naturally inherits the inner product from the original Hilbert space \cite{ave3}\cite{ave4}. Also, in many cases, the self adjoint operators in the original Hilbert space naturally induce physical observables in the physical Hilbert space \cite{ave3}\cite{ave4}. The method is well understood for the cases of compact Lie groups, while there are many unknowns when the group is non-compact, or not a Lie group. Nevertheless, individual successful implementations of group averaging do exist for the latter cases. The construction of knot space is one remarkable example. 

The symmetry group of concern here is the diffeomorphism group on the spatial manifold $M$, denoted by $\mathit{diff_M}$. Here we will use $\hat{\mathbb{P}}$ to denote `projectors' that apply group-averaging actions to the states in a Hilbert space. Notice that for every $\bar\gamma$, there is a subgroup  $\mathit{diff_{M,\bar{\gamma}}}$ of $\mathit{diff_M}$ that leaves the graph invariant, maintaining the \emph{set} of all the edges and their orientations. There is also a subgroup $\mathit{Tdiff_{M,\bar{\gamma}}}$ of $\mathit{diff_{M,\bar{\gamma}}}$ that acts trivially on $\bar \gamma$. We define the graph symmetry group $G_{M,\bar{\gamma}}\equiv  \mathit{diff_{M,\bar{\gamma}}}/ \mathit{Tdiff_{M,\bar{\gamma}}}$, and apply the averaging over $G_{M,\bar{\gamma}}$ to any $\langle S_{\bar{\Gamma}}|\in \mathcal S^*$ as 
\begin{equation}
\begin{split}
 \langle S_{\bar{\Gamma}}|\hat{\mathbb P}_{G_{M,\bar{\gamma}}}\equiv \frac{1}{N_{G_{M,\bar{\gamma}}}} \sum_{\mu \in G_{M,\bar{\gamma}}}\langle S_{\bar{\Gamma}}|\hat{\mu}
\end{split}
\end{equation}
where $N_{G_{M,\bar{\gamma}}}$ is the number of elements in $G_{M,\bar{\gamma}}$, which is finite. Subsequently, we further apply the averaging over $  \mathit{\mathit{diff_M}/ \mathit{diff_{M,\bar{\gamma}}}}$, the group that changes the graph, to $\langle S_{\bar{\Gamma}}|\hat{\mathbb P}_{G_{M,\bar{\gamma}}}$ as
\begin{equation}
\begin{split}
\langle S_{\bar{\Gamma}}|\hat{\mathbb P}_{\mathit{diff_M}}\equiv\langle S_{\bar{\Gamma}}|\hat{\mathbb P}_{G_{M,\bar{\gamma}}}\hat{\mathbb P}_{\mathit{diff_M}/\mathit{diff}_{M,\bar{\gamma}}}\equiv \sum_{\mu \in \mathit{diff_M}/\mathit{diff_{M,\bar{\gamma}}}}\langle S_{\bar{\Gamma}}|\hat{\mathbb P}_{G_{M,\bar{\gamma}}}\hat{\mu}
\end{split}
\end{equation}
The result of $\langle S_{\bar{\Gamma}}|\hat{\mathbb P}_{\mathit{diff_M}}$ is a finite element in $\mathcal S^*$, since $\langle S_{\bar{\Gamma}}|\hat{\mathbb P}_{\mathit{diff_M}}| S_{\bar{\Gamma}'}\rangle$ receive contributions from the finite number of terms from $(2.27)$ where $\mu\bar{\gamma}'$ matches $\bar\gamma$ completely. Further, $\langle S_{\bar{\Gamma}}|\hat{\mathbb P}_{\mathit{diff_M}}$ is $\mathit{diff_M}$ invariant by construction. Note that from $(2.26)$ and $(2.27)$, we have $\langle S_{\bar\Gamma}| \hat{\mathbb P}_{\mathit{diff_M}}=\langle S_{\mu\bar\Gamma}| \hat{\mathbb P}_{\mathit{diff_M}}$ for any $\mu\in \mathit{diff_M}$. Since the resulting state is characterized  by the diffeomorphism class $[\bar{\Gamma}]$ of the embedded colored graph $\bar{\Gamma}$, which is a knot structure, we will write $\langle S_{\bar\Gamma}| \hat{\mathbb P}_{\mathit{diff_M}}\equiv \langle s_{[\bar{\Gamma}]}|$ and call it a \emph{knot state}. The inner product $(2.12)$ induces a finite inner product for knot states through
\begin{equation}
\begin{split}
\langle s_{[\bar{\Gamma}]}| s_{[\bar{\Gamma}']}\rangle
\equiv \langle s_{[\bar{\Gamma}]}| S_{\bar{\Gamma}'}\rangle
= \langle S_{\bar{\Gamma}}|\hat{\mathbb P}_{\mathit{diff_M}}| S_{\bar{\Gamma}'}\rangle
\end{split}
\end{equation}
 With the above inner product, the set of knot states $\{\langle s_{[\bar{\Gamma}]}|\}$ provides an orthonormal basis for knot space $K$ which solves both Gauss and momentum constraints.

\subsubsection{Quantum Geometry in K}

Intuitively, $K$ is just $\mathcal S$ without detailed embedding information, and each knot state is an equivalence class of diffeomorphic spin network states. Therefore, one would expect $K$ to inherit the quantum geometry in $\mathcal S$ that is also ignorant of the embedding details. However, the inheritance is not as trivial as it might seem.

To start with, we study the actions of $\hat{A}_{\bar{S}}$ and $\hat{V}_{\bar{R}}$ on $K$. Recall that the action of $\hat{A}_{\bar{S}}$ on $| S_{\bar{\Gamma}}\rangle$ takes place at the intersection between $\bar{S}$ and $\bar{\Gamma}$. Similarly, the action of $\hat{V}_{\bar{R}}$ on $| S_{\bar{\Gamma}}\rangle$ happens at the nodes of $\bar{\Gamma}$ contained in $\bar{R}$. Based on $(2.24)$, the action of the area and volume operators on a knot state $\langle s_{[\bar{\Gamma}]}|$ is given by their actions on the components $\{\langle S_{\mu\bar{\Gamma}}|\}$ with $\mu\in \mathit{diff_{M}}$. Consider the nodes of $\mu\bar{\Gamma}$ contained in $\bar{R}$, letting $\mu$ runs over all elements in $\mathit{diff_{M}}$. Since the diffeomorphisms move $\bar{\Gamma}$ around freely, the number of nodes of $\mu\bar{\Gamma}$ in $\bar{R}$ (in general) changes with $\mu$ . Thus, $\hat{V}_{\bar{R}}$ acts on different number of nodes for the different members in $\{\langle S_{\mu\bar{\Gamma}}|\}$, resulting in a new set of spin network states that are non-diffeomorphic. A similar situation also happens to the area operator $\hat{A}_{\bar{S}}$. Briefly speaking, $\hat{A}_{\bar{S}}$ and $\hat{V}_{\bar{R}}$ are not operators in $K$, since their actions on a set of diffeomorphic spin network states would result to another set of spin network states that are not diffeomorphic to each other.

Clearly, $\hat{A}_{\bar{S}}$ and $\hat{V}_{\bar{R}}$ are not operators in $K$ because $\bar{S}$ and ${\bar{R}}$ does not transform under $\mathit{diff_{M}}$ and break the symmetry. Therefore, a natural solution would be defining the `dynamical' surfaces $S\equiv \bar S(\bar \Gamma)$ and regions $R\equiv \bar R(\bar \Gamma)$, which satisfy $ \bar S(\mu\bar\Gamma)= \mu'\mu\bar S(\bar \Gamma)$ and  $ \bar R(\mu\bar\Gamma)= \mu'\mu\bar R(\bar \Gamma)$ for any $\mu \in \mathit{diff_M}$, with some $\mu' \in \mathit{Tdiff_{M,{\mu\bar\gamma}}}$. By construction, $\bar S(\bar \Gamma)$ and $\bar \Gamma$ have the same differential-topological relation as  $\bar S(\mu\bar \Gamma)$ and $\mu\bar \Gamma$ do. $\bar R(\bar \Gamma)$ also enjoys the same feature. Then one can define the possible area operator $\hat{A}_{S}$ and volume operator $\hat{V}_{R}$ as
\begin{equation}
\begin{split}
\langle s_{[\bar \Gamma]}| \hat{A}_{S}
\equiv\langle S_{\bar{\Gamma}}|\hat{A}_{\bar S(\bar \Gamma)}\hat{\mathbb P}_{\mathit{diff_M}}
\\\\
\langle s_{[\bar \Gamma]}| \hat{V}_{R}
\equiv\langle S_{\bar{\Gamma}}|\hat{V}_{\bar R(\bar \Gamma)}\hat{\mathbb P}_{\mathit{diff_M}}
\end{split}
\end{equation}
for \emph{any} $\bar \Gamma \in [\bar \Gamma]$. This shows that $\hat{A}_{S}$ and $\hat{V}_{R}$ are now operators in $K$. Generally, one can promote any operator $\hat{O}_{\bar\Omega}$ in $\mathcal S^*$ that is sensitive only to the differential-topological relation between $\bar\Omega \in M$ and every $\bar \Gamma$ into an operator $\hat{O}_{\Omega}$ in $K$. Accordingly, $\Omega \equiv \bar \Omega(\bar \Gamma)$ satisfy $ \bar \Omega(\hat\mu\bar \Gamma)= \mu'\mu \bar \Omega(\bar \Gamma)$ for any $\mu \in \mathit{diff_M}$, and some $\mu' \in  \mathit{Tdiff_{M,{\mu\bar\gamma}}}$. 
\begin{equation}
\begin{split}
 \langle s_{[\bar \Gamma]}| \hat{O}_{\Omega} 
\equiv\langle S_{\bar{\Gamma}}|\hat{O}_{\bar \Omega(\bar{\Gamma})}\hat{\mathbb P}_{\mathit{diff_M}}
\\
\end{split}
\end{equation}
for \emph{any} ${\bar{\Gamma}}\in [\bar \Gamma]$.
From $(2.27)$ and $(2.28)$, $S$, $R$ and $\Omega$ above are equivalently determined by the assignments $\bar S_0=\bar S(\bar \Gamma_0)$, $\bar R_0=\bar R(\bar \Gamma_0)$ and $\bar\Omega_0=\bar \Omega(\bar \Gamma_0)$ given one specific representative $\bar\Gamma_0$ from each $[\bar\Gamma_0]$. Note that the assignment of $\bar S_0$, $\bar R_0$ and $\bar\Omega_0$ are completely arbitrary so far. This arbitrariness indicates the degree of ambiguity that has to be fixed before we can meaningfully describe spatial geometry in $K$. The ambiguity is due to the removal of embedding information from knot states. Note that the definitions $(2.20)$ and $(2.22)$ do not suffer any ambiguity, since $\bar S$  and $\bar R$ are embedded in a common background $M$ with any embedded graph $\bar \gamma$, and the intersections between them are thus clearly given. In knot space $K$, the background $M$ is removed and the reference is missing.

Since $\bar S(\bar \Gamma)$, $\bar R(\bar \Gamma)$ and $\bar \Omega(\bar \Gamma)$ transform consistently with $\bar \Gamma$, we might consider them as genuine dynamical entities. Specifically, any subset of $M$ specified by the values of dynamical scalar fields will have the desired transformation property. This idea of using dynamical fields to specify physical regions of spacetime has a long history and explored in abundant works \cite{kuchar}\cite{kuchartorre}\cite{torre}\cite{rovelli}\cite{torre1}. We will follow through this idea in this thesis and show that it is indeed valid in our context.

In summary, the quantum geometry of $K$ should be based on the area and volume operators of dynamical surfaces $S$ and regions $R$. Without the background, the relation between $S$, $R$ and every $[\bar\Gamma]$ has to be individually specified. Since the goal of this thesis is to examine the semi-classical limit of loop quantum gravity, it is essential to describe locally the spatial geometry. To this end, we will need to define $S$ and $R$ in a physically meaningful way, so they truly define operators corresponding to variables in general relativity.

\subsubsection{Hamiltonian Constraint and Physical Hilbert Space}

The remaining constraint to be imposed to achieve the physical Hilbert space is the Hamiltonian constraint. Adhering to the quantum geometry of space, the Hamiltonian constraint operator $\hat{H}_g(\bar N)$ in loop quantum gravity, with lapse function $\bar N$, is built with a discrete structure. The operator is defined as the limit $\lim_{\epsilon\to 0}\hat{H}_g^{\epsilon}(\bar N)$, while $\hat{H}_g^{\epsilon}(\bar N)$ is the quantized, regularized Hamiltonian constraint acting on $\mathcal S$. The full construction procedure is highly technical, for details please refer to \cite{intro1}\cite{intro}\cite{perez}.

The action of $\hat{H}_g^{\epsilon}(\bar N)$ on a spin network state $| S_{\bar{\Gamma}}\rangle$ is given by a combination of a specific set of flux and holonomy operators, whose embedded surfaces and paths intersect with $\bar\gamma$ in a particular way. The fiducial sizes of the surfaces and paths are $\epsilon^2$ and $\epsilon$, and the expression of $\hat{H}_g^{\epsilon}(\bar N)$ in terms of the corresponding flux and holonomy operators gives a discretized approximation for the classical constraint $H_g(\bar N)$, with $\epsilon$ indicating the fineness.

 The removal of the regulator is a subtle issue, for $\lim_{\epsilon\to 0}\hat{H}_g^{\epsilon}(\bar N)$ does not exist in $\mathcal S$. The reason is that the embedded surfaces and paths for the flux and holonomy operators in $\hat{H}_g^{\epsilon}(\bar N)$ shrink with $\epsilon$. Therefore the embedded colored-graphs for the resulting states also change with $\epsilon$. Since any two states in $\mathcal S$ based on distinct embedded colored-graphs have a zero inner product, $\hat{H}_g^{\epsilon_1}(\bar N) |S_{\bar{\Gamma}}\rangle $ will be orthogonal to $\hat{H}_g^{\epsilon_2}(\bar N) |S_{\bar{\Gamma}}\rangle $ for $\epsilon_2<\epsilon_1$, no matter how close the two values are. However, the two states $\hat{H}_g^{\epsilon_1}(\bar N) |S_{\bar{\Gamma}}\rangle $ and $\hat{H}_g^{\epsilon_2}(\bar N) |S_{\bar{\Gamma}}\rangle$ do belong to the same \emph{diffeomorphism class}, provided $\epsilon_1$ and $\epsilon_2$ are both sufficiently close to zero. That means $\langle s_{[\bar{\Gamma}']}|\lim_{\epsilon\to 0}\hat{H}_g^{\epsilon}(\bar N)|S_{\bar{\Gamma}}\rangle$ for any $S_{\bar{\Gamma}}$ does exist, and thus $\lim_{\epsilon\to 0}\hat{H}_g^{\epsilon}(\bar N)$ converges acting upon $K \subset \mathcal S^*$.

Unlike the cases of Gauss and momentum constraints, there is a large degree of ambiguity in quantizing the Hamiltonian constraint. This is due to the many ways one can regularize the same classical constraint using flux and holonomy variables. Moreover, the different versions of Hamiltonian constraint operators have been shown to be physically distinct. While the quantization ambiguity remains an issue in loop quantum gravity, in this paper we will build our model based on the most standard version described in \cite{intro1}\cite{intro}\cite{perez}, and will involve only the most general features shared by all the other possibilities. To describe the action of $\lim_{\epsilon\to 0}\hat{H}_g^{\epsilon}(\bar N)= \hat{H}_g(\bar N)$ upon $K$, we introduce the following notations (fig.2). For each embedded graph $\bar\gamma$, we label each node with an integer $n$, and each of the edges connected to the node $n$ by an integer pair $(n,i)$.\footnote{Note that each edge has two labels since it contains two nodes.} For $\bar\gamma$, we denote the node $n$ by $\bar v_n^{\bar\gamma}$, and the oriented path starting from $\bar v_n^{\bar\gamma}$ and overlapping exactly with the edge $(n,i)$ by $\bar e_{n,i}^{\bar\gamma}$. Since the number of edges connected to a node varies depending on the node, the range of $i$ depends on $n$. Further, to each pair $(\bar v_n^{\bar\gamma},\bar e_{n,i}^{\bar\gamma})$ we assign an oriented path $\bar e_{(n,i)}^{\bar\gamma}$ that starts from $\bar v_n^{\bar\gamma}$ and lies in $\bar e_{n,i}^{\bar\gamma}$ without covering the other end point. Lastly, to each pair $(\bar e_{n,i}^{\bar\gamma},\bar e_{n,j}^{\bar\gamma})$ we assign an oriented closed path $\bar e_{(n,i,j)}^{\bar\gamma}$ containing the outgoing path $\bar e_{(n,i)}^{\bar\gamma}$, incoming path $(\bar e_{(n,j)}^{\bar\gamma })^{-1}$, and another smooth bridge path, such that it gives the boundary of an embedded surface who has no other intersection with $\bar\gamma$.
\begin{figure}
\begin{center}
\includegraphics[angle=0,width=1.6in,clip=true]{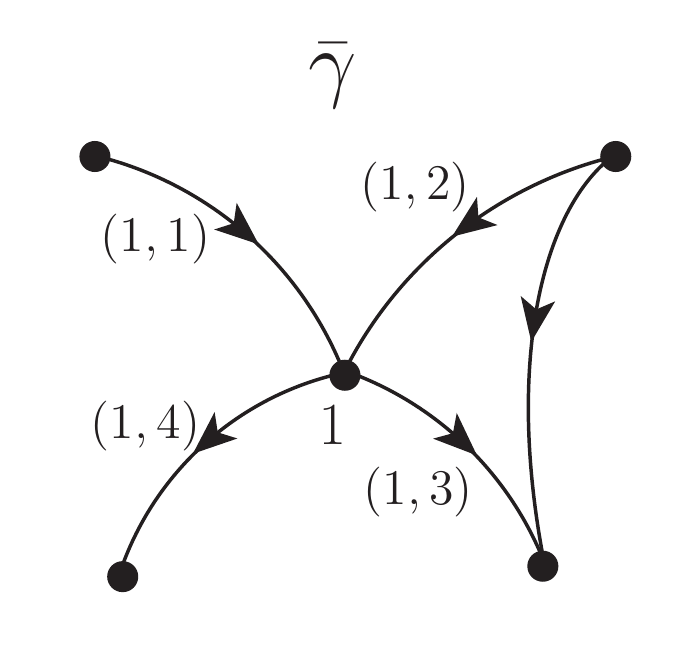}
\includegraphics[angle=0,width=4in,clip=true]{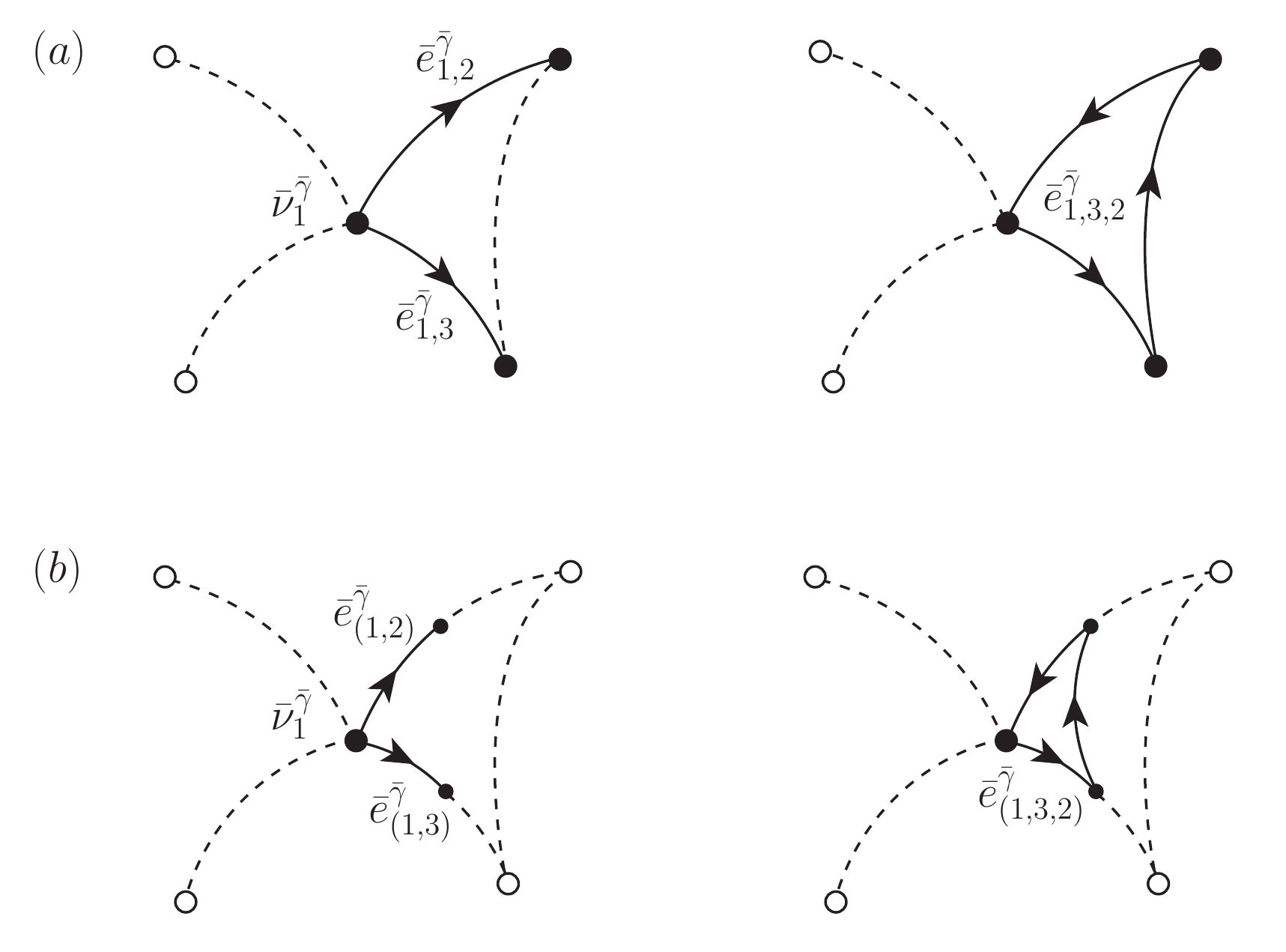}
\begin{spacing}{1}
\caption{The left figure pictures an embedded graph $\bar\gamma$, with one of its nodes labeled by $n=1$ and the four edges connected to this node labeled by $(1,j)$. The figures in $(a)$ demonstrate the corresponding definitions of $\bar v_n^{\bar\gamma}$, $\bar{e}^{\bar\gamma}_{n,j}$ and $\bar{e}^{\bar\gamma}_{n,i,j}$, while the figures in $(b)$ demonstrate the corresponding definitions of $\bar{e}^{\bar\gamma}_{(n,j)}$ and $\bar{e}^{\bar\gamma}_{(n,i,j)}$.}
\end{spacing}
\end{center}
\end{figure}
Using these notations and setting $\kappa \equiv 8\pi G/c^3$, the action of $\hat{H}_g(\bar N)$ upon $K$ is given by $\langle s_{[\bar\Gamma]}|\hat{H}_g(\bar N)|S_{\bar\Gamma'}\rangle$ for any $ s_{[\bar\Gamma]}$ and $S_{\bar\Gamma'}$, with \cite{intro1}\cite{intro}\cite{perez}
\begin{equation}
\begin{split}
\langle s_{[\bar\Gamma]}| \hat{H}_{g}(\bar N)|S_{\bar\Gamma'}\rangle\rule{364pt}{0pt}\\
\equiv \langle s_{[\bar\Gamma]}| \sum_{\bar v_n^{\bar\gamma'}\in \bar\gamma'}\hat{H}_{g}^{(\bar v_n^{\bar\gamma'})}\bar N(\bar v_n^{\bar\gamma'})|S_{\bar\Gamma'}\rangle\rule{312pt}{0pt}\\
\equiv \langle s_{[\bar\Gamma]}|\hat{H}^{E}_{g}(\bar N)|S_{\bar\Gamma'}\rangle  -  \langle s_{[\bar\Gamma]}|(i\hbar\kappa\gamma)^{-5}\frac{1}{48}\frac{(1+\gamma^2)}{2} \sum_{\bar v_n^{\bar\gamma'}\in \bar\gamma'} \bar N(\bar{ v}_n^{\bar\gamma'}) \sum_{i,j,k=1} \text{sgn}\left(\bar e_{(n,i)}^{\bar\gamma'},\bar e_{(n,j)}^{\bar\gamma'},\bar e_{(n,k)}^{\bar\gamma'}\right)\rule{19pt}{0pt}\\ \times\left(\hat{h}^{-1} (\bar e_{(n,i)}^{\bar\gamma'})\right)_{\bar{i}}^{\bar{l}} \left[\left(\hat{h}(\bar e_{(n,i)}^{\bar\gamma'})\right)_{\bar{l}}^{\bar{j}},\left[\hat{H}^{E}_{g}(1),\hat{V}\right] \right]\left(\hat{h}^{-1}( \bar e_{(n,j)}^{\bar\gamma'})\right)_{\bar{j}}^{\bar{p}}\left[\left(\hat{h}(\bar e_{(n,j)}^{\bar\gamma'}\right)_{\bar{p}}^{\bar{k}},\left[\hat{H}^{E}_{g}(1),\hat{V}\right] \right]\rule{20pt}{0pt} \\ \times\left(\hat{h}^{-1} (\bar e_{(n,k)}^{\bar\gamma'}\right)_{\bar{k}}^{\bar{q}}\left[\left(\hat{h}(\bar e_{(n,k)}^{\bar\gamma'}\right)_{\bar{q}}^{\bar{i}},\hat{V}\right]|S_{\bar\Gamma'}\rangle\rule{255pt}{0pt}\\\\
\end{split}
\end{equation}
where $\hat{h}(\bar e) \equiv\hat{h}^{(\frac{1}{2})}(\bar e) $ is the holonomy operator with spin $1/2$, and the Immirzi parameter $\gamma$ is set to be any real number.\footnote{ Notice that part of the quantization ambiguity lies in the possible choices of the spin representation for the holonomy operators in $(2.28)$ through $(2.30)$, which we pick to be $\frac{1}{2}$.} Also, the action of $\hat{H}^{E}_{g}(\bar N)$ on any cylindrical function $\psi_{\bar\gamma,f}$ is given by \cite{intro1}\cite{intro}\cite{perez}
  \begin{equation}
\begin{split}
\hat{H}^{E}_{g}(\bar N)|\psi_{\bar\gamma,f}\rangle\rule{342pt}{0pt}\\
\equiv (i\hbar\kappa\gamma)^{-1}\frac{1}{96}   \sum_{\bar v_n^{\bar\gamma}\in \bar\gamma}\bar N(\bar{ v}_n^{\bar\gamma})\sum_{i,j,k=1} \text{sgn}\left(\bar e_{(n,i)}^{\bar\gamma},\bar e_{(n,j)}^{\bar\gamma},\bar e_{(n,k)}^{\bar\gamma}\right)
       \left( 
      \hat{ h}(\bar e_{(n,i,j)}^{\bar\gamma}) -\hat{ h}^{-1}(\bar e_{(n,i,j)}^{\bar\gamma})
      \right) ^{\bar{i}}_{\bar{j}}\rule{25pt}{0pt} \\ \times\left(\hat{h}^{-1} (\bar e_{(n,k)}^{\bar\gamma})\right)_{\bar{i}}^{\bar{l}} \left[\left(\hat{h}(\bar e_{(n,k)}^{\bar\gamma})\right)_{\bar{l}}^{\bar{j}},\hat{V} \right]|\psi_{\bar\gamma,f}\rangle\rule{220pt}{0pt}
\end{split}
   \end{equation}
where the total volume operator $\hat{V}$ is defined as
\begin{equation}
\begin{split}
 \hat{V}| \psi_{\bar\gamma,f}\rangle\rule{340pt}{0pt}\\
\equiv\sum_{\bar v_n^{\bar\gamma}\in \bar\gamma} \left[ \frac{1}{48} \sum_{i,j,k=1} \text{sgn}\left(\bar e_{(n,i)}^{\bar\gamma},\bar e_{(n,j)}^{\bar\gamma},\bar e_{(n,k)}^{\bar\gamma}\right)  \epsilon^{pqr}  \hat{J}_{p}(\bar e_{n,i}^{\bar\gamma})  \hat{J}_{q}(\bar e_{n,j}^{\bar\gamma}) \hat{J}_{r}(\bar e_{n,k}^{\bar\gamma})\right ]^{\frac{1}{2}}| \psi_{\bar\gamma,f}\rangle \rule{25pt}{0pt}
\end{split}
\end{equation}

Notice that the embedding details for the assignments of $\bar e_{(n,i)}^{\bar\gamma}$ and $\bar e_{(n,i,j)}^{\bar\gamma}$ for any of the $\bar\gamma$ are irrelevant for $(2.29)$, since $\langle s_{[\bar\Gamma]}|$ does not distinguish deformations within the same $\mathit{diff_M}$ class. It is indeed remarkable that $\lim_{\epsilon\to 0}\hat{H}_g^{\epsilon}(\bar N)$ converges upon $\mathit{diff_M}$ invariant states. This fact allows us to impose the Hamiltonian constraint upon knot space. However, $\hat{H}_g(\bar N)$ in general does not preserve $K$, for the same reason  why the operators $\hat A_{\bar S}$ and $\hat V_{\bar R}$ fail to do so. Recalled that $\hat A_{\bar S}$ and $\hat V_{\bar R}$ violate $\mathit{diff_M}$ symmetry because they are smeared over $\bar S$ and $\bar R$, which are not $\mathit{diff_M}$ invariant. Similarly, $\langle s_{[\bar\Gamma]}|\hat{H}_g(\bar N)|S_{\bar\Gamma'}\rangle$ would not be equal to $\langle s_{[\bar\Gamma]}|\hat{H}_g(\bar N)|S_{\mu\bar\Gamma'}\rangle$ if $ \bar N(\mu\bar{ v}_n^{\bar\gamma'})$ differs from $\bar N(\bar{ v}_n^{\bar\gamma'})$. On the other hand, the only reason for the violation is that $\bar N(\text x)$ is generally not $\mathit{diff_M}$ invariant. In fact, one can check that we have $\langle s_{[\bar\Gamma']}| \hat{H}_{g}^{(\bar v_n^{\bar\gamma})}|S_{\bar\Gamma}\rangle=\langle s_{[\bar\Gamma']}| \hat{H}_{g}^{(\mu\bar v_n^{\bar\gamma})}|S_{\mu\bar\Gamma}\rangle$, and thus $\hat{H}_g(1)$ is an operator in $K$.

The physical Hilbert space for loop quantum gravity should consist of quantum states annihilated by the Hamiltonian constraint operator with any lapse function, whose complexity is demonstrated above. Currently, the search of such states remains a pressing challenge for loop quantum gravity, and is tackled with a great variety of approaches \cite{intro1}\cite{perez}\cite{group1}\cite{ave2}. Many proposals are pursued using the Hamiltonian constraint operators with different variations of $(2.29)$. Aside from the canonical Hamiltonian approaches , solving the constraint using path integral formalism in spin foam models also provides important insights from a different point of view \cite{perez}.  In our model, we will modify the Hamiltonian operator into a graph preserving operator in $K$. The much simplified setting will allow us to apply group averaging method, based on certain concrete assumptions, to construct the physical Hilbert space for the model.

\subsection{Loop Quantization with Matter Fields}

The previous sections outlined loop quantization of gravitational fields, which is guided by the background independence of general relativity.  We saw that the procedure leads to the knot space $K$ whose elements are drastically different from Fock states in particle theories. 
We encountered difficulty of describing local geometry in $K$. Tracing the causes of these obstacles, we were led to the classical problem of finding local observables in general relativity. However, in the quantum theory the problem is further sharpened by the need to describe the superposition of background independent states.

In general relativity, the internal coordinates approach to define local observables has been extensively studied \cite{kuchar}\cite{kuchartorre}\cite{torre}\cite{rovelli}\cite{torre1}. The approach attempts to describe the gravitational fields \emph{relatively to} coordinates given by values of dynamical scalar fields. The scalar fields can be a part of the gravitational field or a part of matter fields coupling to the gravitational field. In this setting, background independence is respected because both the gravitational field and the coordinate fields are dynamical entities in the full theory. Indeed, the approach is mere reflection of the daily life reality: we can only measure one physical system by referring to another. However, the precise mathematical prescription for the local observables defined by internal coordinates is subtle, and often involves the intrinsic properties of spacetime. Among the abundant works in this approach, different fields have been used as internal coordinates to describe gravity. The most popular method in cosmology introduces four dust-matter scalar fields coupling to the gravitational field, with one of the fields identified as the clock and the other three being the spatial coordinates \cite{kuchar}\cite{kuchartorre}\cite{torre}. In this method, the scalar fields' degrees of freedom are absorbed to dynamically break diffeomorphism symmetry, resulting in gravitational physical degrees of freedom that live on each matter coordinate point given by the scalar fields. Consequently, when the matter back reactions are ignored, one expects to obtain gauge fixed general relativity from those gravitational local observables.

In the hope of recovering full general relativity in semi-classical limit of our model, we will introduce the dust-matter method in the loop representation. Loop quantization for a system of gravity coupling to general matter fields is itself an extensive body of work \cite{matter1}\cite{matter2}\cite{thiemann}. It has been shown rigorously that loop quantization is applicable to such general systems, and the quantization leads to a knot space including the matter degrees of freedom. In the loop representation, the matter sector of knot states appears extremely different from Fock states in particle theories.  Beyond the Planck scale, a much studied speculation is that the non-perturbative prescriptions of the knot states would be approximated by perturbative quantum field theory on a smooth spacetime background. In our context,  we will simply focus on the basic setting of knot states including matter fields, and also the matter operators acting on such states.

\subsubsection{ Classical Theory with Matter Fields}

The system we consider in this thesis consists of the gravitational field, gauge fields, fermion fields and scalar fields. The matter sector has the usual interactions mediated by gauge fields, and is described by Yang-Mills theory minimally coupled to fermion and scalar fields. Moreover, the gravity and matter sectors are fully coupled, so the theory for the whole system is diffeomorphism invariant. As mentioned previously, Yang-Mills theory shares many critical features with general relativity expressed in Ashtekar formalism. Such features enable a single representation of knot space for the whole system. To facilitate loop quantization, the matter sector will also employ a triad formalism in the classical theory.

 Aside from diffeomorphism symmetry and local $SU(2)$ symmetry from gravitational coupling, the system has additional local gauge symmetry of group $\mathcal G$ due to the matter interactions. In the triad formalism we use half-densitized, Grassmann-valued spinor fields $\xi^{\bar i}_{\bar{\text i}}(\text{x})$ and their conjugate momenta $\pi_{\bar i}^{\bar{\text i}}(\text{x})$ to describe the fermions in the system, where $\bar i$  and ${\bar{\text i}}$ are respectively the (spin $ \frac{1}{2}$) $SU(2)$ and $\mathcal G$ indices. The real scalar fields and their momenta will be described by $\phi _{\text i}(\text{x})$ and $P^{\text i}(\text{x})$ . The matter gauge fields will be described by $\underbar{A}^{\text i}_a(\text{x})$, where ${\text i}$ and $a$ are respectively the adjoint $\mathcal G$  and spatial coordinate indices. Finally, the magnetic fields corresponding to $\underbar{A}^{\text i}_a(\text{x})$ will be denoted as $\underbar{B}^{\text i}_a(\text{x})$. Written in the terms of the specified fields, the Hamiltonian constraint density for the matter sector contains the terms \cite{thiemann}
\begin{equation}
\begin{split}
\\
H_f(\text{x})\equiv \frac{ E^a_i}{2 \sqrt{ \det E}}\left[i \pi^T \tau^i D_a \xi + D_a(\pi ^T \tau_i \xi) + \frac{i}{2} K^i_a  \pi^T \xi +c.c.\right](\text{x})\rule{85pt}{0pt}
\\\\
H_s(\text{x})\equiv \frac{P ^{\text i} P_{\text i}}{2 \kappa \sqrt{ \det E}}(\text{x}) + \frac{1}{2}\left[\frac{ \eta^{jk} E_j^a E_k^b}{\sqrt{ \det E}} (D_a \phi^{\text i})(D_b \phi_{\text i})/\kappa +  \sqrt{ \det E} V(\phi_{\text i} \phi^{\text i})/ (\hbar \kappa^2)\right](\text{x})
\\\\
H_{YM}(\text{x}) \equiv \frac{\eta^{jk} E_{aj} E_{bk}}{2 Q^2 \det E} \left[\underbar{E}^{a\text i}\underbar{E}^b_{\text i} +\underbar{B}^{a\text i} \underbar{B}^b_{\text i}\right](\text{x})\rule{170pt}{0pt}
\end{split} 
\end{equation}
Here $\kappa \equiv 8\pi G/c^3$, $Q$ is the Yang-Mills coupling constant, and the covariant differentiation $D_a$ is taken with respect to the total connection field $A_a +\underbar{A}_a$. Also, $\eta_{jk}$ are the spatial components of the flat Lorenzian metric with signature $(-,+,+,+)$. The generalized magnetic field $\underbar{B}^{\text i}_a$ is defined as:
\begin{equation}
\underbar{B}^{\text i}_a(\text{x})\equiv \frac{1}{2}{\epsilon_{a}}^{bc}(\partial_{b} \underbar{A}^{\text i}_c-\partial_{c} \underbar{A}^{\text i}_b + {f^{\text i}}_{ \text j \text k} \underbar{A}^{\text j}_b \underbar{A}^{\text k}_c)(\text{x})
\nonumber
\end{equation}
where ${f^{\text i}}_{ \text j \text k}$ is the structure constant of $\mathcal G$.
Including the gravitational part, the full Hamiltonian constraint for the system, with a lapse function $\bar N$, is given by the sum of all the sectors $H(\bar N)= H_g(\bar N)+H_f(\bar N)+H_s(\bar N) +H_{YM}(\bar N)$. Here the matte terms are obtained from smearing $(2.34)$ with $\bar N$. 

Aside from the Hamiltonian constraint, the $SU(2)$ Gauss constraint and momentum constraint also acquire new terms because of the inclusion of the matter fields \cite{matter2}\cite{matter1}, although we refrain from listing them in this thesis. Further, there is also an additional Gauss constraint coming from the gauge symmetry of the group $\mathcal G$. The four constraints thus govern the background independent theory of the whole system.

\subsubsection{ $\mathcal S$ and $K$ with Matter Fields}
The classical expressions above suggest that we may generalize the holonomies of $SU(2)$ group to those of $SU(2) \times G$ to incorporate the gauge fields. On the other hand, since the generalized holonomies parallel transport $\xi^{\bar i}_{\bar{\text i}}(\text{x})$ and $\phi_{\text i}(\text{x})$, the fermion and scalar fields should play roles similar to intertwiners in the loop representation. The loop quantization procedure does apply to the classical theory in this way \cite{matter1}\cite{matter2}\cite{thiemann}. The result is the generalized knot space $K$ that solves all but Hamiltonian constraints. Here we will give a brief description of the knot states in $K$.

A spin network state for the system is a wave functional of the configuration fields, constructed from the loop variables $h^{(j)}(\bar{e})$, $\text{h}^{(\text{i})}(\bar{e})$, $\theta^{(d)}(\bar{v})$ and $h^{( \text{k})}(\bar{v})$. Here $\text{h}^{( \text{i})}(\bar{e})$ denotes a $\mathcal G$ holonomy in the assigned representation $ \text{i}$. The scalar fields are described by $h^{(\text{k})}(x) \equiv \exp(\phi (x)^{\text{j}} {\tau^{(\text{k})}}_{\text{j}})$, which are called $point$ $holonomies$ in their representation $\text{k}$. The fermion fields are described by the irreducible tensors $\theta^{(d)}(\bar{v})$, each given by a Grassmann monomial of $\xi^{\bar i}_{\bar{\text i}}(\bar{v})$ of degree $d$ (for details refer to \cite{thiemann}). Constructed from the above variables, each of the spin network states is specified by \cite{matter1}\cite{matter2}\cite{thiemann}: 1)  an embedded graph\footnote{Because of the presence of the matter fields, the graphs for the gauge invariant states no longer have to be closed or connected.} $\bar{\gamma}$; 2) an $SU(2)$ spin representation $j_{i}$ and a $\mathcal{G}$ group representation $\text{j}_i$ assigned to each edge;  3) two generalized Clebsch-Gordan coefficients (intertwiners) $i_{n}$ and $\text{i}_n$ for $SU(2)$ and $\mathcal{G}$ respectively, another $\mathcal{G}$ group representation $\text{k}_n$, and a degree $d_n$ of the Grassmann monomials assigned to each node. Analogous to the pure gravitational case, the wave functional forms of the spin network states are defined with cylindrical functions as
\begin{equation}
\begin{split}
S_{\bar{\gamma},( j_i, i_n, d_n), ( \text{j}_i,\text{i}_n, \text{k}_n )}\equiv \psi_{\bar{\gamma},f_{( j_i, i_n,d _n), ( \text{j}_i,\text{i}_n,  \text{k}_n)}}\rule{190pt}{0pt}\\\\
\psi_{\bar{\gamma},f_{( j_i, i_n,d _n), ( \text{j}_i,\text{i}_n,  \text{k}_n)}}[A,\underbar{A}, \theta, \phi]\rule{262pt}{0pt}\\
\equiv Inv\left\{\bigotimes_n^{N_v} i_n \bigotimes_n^{N_v}\text{i}_n\bigotimes_n^{N_v}\theta^{(d_n)} (\bar{v}_n)\bigotimes_n^{N_v} h^{(\text{k}_n)}(\bar{v}_n) \bigotimes_i^{N_e}{h}^{(j_i)}(\bar{e}_i)  \bigotimes_i^{N_e}\text{h}^{(\text j_i)}(\bar{e}_i)\right\}[A,\underbar{A}, \theta, \phi]\\\\
\end{split}
\end{equation}
where $Inv\{ ...\}$ denotes the $SU(2) \otimes\mathcal{G}$ invariant contraction. Thus by construction, the spin network states are locally $SU(2) \otimes \mathcal{G}$ invariant. The inner product of these states are defined similarly to the previous case, using Haar and  Berezin measures for the bosonic and fermionic sectors. Using such an inner product, all spin net work states $\{S_{\bar{\gamma},( j_i, i_n, d_n), ( \text{j}_i,\text{i}_n, \text{k}_n )}\}$ form a (non-orthogonal) basis for the space of the wave functionals satisfying both of the Gauss constraints. One can construct an orthonormal basis for $\mathcal S$, whose each element is a linear combination of the spin network states differ \emph{only} in fermionic sector. Therefore, an orthonormal basis is given by $\{S_{\bar{\gamma},( j_i, i_n, k_n), ( \text{j}_i,\text{i}_n, \text{k}_n )}\equiv S_{\bar\Gamma}\}$, obtained from recombining $d_n$  into $k_n$. Again, a knot state is obtained by group averaging the states in $\mathcal S$ over diffeomorphism transformations
\begin{equation}
\begin{split}
 \langle S_{\bar \Gamma}| \hat{\mathbb P}_{diff}\equiv \langle s_{[\bar \Gamma]}|\\
\end{split}
\end{equation}

The set of knot states $\{ s_{[\bar \Gamma]}\}$ forms a orthonormal basis for the knot space $K$ for the system.  By construction, $K$ solves the $SU(2)$ Gauss constraint, $\mathcal G$ Gauss constraint, and momentum constraint.

\subsubsection{ Matter Operators in $\mathcal{S}$ and $K$}

In addition to the gravitational flux and holonomy operators given previously, the operators in $\mathcal S$ also include matter operators that act on the spin network states. Similar to the gravitational operators, the matter operators are also smeared with test functions $\text g$, $\text f$, $g$ and $\bar g$ with appropriate group values. From the gauge fields, we have the holonomy operators $\hat{\text{h}}^{(\text j)}({\bar e})_{\text g}$, and the corresponding operators $ \hat{\text{J}}_{\text f}(\bar e)$ that give the flux of the gauge fields analogously to $(2.19)$. From the fermionic monomials of degree one, we have the spinor operators $\hat{\theta}(\bar p)_{g}\equiv\hat{\theta}^{(1)}(\bar p)_{g}$ and their conjugate momenta operators $\hat{\eta}(\bar p)_{\bar g}\equiv i\hat{\theta}^{\dagger}(\bar p)_{\bar g}$. Finally, from the scalar fields we have the point holonomy operators $\hat{h}^{(\text{i})}(\bar p)_{\text g}$ and the conjugate  momenta operators  $\hat{p}_{\text f}(\bar p)$. 

While the physics in the matter sector under loop quantization is an important subject discussed in other works (see, for example, \cite{qft}), the goal of this paper is to explore the semi-classical limit of the \emph{gravity} sector, involving the matter sector only to provide coordinates. Nevertheless, the framework of the matter sector in loop representation is essential for obtaining a faithful extraction of the full theory, leaving out only the matter back reactions. The important feature of the matter sector for our purpose, is that the fermion and scalar operators act on the nodes. 

Four matter scalar fields are required to coordinatize spacetime in the classical theory. Moreover, the triads formalism introduces additional $SU(2)$ components of the gravitational field that need to be described in terms of matter frames. In the quantum theory, such coordinates and frames would be provided by matter operators. Since loop quantum gravity is based on flux and holonomy variables, the required coordinates and frame operators are: 1) four scalar operators for the spacetime coordinates; 2) three $SU(2)$ vector operators and their conjugates, for the description of flux variables; 3) two $SU(2)$ spinor operators and their conjugates for the description of holonomy variables.

Clearly, the requirement can be fulfilled provided enough species of fermion and scalar fields exist. To obtain the coordinate and frame operators in $K$, we will define them with a generic dynamical point $p\equiv \bar p(\bar\Gamma)$ following the notion in section $2.2.5$. From the ingredients provided by the fundamental matter operators listed above, one can easily construct chargeless scalar operators of the form $\hat{\phi}(p)$, chargeless vector operators of the form $\hat{J}(p)^{i}$ and their conjugates, and chargeless spinor operators of the form $\hat{U}(p)^{\bar{j}}$ and their conjugates. For an explicit example, we can set:
\begin{equation}
\begin{split}
\hat{\phi}(p)\equiv \text{g}^{\text{i}\text{j}}\hat{p}_{\text i}\hat{p}_{\text j}( p)\rule{210pt}{0pt}\\\\
\hat{J}(p)^{k} \equiv \hat{\eta}_{\bar j}^{\bar{\text i}}(\tau^k)_{\bar i}^{\bar{j}}\hat {\theta}
^{\bar i}_{\bar{\text i}}(p)\rule{1pt}{0pt};\rule{2pt}{0pt} \hat{\bar J}(p)_{k} \equiv \hat{\eta}_{\bar j}^{\bar{\text i}\dagger}(\tau_k)_{\bar i}^{\bar{j}}\hat {\theta}^{\bar i \dagger}_{\bar{\text i}}(p)\rule{2pt}{0pt};\rule{4pt}{0pt} 
\hat{U}(p)^{\bar{k}}\equiv ( C ^{\bar{\text i},\bar{\text j}}_{\bar i,\bar j})^{\bar{k}}\hat {\theta}^{\bar i}_{\bar{\text i}}\hat {\theta}^{\bar j}_{\bar{\text j}} (p)\rule{1pt}{0pt};\rule{2pt}{0pt} \hat{\bar U}(p)_{\bar{k}}\equiv ( C ^{\bar{\text i},\bar{\text j}}_{\bar i,\bar j})_{\bar{k}}\hat {\theta}^{\bar i \dagger}_{\bar{\text i}}\hat {\theta}^{\bar j \dagger}_{\bar{\text j}} (p)
\nonumber
\end{split}
\end{equation}
Here $\text{g}^{\text{i}\text{j}}$ is the adjoint-$\mathcal G$ Killing-Cartan metric, and $( C ^{\bar{\text i},\bar{\text j}}_{\bar i,\bar j})^{\bar{k}}=( C ^{\bar{\text i},\bar{\text j}}_{\bar i,\bar j})_{\bar{k}}$ is the Clebsch-Gordan coefficient that projects the direct spinor products into a $\mathcal G$ scalar and $SU(2)$ spinor representation. It should be clear that there are many other ways to obtain the chargeless scalar and current operators, as long as the results have the the desired transformation properties. As mentioned in $2.2.5$, there is excessive freedom in the definition of $p$ due to $\mathit{diff_M}$ symmetry.   However, as we will describe in detail, the symmetry between different choices of $p$ would be dynamically broken by the quantum state of the system through the coordinate and frame operators, in an analogy with the classical case.

Lastly, the matter Hamiltonian constraint operator $\hat{H}_m(\bar N)\equiv\hat{H}_f(\bar N)+\hat{H}_s(\bar N)+\hat{H}_{YM}(\bar N)$  has been also constructed in the loop representation \cite{thiemann}. Similar to the construction of $\hat H_g(\bar N)$, the quantization of $H_m(\bar N)$ starts from the regularization of $(2.32)$ using the matter loop variables, and results in the well-defined operator $\hat{H}_m(\bar N)$ acting upon $K$ after the removal of the regulator. Following the notation used in $(2.31)$-$(2.33)$ the action of $\hat{H}_m(\bar N)$ on a knot state  has the form
\begin{equation}
\begin{split}
  \langle s_{[\bar \Gamma]}|\hat{H}_{m}(\bar N)|S_{\bar \Gamma'}\rangle
\equiv  \langle s_{[\bar\Gamma]}| \sum_{\bar v_n^{\bar\gamma'}\in \bar\gamma'}\hat{H}_{m}^{(\bar v_n^{\bar\gamma'})}\bar N(\bar v_n^{\bar\gamma'})|S_{\bar\Gamma'}\rangle\\
\end{split}
\end{equation}
where the implicit form of $\hat{H}_{m}^{(\bar v_n^{\bar\gamma'})}$ is given by \cite{thiemann}
\begin{equation}
\begin{split}
 \langle s_{[\bar\Gamma]}|\hat{H}_{m}^{(\bar v_n^{\bar\gamma'})}|S_{\bar\Gamma'}\rangle\rule{338pt}{0pt}\\
\equiv \langle s_{[\bar \Gamma']}|\hat{H}_{m}^{(\bar v_n^{\bar\gamma})}\left(\hat{J}_{i}(e_{n,i}^{\bar\gamma}),\hat{h}(e_{(n,i)}^{\bar\gamma})_{\bar{l}}^{\bar{j}},\hat{h}( \bar e_{(n,i,j)}^{\bar\gamma})_{\bar{l}}^{\bar{j}},\hat{\text{J}}_{\text i}(\bar e_{n,i}^{\bar\gamma}), \hat{\text{h}}^{(\text j)}(\bar e_{(n,i)}^{\bar\gamma})^{\bar{\text i}}_{\bar{\text j}},\hat{\text{h}}^{(\text j)}( \bar e_{(n,i,j)}^{\bar\gamma})^{\bar{\text i}}_{\bar{\text j}},\rule{50pt}{0pt}\right.\\ \left.\hat {\theta} (\bar v_n^{\bar\gamma})^{\bar i}_{\bar{\text i}}, \hat {\theta} (\bar e_{(n,i)(1)}^{\bar\gamma})^{\bar i}_{\bar{\text i}}, \hat{\eta} (\bar v_n^{\bar\gamma})_{\bar i}^{\bar{\text i}},\hat{\eta} (\bar e_{(n,i)(1)}^{\bar\gamma})_{\bar i}^{\bar{\text i}}, \hat{h}^{(\text{i})}(\bar v_n^{\bar\gamma})^{\bar{\text i}}_{\bar{\text j}}, \hat{h}^{(\text{i})}(\bar e_{(n,i)(1)}^{\bar\gamma})^{\bar{\text i}}_{\bar{\text j}},\hat{p}_{\text i}(\bar v_n^{\bar\gamma}),\hat{p}_{\text i}(\bar e_{(n,i)(1)}^{\bar\gamma}) \right)
|S_{\bar \Gamma}\rangle
\\   \\                                                
\end{split}
\end{equation}
where $\bar e_{(1)}$ is again the target point of $\bar e$. We will skip the detailed expression is  in this paper. Nevertheless, from the operators involved, we know that the action of $\hat{H}_{m}^{(\bar v_n^{\bar\gamma})}$ changes the graph and representations (of both $SU(2)$ and $\mathcal{G}$) of the state. Just as the case  of $\hat H_g(\bar N)$, we have $\langle s_{[\bar\Gamma']}| \hat{H}_{m}^{(\bar v_n^{\bar\gamma})}|S_{\bar\Gamma}\rangle=\langle s_{[\bar\Gamma']}| \hat{H}_{m}^{(\mu\bar v_n^{\bar\gamma})}|S_{\mu\bar\Gamma}\rangle$. Therefore, $\hat{H}_{m}(\bar N)$ does not preserve $K$ only because $\bar N$ is generally not $\mathit{diff_M}$ invariant.

\section{The Graph-Preserving  Model}
We start with the Hamiltonian constraint operator for our model, which is modified from the standard self-adjoint form 
\begin{equation}
\hat H(\bar N)\equiv\frac{1}{2}(\hat H_g(\bar N)+\hat H_m(\bar N))+\frac{1}{2}(\hat H_g(\bar N)+\hat H_m(\bar N))^{\dagger}
\nonumber
\end{equation}
where each term is described in $(2.30)$-$(2.32)$ and $(2.36)$-$(2.37)$. Recall that the standard constraint operator does not preserve $K$ due to the fact that $\bar N(\text x)$ is not $\mathit{diff_M}$ invariant. In our model, we will replace the embedded lapse function $\bar N(\text x)$ with the dynamical lapse function $N_p(p_k)$ defined on a set of dynamical spatial points $p\equiv\{p_k\}$, where the integer $k$ ranges to infinity.\footnote{The spatial manifold $M$ contains uncountably many spatial points, but we use only countably infinite set $\{p_k\}$ in correspondence to the discrete structure of $K$.} Using the notion introduced in section $2.2.5$, the set of dynamical nodes $\{p_k\}$ consists of $p_k \equiv  \bar p_k(\bar\Gamma )$  satisfying $\bar p_k(\mu\bar\Gamma)=\mu' \mu \bar p_k(\bar \Gamma)$ for any $\mu \in \mathit{diff_M}$ and some $\mu' \in  \mathit{Tdiff_{M,{\mu\bar\gamma}}}$. Naturally we require that for each $\bar\Gamma$ and $n$, $\bar v_n^{\bar\gamma}=\bar p_k(\bar\Gamma )$ holds for exactly one $k$ value.\footnote{Note that this is obviously true had we taken $\{\bar p_k(\bar\Gamma)\}$ for any $\bar\Gamma$ to be the set of $all$ $points$ in $M$. Here we define $\{\bar p_k(\bar\Gamma)\}$ to be a countably infinite subset of $M$, but insist on maintaining the condition.} Notice that there are infinitely many distinct sets of dynamical spatial points satisfying the above, and we encountered exactly the same ambiguity resulting from the background independence in section $2.2.5$. Respecting the background independence, we will consider all lapse functions $N_p$ based on all possible $p$. The  Hamiltonian constraint $\hat{H}^{\star}(N_p)$ in $K$ is defined by
\begin{equation}
\begin{split}
\langle s_{[\bar\Gamma]}| \hat{H}^{\star}(N_p)
\equiv\langle s_{[\bar\Gamma]}| \left[\frac{1}{2}(\hat{H}^{\star}_g(N_p)+\hat{H}^{\star}_m(N_p))+\frac{1}{2}(\hat{H}^{\star}_g(N_p)+\hat{H}^{\star}_m(N_p))^{\dagger}\right]\\\\
\end{split}
\end{equation}
where we have
\begin{equation}
\begin{split}
\langle s_{[\bar\Gamma]}|(\hat{H}^{\star}_g(N_p)+\hat{H}^{\star}_m(N_p))
\equiv \langle s_{\bar\Gamma}| \sum_{\bar v_n^{\bar\gamma}\in \bar\gamma}(\hat{H}_{g}^{(\bar v_n^{\bar\gamma})} +\hat{H}_{m}^{(\bar v_n^{\bar\gamma})} )N_p(  p_k|_{ \bar p_k(\bar\Gamma )=\bar v_n^{\bar\gamma}})\hat{\mathbb P}_{diff}\\
\end{split}
\end{equation}
for any $\bar\Gamma \in [\bar\Gamma]$. Thus $\hat{H}^{\star}(N_p)$ is now an operator in knot space. It is important to note that the classical counterpart ${H}_{g}(N)$ of $\hat{H}_g^{\star}(N_p)$ is different from ${H}_{g}(\bar N)$, which is the counterpart of $\hat{H}_{g}(\bar N)$. In the classical theory, we would expect $\{{H}_{g}(N), M_g(\bar V)\}=0$, since $N$ is now dynamical and transforms as a scalar field under the diffeomorphisms generated by $M_g(\bar V)$. On the other hand, the algebra $(2.6)$ applies to ${H}_{g}(\bar N)$, with $\bar N$ being a non-dynamical Lagrangian multiplier.

Referring back to $(2.30)$-$(2.32)$ and $(2.36)$-$(2.37)$, we note that $\hat{H}^{\star}(N_p)$ is graph changing since $\hat{H}_g^{(\bar v_n^{\bar\gamma})}$ and $\hat{H}_m^{(\bar v_n^{\bar\gamma})}$ contains the holonomy operators based on $\bar e_{(n,i,j)}^{\bar\gamma}$ that is not contained in $\bar\gamma$. To obtain a graph preserving Hamiltonian constraint operator for the model, we now replace $\bar e_{(n,j)}^{\bar\gamma }$ with the full path $\bar e_{n,j}^{\bar\gamma }$ that coincides with the edge (see fig.2). Also we replace $\bar e_{(n,i,j)}^{\bar\gamma}$ with $\bar e_{n,i,j}^{\bar\gamma}$, a closed oriented path covered by $\bar\gamma$, with the outgoing $\bar e_{n,i}^{\bar\gamma }$ and incoming $\bar e_{n,j}^{\bar\gamma }$, which contains a minimal number of edges of $\bar\gamma$.\footnote{ If such a path is not unique, then $e_{n,i,j}^{\bar\gamma}$ represent averaging over all the minimal closed paths that satisfy the requirements.} Applying the substitutions to $\hat{H}^{\star}(N_p)$, one obtains the graph preserving Hamiltonian constraint operator 
\begin{equation}
\hat H( N_p)\equiv\frac{1}{2}(\hat H_g(N_p)+\hat H_m(N_p))+\frac{1}{2}(\hat H_g(N_p)+\hat H_m(N_p))^{\dagger}
\nonumber
\end{equation}
 which is defined in the subspace $K_{\mathcal{T}}$ of $K$  based on a single network topology $\mathcal{T}$. 

\begin{figure}
\begin{center}
\includegraphics[angle=0,width=4in,clip=true]{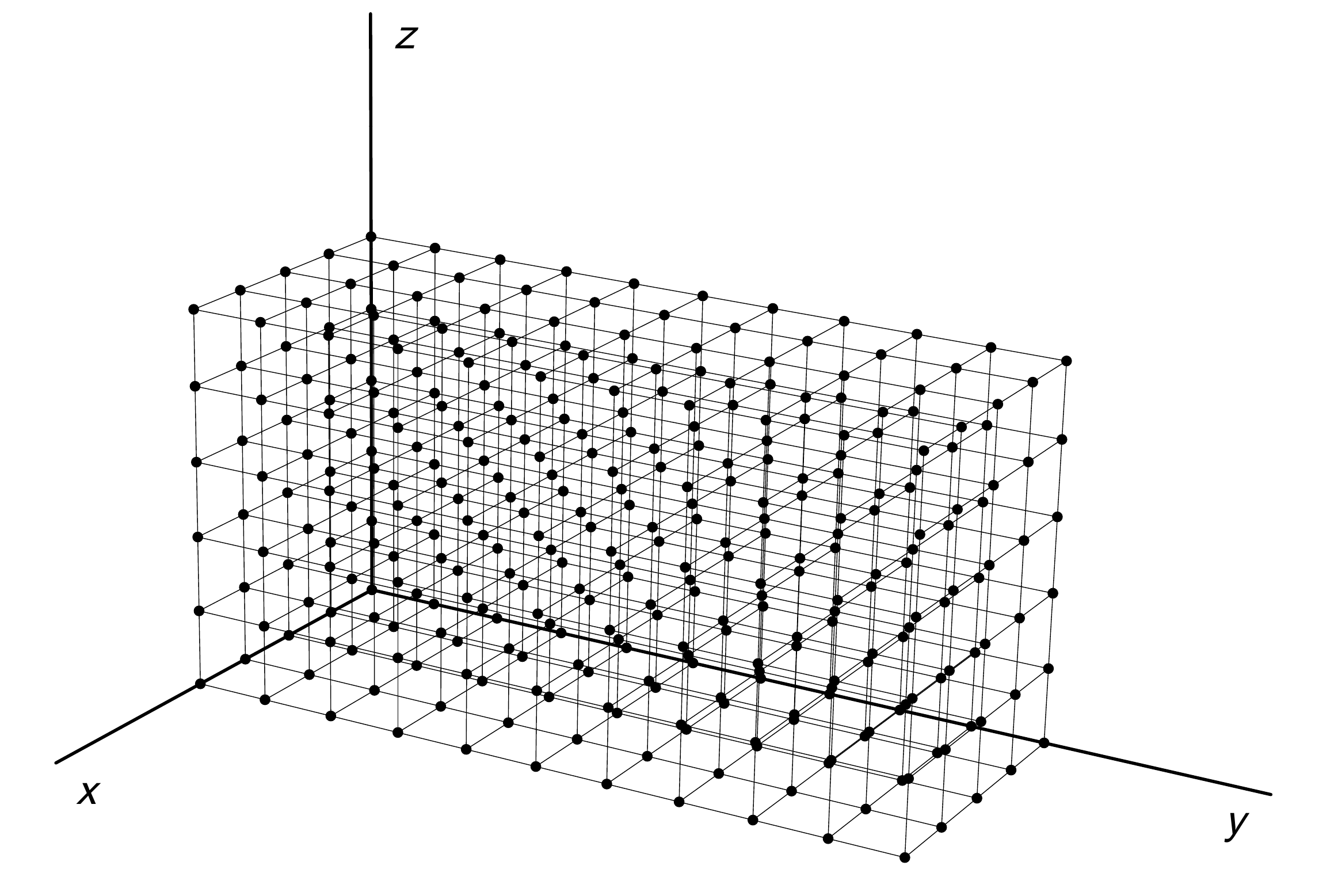}
\begin{spacing}{1}
\caption{The lattice rectangular prism $\bar I_{\mathbb{Z}}^3 \subset \mathbb{R}^3$ with the vertices represented by the black dots at the locations $\{X_n\}$, and the links represented by the lines between two adjacent vertices. The lattice torus $\mathcal{T}_{torus}$ for $\mathfrak{K}$ can be obtained by identifying the opposite boundary faces of $\bar I_{\mathbb{Z}}^3$. As an illustration for chapter 8, the construction also assigns the location $X_n$ of a vertex to the corresponding vertex $v^*_n$ of the lattice torus as its coordinate value.}
\end{spacing}
\end{center}
\end{figure}

Consider a lattice rectangular prism in $\mathbb{R}^3$ with a large number of vertices. Identifying the correspondent boundary vertices in the opposite faces of the rectangular prism, one obtains a lattice torus $\mathcal{T}_{torus}$ with $N_v$ nodes (fig.3). By construction, $\mathcal{T}_{torus}$ has locally cubical lattice topology with six neighboring nodes around each of its nodes. With the graph-preserving Hamiltonian constraint operator, from now on we will consider only the subspace $K_{\mathcal{T}_{torus}}\equiv \mathfrak K $ for simplicity of the topology. To explicitly write down the action of $\hat{H}(N_p)$ on $ \mathfrak K $, we will re-express the operator using the dynamical nodes and paths instead of the embedded ones. Define a set of dynamical nodes $\{ v_m \equiv \bar v_m(\bar\Gamma )\}$, with $m$ ranges from $1$ to $N_v$, satisfying $\bar v_m(\bar\Gamma )=\bar v_{n(m)}^{\bar\gamma}$ such that $n(m)$ is one-to-one. Corresponding to the given $\{v_m\}$, define the dynamical paths $\{ e_{m,j} \equiv \bar e_{m,j}(\bar\Gamma )\}$ satisfying $\bar e_{m,j}(\bar\Gamma )=\bar e_{n(m),k(j)}^{\bar\gamma }$ and $\bar v_m(\bar\Gamma )=\bar v_{n(m)}^{\bar\gamma}$ such that $j(k)$ is one-to-one. The  dynamical closed paths $e_{m,i,j}=\bar e_{m,i,j}(\bar\Gamma)$ may then be defined accordingly. Note that in this case the minimal loop $\bar e_{n,i,j}(\bar\Gamma)$ covers exactly four edges because of the local cubical structure. The action of $\hat{H}_g(N_p)$ on any $\langle s_{[\bar\Gamma]}|\in\mathfrak K$ can be re-expressed as
\begin{equation}
\begin{split}
\langle s_{[\bar\Gamma]}|\hat{H}_g(N_p)\rule{462pt}{0pt}\\
\equiv \langle S_{\bar\Gamma}| \hat{H}_{g(\bar\Gamma)}(N_p)\hat{\mathbb P}_{diff}\rule{427pt}{0pt}\\
=\langle S_{\bar\Gamma}|\hat{H}^{E}_{g(\bar\Gamma)}(N_p)+(i\hbar\kappa\gamma)^{-5}\frac{(1+\gamma^2)}{96} \sum_{ v_m}  N_p(  p_k|_{ \bar p_k(\bar\Gamma )=\bar v_m(\bar\Gamma)}) \sum_{i,j,k=1} \text{sgn}\left(\bar e_{m,i}(\bar\Gamma ),\bar e_{m,j}(\bar\Gamma ),\bar e_{m,k}(\bar\Gamma )\right)\rule{57pt}{0pt}\\ \times\left(\hat{h}^{-1} (\bar e_{m,i}(\bar\Gamma) )\right)_{\bar{i}}^{\bar{l}} \left[\left(\hat{h}(\bar e_{m,i}(\bar\Gamma ))\right)_{\bar{l}}^{\bar{j}},\left[\hat{H}^{E}_{g(\bar\Gamma)}(1),\hat{V}_{(\bar\Gamma)}\right] \right]\rule{224pt}{0pt}\\\times\left(\hat{h}^{-1}(\bar e_{m,j}(\bar\Gamma ))\right)_{\bar{j}}^{\bar{p}}\left[\left(\hat h(\bar e_{m,j}(\bar\Gamma ))\right)_{\bar{p}}^{\bar{k}},\left[\hat{H}^{E}_{g(\bar\Gamma)}(1),\hat{V}_{(\bar\Gamma)}\right] \right]\rule{223pt}{0pt}\\\times\left(\hat{h}^{-1} (\bar e_{m,k}(\bar\Gamma )\right)_{\bar{k}}^{\bar{q}}\left[\left(\hat{h}(\bar e_{m,k}(\bar\Gamma )\right)_{\bar{q}}^{\bar{i}},\hat{V}_{(\bar\Gamma)}\right]\hat{\mathbb P}_{diff}\rule{260pt}{0pt}\\\\
\end{split}
\end{equation}
for any $\bar\Gamma \in[\bar\Gamma]$, where we have
\begin{equation}
\begin{split}
\langle\psi_{\bar\gamma,f}|\hat{H}^{E}_{g(\bar\Gamma)}(N_p)\rule{355pt}{0pt}\\
\equiv\langle\psi_{\bar\gamma,f}| (i\hbar\kappa\gamma)^{-1}\frac{1}{96}\sum_{ v_m}   N_p(  p_k|_{ \bar p_k(\bar\Gamma )=\bar v_m(\bar\Gamma)})  \sum_{i,j,k=1} \text{sgn}\left(\bar e_{m,i}(\bar\Gamma ),\bar e_{m,j}(\bar\Gamma ),\bar e_{m,k}(\bar\Gamma )\right)\rule{60pt}{0pt}\\
      \times \left(\hat{ h}(\bar e_{m,i,j}(\bar\Gamma)) -\hat{ h}^{-1}(\bar e_{m,i,j}(\bar\Gamma))
      \right) ^{\bar{i}}_{\bar{j}} \left(\hat{h}^{-1} (\bar e_{m,k}(\bar\Gamma ))\right)_{\bar{i}}^{\bar{l}}\cdot \left[\left(\hat{h}(\bar e_{m,k}(\bar\Gamma ))\right)_{\bar{l}}^{\bar{j}},\hat{V}_{(\bar\Gamma)} \right]\rule{25pt}{0pt}\\\\
\end{split}
  \end{equation}
Here, the total volume operator $\hat{V}_{(\bar\Gamma)}$ is defined as
\begin{equation}
\begin{split}\\
\langle\psi_{\bar\gamma,f}| \hat{V}_{(\bar\Gamma)}\rule{395pt}{0pt}\\
\equiv\langle\psi_{\bar\gamma,f}| \sum_{ v_m}\left [\frac{1}{48}\sum_{i,j,k=1} \text{sgn}\left(\bar e_{m,i}(\bar\Gamma ),\bar e_{m,j}(\bar\Gamma ),\bar e_{m,k}(\bar\Gamma )\right) \epsilon^{pqr}\hat{J}_{p}(\bar e_{m,i}(\bar\Gamma ))\hat{J}_{q}(\bar e_{m,j}(\bar\Gamma ))  \hat{J}_{r}(\bar e_{m,k}(\bar\Gamma ))\right]^{\frac{1}{2}}\\\\
\end{split}
\end{equation}
Note that the particular choices for $\{v_m\}$ and $\{ e_{m,j}\}$ are irrelevant in $(3.3)$-$(3.5)$ since they serve as dummy variables. The explicit form of $\hat{H}_m(N_p)$ acting on $s_{[\bar\Gamma]}$ has a similar structure as above, and in brief we have
\begin{equation}
\begin{split}
 \langle s_{[\bar\Gamma]}|\hat{H}_{m}(N_p)\rule{380pt}{0pt}\\\\
\equiv \langle S_{\bar\Gamma}| \hat{H}_{m(\bar\Gamma)}(N_p)\hat{\mathbb P}_{diff}\rule{350pt}{0pt}\\\\
\equiv\langle S_{\bar\Gamma}| \sum_{ v_m} N_p(  p_k|_{ \bar p_k(\bar\Gamma )=\bar v_m(\bar\Gamma)})\rule{320pt}{0pt}\\
\hat{H}_{m}^{(\bar{ v}_m(\bar\Gamma ))}\left(\hat{J}_{i}(\bar e_{m,i}(\bar\Gamma )),\hat{h}(\bar e_{m,i}(\bar\Gamma ))_{\bar{l}}^{\bar{j}},\hat{h}( \bar e_{m,i,j}(\bar\Gamma))_{\bar{l}}^{\bar{j}},\hat{\text{J}}_{\text i}(\bar e_{m,i}(\bar\Gamma )), \hat{\text{h}}^{(\text j)}(\bar e_{m,i}(\bar\Gamma ))^{\bar{\text i}}_{\bar{\text j}}, \hat{\text{h}}^{(\text j)}(\bar e_{m,i,j}(\bar\Gamma))^{\bar{\text i}}_{\bar{\text j}}\right.\\ \hat {\theta} (\bar{ v}_m(\bar\Gamma ))^{\bar i}_{\bar{\text i}}, \hat {\theta} (\bar e_{m,i}(\bar\Gamma )_{(1)})^{\bar i}_{\bar{\text i}},\hat{\eta} (\bar{ v}_m(\bar\Gamma ))_{\bar i}^{\bar{\text i}},\hat{\eta} (\bar e_{m,i}(\bar\Gamma )_{(1)})_{\bar i}^{\bar{\text i}}, \hat{h}^{(\text{i})}(\bar{ v}_m(\bar\Gamma ))^{\bar{\text i}}_{\bar{\text j}},\hat{h}^{(\text{i})}(\bar e_{m,i}(\bar\Gamma )_{(1)})^{\bar{\text i}}_{\bar{\text j}},\\\left.\phantom{\hat{\text{J}}_{\text i}}\hat{p}_{\text i}(\bar{ v}_m(\bar\Gamma )),\hat{p}_{\text i}(\bar e_{m,i}(\bar\Gamma )_{(1)}) \right)\hat{\mathbb P}_{diff}\\
\\
\end{split}
\end{equation}
for any $\bar\Gamma \in[\bar\Gamma]$.

Observe that in $(3.3)$-$(3.6)$, the operators $\hat{H}_{g(\bar\Gamma)}(N_p)$ and $\hat{H}_{m(\bar\Gamma)}(N_p)$ act on $\langle S_{\bar\Gamma}|$ using the operators defined with the embedded paths and points that lie inside of $\bar\Gamma$, without disturbing its topology. Consequentially, the states $\langle s_{[\bar\Gamma]}|\hat{H}_g(N_p)$ and $ \langle s_{[\bar\Gamma]}|\hat{H}_{m}(N_p)$ will still be based on $\mathcal{T}_{torus}$ and belong to $\mathfrak{K}$. Therefore,
we can construct our model upon $\mathfrak K$ using the self-adjoint Hamiltonian constraint given by
\begin{equation}
\begin{split}
\langle s_{[\bar\Gamma]}|\hat{H}(N_p)\equiv\langle S_{\bar\Gamma}|\hat{H}_{(\bar\Gamma)}(N_p)\hat{\mathbb P}_{diff}\equiv\langle s_{[\bar\Gamma]}|\frac{1}{2} (\hat{H}_g(N_p) +\hat{H}_m(N_p)) + \frac{1}{2}(\hat{H}_g(N_p) +\hat{H}_m(N_p))^\dagger\\\\
\end{split}
\end{equation}
The physical Hilbert space for the model will be the space annihilated by $\hat{H}(N_p)$ for all $N_p$ based on all $p$.

\section{Physical Hilbert Space}

The idea of applying group averaging method to solve the Hamiltonian constraint in quantum gravity is an ongoing project \cite{ave1}\cite{ave2}. Specifically, the possibility of solving the standard Hamiltonian constraint $(2.31)$-$(2.33)$ in loop quantum with this method has been broadly discussed. While difficulties due to the complexity of diffeomorphism group still await to be overcome for the full theory \cite{ave3}\cite{ave2}, the procedure has been carried out rigorously in the minisuperspace models \cite{ave1}\cite{marolf1}\cite{marolf2}. The graph-changing operations are frozen out in $\hat{H}(N_p)$ for our model, and such simplicity allows one to solve the constraint with group averaging method, based on few concrete assumptions. In the following, I apply this method to the graph-preserving model in close analogy to Marolf's approach \cite{marolf1}\cite{marolf2} for the minisuperspace models.

Consider the set of unitary operators $\{\exp(i\hat{H}(N_p))\}$ given by all lapse functions $\{N_p\}$ based on all possible $p$. To solve the Hamiltonian constraint, we look for an object that is invariant under the action of this set of operators. Following the experience in $(2.2.4)$, we look for the group generated by $\{\exp(i\hat{H}(N_p))\}$. Explicitly, we consider
\begin{equation}
\left\{\prod_{k=1}^{k_{{}_{max}}}\exp(i\hat{H}(N_{p_k}))\right\}\bigg{\slash _{{=}_{{}_{\mathfrak{K}}}}} \equiv \left\{ \hat{U}(g)\right\}_{g\in G}
\end{equation}
where $N_{p_k}$ for each $k$ is a member of all lapse functions $\{N_p\}$ based on all possible $p$, and $\slash _{{=}_{{}_{\mathfrak{K}}}}$ means that we identify two expressions if they give the same operator in $\mathfrak{K}$. $G$ denotes the group faithfully represented by the set of unitary operators generated in this manner. Note that in our model, $\exp(i\hat{H}(N_{p_1}))\exp(i\hat{H}(N_{p_2}))$ is not equal to $\exp(i\hat{H}(N_{p_1}+N_{p_2}))$. This non-Abelean nature of $G$ is due to the fact that the actions of $\hat{H}(N_p)$ on two adjacent nodes of  a knot state do not commute, as a consequence of the graph-preserving modification. Assuming the existence of a left and right invariant measure $dg$ for the group $G$, then the desired invariance can be achieved by the group averaging operator over $G$
\begin{equation}
\hat{\mathbb{P}} \equiv \int_G dg \hat{U}(g) 
\end{equation}
Recall that the group averaging operator $\hat{\mathbb{P}}_{diff}$ maps spin network states $\{|S_{\bar\Gamma}\rangle \in\mathcal S\}$ into knot states $\{\langle S_{\bar\Gamma}|\hat{\mathbb{P}}_{diff} \in\mathcal S^*\}$ that span $K$. In minisuperspace models, there are group averaging operators $\hat{\mathbb{P}}_{mini}$ (corresponding to $\hat{\mathbb{P}}$ in our model) generated by the symmetrically reduced Hamiltonian constraint operators. Like $\hat{\mathbb{P}}_{diff}$, $\hat{\mathbb{P}}_{mini}$ maps the dual states of the states in kinematical Hilbert space into the physical states that lie in the dual space. Here we assume that the same applies to our model, that $\hat{\mathbb{P}}$ maps from knot states $\{\langle s_{[\bar\Gamma]}|\in\mathfrak K\}$ into $\{\hat{\mathbb P}|s_{[\bar\Gamma]}\rangle\in \mathfrak{K}^*\}$ that span the physical Hilbert space.

Since $\hat{\mathbb{P}}$ is invariant under the action of $G$, so is the element $\hat{\mathbb{P}}|\psi\rangle\in \mathfrak{K}^*$ with $\langle\psi|\in \mathfrak K$. Further, the inner product between any two states $|\Psi_1\rangle=\hat{\mathbb{P}}|\psi_1\rangle$ and $| \Psi_2\rangle=\hat{\mathbb{P}}|\psi_2\rangle$ may be defined as
\begin{equation}
\langle\Psi_1|\Psi_2\rangle\equiv\langle\psi_1|\Psi_2\rangle= \langle\psi_1|\hat{\mathbb{P}}|\psi_2\rangle \\
\end{equation}
To be clear, the assumptions being made here are: (1) the existence of the left and right invariant measure $dg$; (2) the operator $\hat{\mathbb{P}}$ maps $\{\langle s_{[\bar\Gamma]}|\in\mathfrak K\}$ into $\{\hat{\mathbb P}|s_{[\bar\Gamma]}\rangle\in \mathfrak{K}^*\}$. These are known to hold in minisuperspace models \cite{ave1}\cite{marolf1}\cite{marolf2}, but remain to be proven in the model.
Finally, under these assumptions we obtain the physical Hilbert space $\mathbb H$ that is spaned by $\{\hat{\mathbb P}|s_{[\bar\Gamma]}\rangle\}$, which solves the Hamiltonian constraint $\hat{H}(N_p)$.

\section{Local Observables}

Now we wish to find the local observables in the physical Hilbert space $\mathbb{H}$ that have classical interpretations, in order to explore the semi-classical region of the model. In the classical case, the values of  matter fields can be treated as internal coordinates that enable the description of local observables \cite{kuchar}\cite{torre}. In the quantum theory, the same idea has been applied in different approaches to a varying extent. In particular, a variety of minisuperspace quantum cosmology models have successfully reproduced correct low energy limits with observables defined by matter clocks \cite{lqc1}\cite{lqc2}.

 In this model, we aim to construct the observables that are completely localized in spacetime in order to make connections to full general relativity. The strategy is to first identified spatially local operators in $\mathfrak K$ using three scalar matter operators to assign the positions, then promote them into localized observables in $\mathbb H$ using another matter scalar operator to assign the time. Note that the dimension of the semi-classical spacetime is thus put in by hand in the model.

For each set of dynamical nodes $\{v_m\}$ we will construct a set of self-adjoint, commuting matter operators, containing the scalar operators $\{ \hat\phi^{0}(v_m),\hat\phi^{1}(v_m), \hat\phi^{2}(v_m), \hat\phi^{3}(v_m)\}$, current operators $\{\hat{V^{i}_{I}}(v_m), \hat{U}^{\bar{i}}_{\bar{I}}(v_m)\}$ and conjugate current operators $\{\hat{\bar{V}}^{I}_{i}(v_m),\hat{\bar{U}}^{\bar{I}}_{\bar{i}} (v_m)\}$ ($I=1,2, 3$ for the vector currents; $\bar{I}=1,2$ for the spinor currents). The three scalar operators $(\hat{\phi}^{1}(v_m), \hat{\phi}^{2}(v_m), \hat{\phi}^{3}(v_m))\equiv \hat{\phi} (v_m)$ will be the spatial coordinate operators, while the operators $\{\hat{V^{i}_{I}}(v_m), \hat{U}^{\bar{i}}_{\bar{I}}(v_m)\}$ and $\{\hat{\bar{V}}^{I}_{i}(v_m), \hat{\bar{U}}^{\bar{I}}_{\bar{i}} (v_m)\}$ will be the spatial frame operators, and $\hat\phi^{0}(v_m)$ will serve as the clock.
In our model, we use the fermionic sector to construct the spatial frame operators, and the bosonic sector to construct the spatial coordinate and clock operators. Specifically, we will set the matter gauge group to be a direct product of $N_{\mathcal G}$ unitary groups, so that $\mathcal G\equiv \bigotimes_N^{N_{\mathcal G}} \mathcal G_N$ and $\text{ k}_n = ( \text{ k}_n^1, \text{ k}_n^2,....,\text{ k}_n^{N_{\mathcal G}})$ in $(2.35)$. Then, define the spatial coordinate:
\begin{equation}
 \hat\phi(v_n)\equiv\left(\phi^1( \hat{\text{ k}}^N{(v_n)}),\phi^2( \hat{\text{ k}}^N{(v_n)}),\phi^3(\hat{\text{ k}}^N{(v_n)})\right)
\nonumber
\end{equation}
 which would be self-adjoint and diagonal in the knot states basis for any real functions $(\phi^1,\phi^2,\phi^3)$. Also, assuming there are $N_f \geq 5$ species of fermions in the matter sector, we denote each species by $\hat{\theta}_J$ $(J=1,2,...,N_f)$.
In this setting, we will use the examples given in section $2.3.3$ for the current operators:
\begin{equation}
\hat{J}(v_m)^{k}_{I} \equiv (\hat{\eta}_I)_{\bar j}^{\bar{\text i}}(\tau^k)_{\bar i}^{\bar{j}}(\hat {\theta}_I )
^{\bar i}_{\bar{\text i}}(v_m)\rule{2pt}{0pt};\rule{10pt}{0pt}
\hat{U}(v_m)^{\bar{k}}_{\bar{I}}\equiv ( C ^{\bar{\text i},\bar{\text j}}_{\bar i,\bar j})^{\bar{k}}(\hat {\eta}_{\bar I+3})^{\bar i}_{\bar{\text i}}(\hat {\theta}_{\bar I+3})^{\bar j}_{\bar{\text j}}(v_m)
\nonumber
\end{equation}
In classical Hamiltonian general relativity, we obtain $\mathit{diff_M}$ invariant local variables by smearing the field variables with delta functions of the spatial matter coordinates. Setting $\hat{O} (v_m)$ and $\hat{O'}(e_{n,i})$ to be some gravitational operators in $\mathfrak K$, we accordingly smear them over all the dynamical nodes with a regularized delta function ${\delta}^{\epsilon}( {\phi}(\bar v_n)- X)$ to obtain the quantum analogies. Separately, $\hat{O} (v_m)$ will be smeared with ${\delta}^{\epsilon}( {\phi}(\bar v_n)- X)$ that picks up the node with coordinate value $X$, and $\hat{O'}(e_{n,i})$ will be smeared with ${\delta}^{\epsilon}( {\phi}(\bar v_n)- X)$ and ${\delta}^{\epsilon}( {\phi}(\bar v_n)- (X+\Delta X))$ that pick up the path connecting the coordinate values $X$ and $X+\Delta X$. Here we choose ${\delta}^{\epsilon}$ to be a Gaussian distribution with width $\epsilon$, which is finite in correspondence to the discrete structure of $\mathfrak{K}$. Explicitly, for any $ s_{[\bar\Gamma]}\in \mathfrak K$
\begin{equation}
\begin{split}
\langle s_{[\bar\Gamma]}|\hat{O}(X) \equiv \langle S_{\bar\Gamma}|\sum_{n} \hat{O}(\bar v_n(\bar\Gamma)) \det(\Delta \hat{\phi}(\bar v_n(\bar\Gamma)))\hat{\delta}^{\epsilon}( \hat{\phi}(\bar v_n(\bar\Gamma))- X) \hat{\mathbb P}_{diff}\\
\langle s_{[\bar\Gamma]}|\hat{O'}(e_{X,\Delta X})\equiv\langle S_{\bar\Gamma}| \sum_{n,i} \hat{O}(\bar e_{n,i}(\bar\Gamma))\det(\Delta \hat{\phi}(\bar v_n(\bar\Gamma))) \det(\Delta \hat{\phi}(\bar v_{(n,i)}(\bar\Gamma)))\\\times\hat{\delta}^{\epsilon}(\hat{\phi}(\bar v_n(\bar\Gamma))-X) \hat{\delta}^{\epsilon}(\hat{\phi}(\bar v_{(n,i)}(\bar\Gamma))-(X+\Delta X))\hat{\mathbb P}_{diff}\\\\
\end{split}
\end{equation}
where the dynamical nodes $\{v_{(n,i)}\}$ are defined by $\bar v_{(n,i)}(\bar \Gamma)\equiv { \bar e_{n,i}}(\bar\Gamma)_{(1)}$ with $\bar e_{(1)}$ being the target point of $\bar e$. In correspondence to the classical expressions, where the smearing is performed through integrations over a coordinate space, the discretized version $(5.1)$ of smearing is summed over the small coordinate volume elements. The coordinate volume element operators are given by
\begin{equation}
\begin{split}
\langle\psi_{\bar\gamma,f}|\det(\Delta \hat{\phi}(\bar v_n(\bar \Gamma)))\rule{260pt}{0pt}\\
\equiv \langle\psi_{\bar\gamma,f}|\sum_{(i,j,k)}\text{sgn}\left(\bar e_{n,i}(\bar \Gamma),\bar e_{n,j}(\bar \Gamma),\bar e_{n,k}(\bar \Gamma)\right)  \det(\Delta \hat{\phi}_{\bar e_{n,i}(\bar \Gamma)}, \Delta \hat{\phi}_{\bar e_{n,j}(\bar \Gamma)}, \Delta \hat{\phi}_{\bar e_{n,k}(\bar \Gamma)})\\
\end{split}
\end{equation}
Here, the coordinate difference operator $\Delta \hat{\phi}_{\bar e_{n,i}}(\bar \Gamma)$ is given by
\begin{equation}
\begin{split}
\langle\psi_{\bar\gamma,f}|\Delta \hat{\phi}_{\bar e_{n,i}(\bar \Gamma)}\equiv\langle\psi_{\bar\gamma,f}| [\hat{\phi}(\bar v_{(n,i)}(\bar \Gamma)) - \hat{\phi}(\bar v_n(\bar \Gamma))]
\nonumber
\end{split}
\end{equation}
 It is crucial to note that the spatially localized operators obtained in $(5.1)$ do not depend on the choices of $\{v_m\}$ and $\{e_{n,i}\}$ which serve as dummy variables. Instead, the operators are localized by the spatial  matter coordinates.

The classical gravitational fields in Ashtekar formalism are $SU(2)$ tensors, and their scalar products with appropriate $SU(2)$ matter currents provide the invariant components. In other words, the physical components of the gravitational fields can be described with respect to the physical frames of the matter currents. In the model, we use $\{\hat{V^{i}_{I}}(v_m), \hat{U}^{\bar{i}}_{\bar{I}}(v_m)\}$ and $\{\hat{\bar{V}}^{I}_{i}(v_m),\hat{\bar{U}}^{\bar{I}}_{\bar{i}} (v_m)\}$ to extract the physical components. Explicitly, for any $ s_{[\bar\Gamma]}\in \mathfrak K$
\begin{equation}
\begin{split}
\langle s_{[\bar\Gamma]}|\hat{J}( e_{n, j})_{I} \equiv\langle S_{\bar\Gamma}| \hat{V}(\bar v_n(\bar \Gamma))^{i}_{I}\hat{J}(\bar e_{n, j}(\bar \Gamma))_{i}\hat{\mathbb P}_{diff}\rule{65pt}{0pt}\\\\
\langle s_{[\bar\Gamma]}|\hat{h}(e_{n,k})^{\bar{I}}_{\bar{J}} \equiv \langle S_{\bar\Gamma}| \hat{\bar{U}}(\bar v_{(n,k)}(\bar \Gamma))^{\bar{I}}_{\bar{i}} \hat{h}(\bar e_{n,k}(\bar \Gamma))^{\bar{i}}_{\bar{j}}\hat{U}(\bar v_n(\bar \Gamma))^{\bar{j}}_{\bar{J}}\hat{\mathbb P}_{diff}
\end{split}
\end{equation}
 Using $(5.1)$, we obtain the spatially localized operators as $\hat{J}(e_{X, \Delta X})_{I}$ and $\hat{h}(e_{X, \Delta X})^{\bar{I}}_{\bar{J}}$. Notice that while these operators are well defined in $\mathfrak K$, how well they describe the physics depends on how well the matter coordinates behave.

The next step is to promote the spatially localized operators into fully localized observables in $\mathbb{H}$ by using the clock field operator. Each state in $\mathbb {H}$ represents a quantum spacetime, and each state in $\mathfrak K$ gives a spatial quantum geometry. Therefore, the temporal localization in our model is equivalent to a foliation of a quantum spacetime $|\Psi \rangle\in \mathbb H$ into the spatial slices $\{\langle\psi_T| \in \mathfrak K\}$ labeled by the clock time $T$. Correspondingly, a localized observable $\hat{O}(X,T)$ will first project a state $|\Psi\rangle \in \mathbb{H}$ into a state $|\psi_T \rangle\equiv \hat{\Pi}_T|\Psi\rangle\in\mathfrak K^*$, which is the dual state of $\langle\psi_T|\in \mathfrak K$. Then, it will  act upon $|\psi_T \rangle$ by the spatially localized operator $\hat{O}(X)$. Finally, it will apply the group averaging operator $\hat{\mathbb{P}}$ to bring the result back into $\mathbb{H}$. With the clock operator $\hat{\phi}^{0}(v_n)$ that has continuous spectrum in $\mathfrak K$, the localized observables described above have the explicit forms:
\begin{equation}
\hat{O}(X, T)\equiv\hat{\mathbb{P}}\hat{O}(X)  \hat{\Pi}_T\\
\end{equation}
where $\hat{\Pi}_T$ is the operator that projects onto the eigenstates of $\hat{\phi}^{0}(v_n)$ with the eigenvalues lying within a range $\epsilon$ around $T$. In our model, the action of $\hat{\Pi}_T$ on $\Psi \in \mathbb H$ is given by
\begin{equation}
\langle s_{[\bar\Gamma]}|\hat{\Pi}_T|\Psi\rangle\equiv\langle S_{\bar\Gamma}|\hat{\Pi}_{T(\bar\Gamma)}\hat{\mathbb P}_{diff}|\Psi\rangle\equiv \langle S_{\bar\Gamma}|sym\left\{\prod_n \hat{\nu}_{\phi^0} (\bar v_n(\bar\Gamma)) \hat{\delta}^{\epsilon}(\hat{\phi}^0(\bar v_n(\bar\Gamma))- T)\right\}\hat{\mathbb P}_{diff}|\Psi\rangle
\end{equation}
for any $s_{[\bar\Gamma]}\in \mathfrak K$ and $\bar\Gamma \in [\bar\Gamma]$, where $\hat{\nu}_{\phi^0} (v_n)$ is the Hamiltonian speed of the clock
\begin{equation}
\langle S_{\bar\Gamma} |\hat{\nu}_{\phi^0} (\bar v_n(\bar\Gamma)) \equiv\langle S_{\bar\Gamma}| \frac{ i}{\hbar}\left[\hat{\phi}^0(\bar v_n(\bar\Gamma)), \hat{H}_{(\bar \Gamma)} (1) \right]
\end{equation}
 Also, since in general the operators $\hat{\nu}_{\phi^0} (\bar v_n(\bar\Gamma))$ with different $n$ do not commute, we fix the ordering ambiguity by applying the symmetrization $sym\{...\}$  in the ordering of $n$.
Clearly, the choice of $\{v_n\}$ is again irrelevant. The classical counterpart of $\hat{\nu}_{\phi^0} (\bar v_n(\bar\Gamma))$ is ${\nu}_{\phi^0}(\text x)\equiv\{\phi^0(\text x), H(\bar N=1)\}$. When all fields are on-shell, $H(\bar N=1)$ generates diffeomorphisms given by unit normal flows orthogonal to spatial slices, which transform $\phi^0(\text x)\equiv\phi^0(\text x,0)$ into $\phi^0(\text x,t)$. Therefore, when all the constraints are satisfied, ${\nu}_{\phi^0} (\text x)$ is simply the speed $\partial_t \phi^0(\text x,t)|_{t=0}$ of the clock with respect to the proper time carried by the flow.

Finally, the localized gravitational observables are constructed by promoting $\hat{J}(e_{X, \Delta X})_{I}$ and $\hat{h}(e_{X, \Delta X})^{\bar{I}}_{\bar{J}}$ using $(5.4)$ into $\hat{J}(e_{X, \Delta X}, T)_{I}$ and $\hat{h}(e_{X, \Delta X}, T)^{\bar{I}}_{\bar{J}}$.

\section{Matter Coordinate Conditions}
In order to give sensible descriptions of the physics, the localized observables must refer to reasonably well-behaved internal coordinates. This requirement is to be met by the choices of the scalar fields, current frames, and the quantum state together.

\subsection{ Conditions on the Clock }

As mentioned in the previous chapter, the purpose of the clock field $\hat{\phi}^0$ is to physically define a slicing for a quantum spacetime. To explore the dynamics in the model, $\hat{\phi}^0$ has to provide a slicing that reveals the causal evolution of the system. In classical general relativity, such a clock specifies an ADM foliation, in which each slice provides the full information of the whole spacetime. In other words, the field values on a slice at any clock time can be used to reconstruct the whole spacetime.

In the setting of our model,  the two spatial slices given by two distinct clock times $T_1$ and $T_2$  are described by $\hat{\Pi}_{T_1}|\Psi\rangle$ and $\hat{\Pi}_{T_2}|\Psi\rangle$. Analogous to the condition above in the classical case, the two spatial slices defined by the ideal clock $\hat{\phi}^0$ should contain the same information about the space time $|\Psi\rangle$, up to some quantum fluctuations. In other words, the group average operator should be able to recover the spacetime from either of the slices
\begin{equation}
\hat{ \mathbb{P}}\hat{\Pi}_{T_1}|\Psi\rangle+O(\hbar) = \hat{\mathbb{P}}\hat{\Pi}_{T_2}|\Psi\rangle+O(\hbar)=|\Psi\rangle
\end{equation}
Notice that the condition involves not only $\hat{\phi}^0$ , but also $|\Psi\rangle$.  This agrees with the classical theory in which the proper clocks are chosen according to the state of the whole system.

Using the definition $(5.4)$, the condition $(6.1)$ implies
\begin{equation}
\begin{split}
\hat{\mathbb{P}}\hat{O}_1 (X) \hat{O}_2 (X) \hat{\Pi}_{T_1} |\Psi\rangle\rule{65pt}{0pt}\\
=\hat{\mathbb{P}}\hat{O}_1 (X) \hat{O}_2 (X) \hat{\Pi}_{T_1} \hat{\mathbb{P}} \hat{\Pi}_{T_1} |\Psi\rangle+O(\hbar)\\ 
=\hat{\mathbb{P}}\hat{O}_1 (X)  \hat{\Pi}_{T_1} \hat{\mathbb{P}} \hat{O}_2 (X) \hat{\Pi}_{T_1} |\Psi\rangle+O(\hbar)\\
= \hat{O}_1 (X,T_1) \hat{O}_2 (X, T_1)  |\Psi\rangle+O(\hbar)\rule{18pt}{0pt}
\end{split}
\end{equation}
More generally, denoting $\Phi(O_1, O_2, ....,O_N)$ as an ordered function of $N$ linear operators, we have
\begin{equation}
\begin{split}
\hat{\mathbb{P}} \Phi(\hat{O}_1 (X), \hat{O}_2 (X),...., \hat{O}_N (X)) \hat{\Pi}_{T_1} |\Psi\rangle]\rule{60pt}{0pt}\\
=  \Phi(\hat{O}_1 (X, T_1), \hat{O}_2 (X, T_1),...., \hat{O}_N (X, T_1)) |\Psi\rangle+O(\hbar)
\end{split}
\end{equation}

 In the context of this thesis, we will need to consider the interactions applied to the clock. For later use in describing this interaction, we introduce the operator $\hat\Phi_{int}(v_n)$:
\begin{equation}
\begin{split}
 \langle s_{[\bar \Gamma]}| \hat\Phi_{int}(v_n)\equiv
\langle S_{\bar \Gamma}| \frac{ i}{\hbar}\hat{\nu}^{-1}_{\phi^0}(\bar v_n(\bar\Gamma))\left[\hat{\nu}_{\phi^0} (\bar v_n(\bar\Gamma)), \hat{H}_{(\bar \Gamma)}(1) \right]\hat{\mathbb{P}}_{diff}\\
\end{split}
\end{equation}
where $\hat{\nu}_{\phi^0} (\bar v_n(\bar\Gamma))$ and $\hat{\nu}^{-1}_{\phi^0}(\bar v_n(\bar\Gamma))$ are the Hamiltonian speed defined in $(5.6)$ and its inverse. Note that $\hat\Phi_{int}(v_n)$ contains mainly the `acceleration' of the clock, which is given by the commutator in $(6.4)$.

\subsection{ Conditions on the Rulers and Frames }
In addition to a good clock, we need good spatial coordinates and frames to physically specify positions and directions at a given clock time. In classical general relativity, ideal spatial coordinates for a certain foliation should provide  local diffeomorphisms between $\mathbb{R}^3$ and the space. In other words, the coordinatization should be locally one to one  and  preserve smoothness and adjacentness.
In loop quantum gravity, the states of spatial slices are described by knot states with discrete structure. Therefore, in the model the
matter spatial coordinates should locally provide a `discretized diffeomorphism' between the quantum space and a set of coordinate values.

 Before imposing the spatial coordinate conditions in the model, we need to show how the coordinatization can occur in the first place. To start with, we observe that by applying to $\hat{\Pi}_{T_1}|\Psi\rangle$ all arbitrary functions $\Phi( \hat{J}(e_{X, \Delta X})_{I} ,\hat{h}(e_{X, \Delta X})^{\bar{I}}_{\bar{J}},\hat{h}^{\dagger}(e_{X, \Delta X})^{\bar{I}}_{\bar{J}})$ of the set of spatially localized operators, one obtains a set of states constituting a linear subspace of $\mathfrak K^*$. This subspace $\{\Phi( \hat{J}(e_{X, \Delta X})_{I} ,\hat{h}(e_{X, \Delta X})^{\bar{I}}_{\bar{J}},\hat{h}^{\dagger}(e_{X, \Delta X})^{\bar{I}}_{\bar{J}}) \hat{\Pi}_{T_1}|\Psi\rangle \}\equiv\mathfrak K^*_{\Psi, T_1}$ will be called the relevant subspace for $\hat{\Pi}_{T_1}|\Psi\rangle$. Recall that there are a great many distinct sets of dynamical nodes and dynamical edges compatible with them. This ambiguity is a manifestation of $\mathit{diff_M}$ symmetry in the quantum theory, which involves the superpositions of the knot states that are defined background independently. Here we impose a requirement on the state $|\Psi\rangle$ such that the corresponding $\mathfrak K^*_{\Psi, T_1}$ dynamically break the symmetry, picking up a preferred $\{v^*_ n\}$ and $\{e^*_{n,i}\}$ by being an approximate eigenspace of $\{\hat{\phi}(v^*_{{n}})\}$ and $\{\hat{\phi}( v^*_{(n,i)})\}$, where we again define $v^*_{(n,i)} (\bar\Gamma) \equiv{ \bar e^*_{n,i}}(\bar\Gamma)_{(1)}$. In other words, we require $|\Psi\rangle$ to satisfy
\begin{equation}
\begin{split}
\det(\Delta \hat{\phi}(v^*_{{n}})) \hat{\delta}^{\epsilon}( \hat{\phi}(v^*_{{n}})- X_{ m}) |\psi\rangle
= \delta_{{m},{n}}|\psi\rangle + O(\hbar)\rule{220pt}{0pt}\\\\
 \det(\Delta \hat{\phi}(v^*_n)) \det(\Delta \hat{\phi}(v^*_{(n,i)}))\hat{\delta}^{\epsilon}(\hat{\phi}(v^*_n)-X) \hat{\delta}^{\epsilon}(\hat{\phi}( v^*_{(n,i)})-X_m-\Delta X_{m,j})|\psi\rangle
= \delta_{{m},{n}} \delta_{{i},{j}}|\psi\rangle+ O(\hbar)\rule{0pt}{0pt}\\\\
\end{split}
\end{equation}
for any $|\psi\rangle\in\mathfrak K^*_{\Psi, T_1}$. The appearance of $\det(\Delta \hat{\phi}(v^*_n))$  is needed for the same reason explained after $(5.1)$. When this is satisfied, the $X_{{m}}$ is assigned to $v^*_ m$ and the coordinate gap $\Delta X_{{m,j}}$ is assigned to $e^*_{m,j}$.
Since $\hat\phi(v^*_n)$ is diagonal in the orthonormal basis of knot states, the condition $(6.5)$ is equivalent to
\begin{equation}
\begin{split}
\langle s_{[\bar{\Gamma}]}|\psi\rangle\det(\Delta \hat{\phi}(v^*_{{n}})) \hat{\delta}^{\epsilon}( \hat{\phi}(v^*_{{n}})- X_{ m})|s_{[\bar{\Gamma}]}\rangle
= \langle s_{[\bar{\Gamma}]}|\psi\rangle\delta_{{m},{n}}|s_{[\bar{\Gamma}]}\rangle+ O(\hbar)\rule{120pt}{0pt}\\\\
\langle s_{[\bar{\Gamma}]}|\psi\rangle\det(\Delta \hat{\phi}(v^*_{(n,i)}))\hat{\delta}^{\epsilon}(\hat{\phi}(v^*_n)-X) \hat{\delta}^{\epsilon}(\hat{\phi}( v^*_{(n,i)})-(X_m+\Delta X_{m,j}))|s_{[\bar{\Gamma}]}\rangle
= \langle s_{[\bar{\Gamma}]}|\psi\rangle\delta_{{m},{n}} \delta_{{i},{j}}|\psi\rangle \\+ O(\hbar)\rule{50pt}{0pt}\\
\end{split}
\end{equation}
for any dual knot state $|s_{[\bar{\Gamma}]}\rangle$ and any $|\psi\rangle\in\mathfrak K^*_{\Psi, T_1}$.\footnote{Condition $(6.6)$ implies that the coordinate gaps between adjacent nodes are comparable to the width $\epsilon$ of  $\delta_{\epsilon}$, such that there is only one dynamical node corresponding to each coordinate value. This set-up is only for simplicity. Coarse grained versions of $(6.6)$ can be imposed when $\epsilon$ is  much bigger than the coordinate gaps, without effecting the following arguments about the semi-classical behavior of the model.} According to $(6.6)$, the condition $(6.5)$ is also equivalent to the requirement that
\begin{equation}
\begin{split}
\langle s_{[\bar{\Gamma}]}|\psi\rangle\hat{\text{ k}}^N( v_n)|s_{[\bar{\Gamma}]}\rangle \equiv \langle s_{[\bar{\Gamma}]}|\psi\rangle\text{ k}^N_{[\bar{\Gamma}]} ( v_n)|s_{[\bar{\Gamma}]}\rangle\rule{3pt}{0pt}; \rule{197pt}{0pt}\\\\
\left(\phi^1(  \text{ k}^N _{[\bar{\Gamma}]}( v^*_n)),\phi^2( \text{ k}^N _{[\bar{\Gamma}]}( v^*_n)),\phi^3(\text{ k}^N _{[\bar{\Gamma}]}( v^*_n))\right)=X_n+O(\hbar)\rule{140pt}{0pt}\\\\
\left(\phi^1(  \text{ k}^N_{[\bar{\Gamma}]} ( v^*_{(n,i)})),\phi^2( \text{ k}^N_{[\bar{\Gamma}]} (  v^*_{(n,i)})),\phi^3(\text{ k}^N_{[\bar{\Gamma}]} (v^*_{(n,i)}))\right)=X_n+\Delta X_{n,i}+O(\hbar)\rule{62pt}{0pt}\\
\end{split}
\end{equation}
for any dual knot state $|s_{[\bar{\Gamma}]}\rangle$ and any $|\psi\rangle\in\mathfrak K^*_{\Psi, T_1}$. From $(6.7)$ we immediately see that $\{v^*_{(m,i)}\} = \{v^*_n\}$ and $\{(X_m+\Delta X_{m,i})\}= \{X_{{n}}\}$.
The physical meaning of $(6.5)$-$(6.7)$ is crucial: since the matter spatial coordinates are what we refer to in spatial measurements, the preferred dynamical nodes $\{v^*_n\}$ represent the \emph{physical spatial points} we observe. Consequentially, there is also a set of preferred dynamical spatial points $p^*\equiv \{p^*_m\}$ agreeing with $\{v^*_n\}$, such that $ \bar p^*_m(\bar\Gamma)=\bar v^*_m(\bar\Gamma)$ for $m \le N_v$. It should be emphasized that $\{v^*_n\}$ and $p^*$ are induced by $\mathfrak K^*_{\Psi, T_1}$, so they depend on the spatial quantum state $\hat{\Pi}_{T_1}|\Psi\rangle$. For this reason, as one can see in $(6.6)$ that $\bar v^*_n(\bar\Gamma)$ can be arbitrary for $| s_{[\bar{\Gamma}]}\rangle$ that is orthogonal to $\mathfrak K^*_{\Psi, T_1}$.

The spatial coordinate condition can now be imposed on the map $( v^*_n, e^*_{n,i}) \to (X_n, \Delta X_{n,i})$. Recall that $\mathfrak K$ is based on the lattice torus $\mathcal{T}_{torus}$. Analogous to the coordinate maps for a torus manifold, the map $( v^*_n, e^*_{n,i}) \to (X_n, \Delta X_{n,i})$ in large scale should appear to be smooth around most of the physical spatial points. On the other hand, since there is no global diffeomorphism between $\mathbb{R}^3$ and a 3-torus, we expect the map to appear discontinuous at some specific set of the points $\{v^*_{n_b} \}\subset \{ v^*_{n}\}$. To approximate local diffeomorphisms between the physical spatial points and $\mathbb{R}^3$, the model imposes the spatial coordinate conditions as the following. For every $ e^*_{n.i}$ with $\bar e^*_{n.i}(\bar\Gamma)\cap \{\bar v^*_{n_b}(\bar\Gamma)\}  =\emptyset $, we demand $|\Delta X_{n',i'}| \leq d$, where $d$ represents  a small coordinate gap, such that the map appears continuous in large scales everywhere except $\{v^*_{n_b} \}$. For any $m$ the set $\{\Delta X_{m,i}\}|_{\bar e^*_{m.i}(\bar\Gamma)\cap \{\bar v^*_{n_b}(\bar\Gamma)\}  =\emptyset}$  should define a parallelepiped in $\mathbb{R}^3$ up to an error of $O(d)$, such that the map appears smooth in large scale everywhere except $\{v^*_{n_b} \}$. Notice that once $(6.7)$ is satisfied the coordinate conditions can be achieved easily through a suitable redefinition of the scalar functions $(\phi^1,\phi^2,\phi^3)$.

In $(5.3)$, we defined the physical $SU(2)$ components of the tensorial observables with respect to the matter current frames. Obviously, valid frames for this purpose should be physically orthonormal to each other. Explicitly, we require that for any $|\psi\rangle\in\mathfrak K^*_{\Psi, T_1}$ 
\begin{equation}
\begin{split}
(\hat O^{j}_{i}\hat{V}^{i}_{I}\hat{\bar{V}}^{I}_{j})(X_{ n})|\psi\rangle=\hat O^{i}_{i}(X_{ n})|\psi\rangle + O(\hbar)\rule{2pt}{0pt}\\
(\hat{ O'}^{\bar{j}}_{\bar{i}} \hat{U}^{\bar{i}}_{\bar{I}} \hat{\bar{U}}^{\bar{I}}_{\bar{j}})(X_{ n}) |\psi\rangle= \hat{ O'}^{\bar{i}}_{\bar{i}}(X_{ n})|\psi\rangle+ O(\hbar)\\
(\epsilon^{ijk}\hat{\bar{V}}^{I}_{i} \hat{\bar{V}}^{J}_{j}\hat{\bar{V}}^{K}_{k} )(X_{ n}) |\psi\rangle=\epsilon^{IJK}|\psi\rangle + O(\hbar)
\end{split}
\end{equation}
for any gravitational operators $\hat O^{j}_{i}(v_n)$ and $\hat{ O'}^{\bar{j}}_{\bar{i}}(v_n)$. Matter frames that satisfy these conditions are orthonormal  in terms of the spatial geometry given by the gravitational sector of the state $|\psi\rangle$.

With the help of the appropriate matter spatial coordinates and frames, any operators acting on the space $\mathfrak K^*_{\Psi, T_1}$ can be re-expressed in terms of the spatially local operators. It is through these new expressions we will decipher the physical meanings of the abstract knot states. From now on, we denote $X$ and $(X, \Delta X)$ to be variables taking values from $\{X_n\}$ and $\{(X_n, \Delta X_{n,i})\}$ respectively.

To start with, we assign physical lapse functions referring to the spatial coordinates. For every function $\mathcal{N}(X)$, there is a correspondent lapse function $N^{\small \mathcal{N}}_{p^*}$ with $N^{\small \mathcal N}_{p^*}(p^*_m) \equiv \mathcal{N}(X_{m})$ for $ m\le N_v$. Therefore, given any function $\mathcal{N}(X)$, one can mimic the Hamiltonian constraint operator with the lapse function $N^{\small \mathcal{N}}_{p^*}$ using the spatially local operators.
\begin{equation}
\begin{split}
 \hat{\mathcal H}_{g}(\mathcal{N})\rule{440pt}{0pt}\\
\equiv \hat{\mathcal H}^{E}_{g}(\mathcal{N}) - (i\hbar\kappa\gamma)^{-5}\frac{1}{48}\frac{(1+\gamma^2)}{2} \sum_{X}\mathcal{N}(X)\sum_{\Delta X, \Delta Y, \Delta Z} \text{sgn}\left(e_{X, \Delta X},e_{X, \Delta Y},e_{X, \Delta Z}\right)\rule{110pt}{0pt}\\\times \left(\hat{h}^{-1} (e_{X,  \Delta X})\right)_{\bar{I}}^{\bar{L}} \left[\left(\hat{h}(e_{X,  \Delta X})\right)_{\bar{L}}^{\bar{J}},\left[\hat{\mathcal H}^{E}_{g}(1),\hat{\mathcal V}\right] \right]\left(\hat{h}^{-1} (e_{X, \Delta Y})\right)_{\bar{J}}^{\bar{P}}\left[\left(\hat{h}(e_{X, \Delta Y})\right)_{\bar{P}}^{\bar{K}},\left[\hat{\mathcal H}^{E}_{g}(1),\hat{\mathcal V}\right] \right]\rule{20pt}{0pt} \\\times \left(\hat{h}^{-1} (e_{X, \Delta Z})\right)_{\bar{K}}^{\bar{Q}}\left[\left(\hat{h}(e_{X, \Delta Z})\right)_{\bar{Q}}^{\bar{I}},\hat{\mathcal V}\right] \rule{295pt}{0pt}\\\\
\end{split}
\end{equation}
where $\hat{\mathcal H}^{E}_{g}(\mathcal N)$ is
  \begin{equation}
\begin{split}
 \hat{\mathcal H}^{E}_{g}(\mathcal{N})\rule{406pt}{0pt}\\
\equiv(i\hbar\kappa\gamma)^{-1} \frac{1}{96}  \sum_{X}\mathcal{N}(X)\sum_{\Delta X, \Delta Y, \Delta Z} \text{sgn}\left(e_{X, \Delta X},e_{X, \Delta Y},e_{X, \Delta Z}\right)
       \left(   \hat{ h}(e_{X,  \Delta X, \Delta Y}) -\hat{ h}^{-1}(e_{X,  \Delta X, \Delta Y})
      \right) ^{\bar{I}}_{\bar{J}} \\ \times\left(\hat{h}^{-1} (e_{X, \Delta Z})\right)_{\bar{I}}^{\bar{L}}\cdot \left[\left(\hat{h}(e_{X, \Delta Z})\right)_{\bar{L}}^{\bar{J}}, \hat{\mathcal V} \right]\rule{256pt}{0pt}\\\\
\end {split}
\end{equation}
and
\begin{equation}
\begin{split}
 \hat{\mathcal V}= \sum_{X} \left[\frac{1}{48}\sum_{ \Delta X, \Delta Y, \Delta Z} \text{sgn}\left(e_{X, \Delta X},e_{X, \Delta Y},e_{X, \Delta Z}\right) \hat{\epsilon}^{PQR}\hat{J}(e_{X, \Delta X})_{P} \hat{J}(e_{X, \Delta Y})_{Q} \hat{J}(e_{X, \Delta Z})_{R}\right]^{\frac{1}{2}}\\\\
\end{split}
\end{equation} 
Clearly, the operators $\hat{H}_{g}( N^{\small \mathcal{N}}_{p^*})$ and $\hat{\mathcal H}_{g}(\mathcal{N})$ are different operators in the full space $\mathfrak K$. However, thank to the coordinate and frame conditions $(6.5)$ and $(6.8)$, for any $|\psi\rangle\in\mathfrak K^*_{\Psi, T_1}$ we have
\begin{equation}
\hat{H}_{g}(N^{\small \mathcal{N}}_{p^*})|\psi\rangle= \hat{\mathcal H}_{g}(\mathcal{N})|\psi\rangle+O(\hbar)
\end{equation}
Therefore, $\hat{H}_{g}(N^{\small \mathcal{N}}_{p^*})$  and $\hat{\mathcal H}_{g}(\mathcal{N})$ act equally on any $ \hat{\Pi}_{T_1}|\Psi\rangle\in\mathfrak K^*_{\Psi, T_1}$ up to quantum fluctuation. 

Set $\hat{\alpha}(e_{Y, \Delta Y})$ to be the operator obtained from localizing $\hat{\alpha}(e_{n,i})$, which is defined by the following. Each knot states $\langle s_{[\bar \Gamma]}| \in \mathfrak{K}$ is an eigenstate of $\hat{\alpha}(e_{n,i})$ with an eigenvalue $+1$ or $-1$. The knot state is assigned the eigenvalue $+1$ if the embedded edge of $\bar \Gamma$ overlapping with $\bar e_{n,i}(\bar \Gamma)$ has the same orientation as $\bar e_{n,i}(\bar \Gamma)$, and the eigenvalue $-1$ if the orientations are opposite. When acting on any $|\psi\rangle\in\mathfrak K^*_{\Psi, T_1}$, the spatially local operators also satisfy the equations:
\begin{equation}
\begin{split}
\hat{J}^\dagger(e_{X, \Delta X})_{I}|\psi\rangle=\hat{J}(e_{X, \Delta X})_{I} |\psi\rangle+O(\hbar)\rule{209pt}{0pt}\\\\
\left[\hat{J}(e_{X, \Delta X})_{I}, \hat{h}(e_{Y, \Delta Y})^{\bar{I}}_{\bar{J}}\right]|\psi\rangle= \delta_{X,Y} \delta_{\Delta X,\Delta Y}i l_p^2\gamma  ({\tau _I})^{\bar{K}}_{\bar{J}} \hat{h}(e_{Y, \Delta Y})^{\bar{I}}_{\bar{K}}|\psi\rangle\rule{87pt}{0pt}\\ - \delta_{X,Y+ \Delta Y} \delta_{-\Delta X,\Delta Y}i l_p^2 \gamma({\tau _I})^{\bar{I}}_{\bar{K}} \hat{h}(e_{Y, \Delta Y})^{\bar{K}}_{\bar{J}}|\psi\rangle+O( l_p^2 \hbar)\\\\
\left[\hat{J}(e_{X, \Delta X})_{I}, \hat{h}^{\dagger}(e_{Y, \Delta Y})^{\bar{I}}_{\bar{J}}\right]|\psi\rangle= \delta_{X,Y} \delta_{\Delta X,\Delta Y}i l_p^2 \gamma ({\tau^* _I})^{\bar{K}}_{\bar{J}} \hat{h}^{\dagger}(e_{Y, \Delta Y})^{\bar{I}}_{\bar{K}}|\psi\rangle\rule{79pt}{0pt}\\ - \delta_{X,Y+ \Delta Y} \delta_{-\Delta X,\Delta Y}i l_p^2\gamma ({\tau^*_I})^{\bar{I}}_{\bar{K}} \hat{h}^{\dagger}(e_{Y, \Delta Y})^{\bar{K}}_{\bar{J}}|\psi\rangle+O( l_p^2 \hbar)\\\\
\left[\hat{h}(e_{X, \Delta X})^{\bar{K}}_{\bar{L}}, \hat{h}(e_{Y, \Delta Y})^{\bar{I}}_{\bar{J}}\right] |\psi\rangle= 0+O( l_p^2  \hbar)\rule{203pt}{0pt}\\\\
\left[\hat{h}(e_{X, \Delta X})^{\bar{K}}_{\bar{L}}, \hat{h}^\dagger(e_{Y, \Delta Y})^{\bar{I}}_{\bar{J}}\right] |\psi\rangle= 0+O( l_p^2  \hbar)\rule{200pt}{0pt}\\\\
\left[\hat{J}(e_{X, \Delta X})_{I}, \hat{J}(e_{Y, \Delta Y})_{J}\right]|\psi\rangle= \delta_{X,Y} \delta_{\Delta X,\Delta Y}i l_p^2\gamma{\epsilon_{IJ}}^K \hat{J}(e_{Y, \Delta Y})_{K} \hat{\alpha}(e_{Y, \Delta Y})|\psi\rangle+O( l_p^2  \hbar)\\\\
\end{split}
\end{equation}

Again, it should be emphasized that the equations in $(6.13)$ are not equations of the operators alone. The equalities are derived from the conditions $(6.6)$ and $(6.7)$ satisfied by the state $ \hat{\Pi}_{T_1}|\Psi\rangle$, so they arise from the physical conditions of the system.

\section{Coherent States}
We have required the state $|\Psi\rangle$ to satisfy quantum coordinate conditions, such that the set of local observables $\{\hat{J}(e_{X, \Delta X}, T)_{I}\}$ and $\{\hat{h}(e_{X, \Delta X}, T)^{\bar{I}}_{\bar{J}}\}$ are expected to give a meaningful description on the gravitational sector, with $T$ around the moment $T_1$ . Since our goal is to obtain the semi-classical limit, specific conditions of coherence should be imposed on $|\Psi\rangle$. As in the case of quantum cosmology, we need semi-classical conditions on the gravitational local observables and  the momentum of the clock. Therefore, the state $|\Psi\rangle$ is required to satisfy
\begin{equation}
\begin{split}
\hat{J}(e_{X, \Delta X}, T_1)_{I}|\Psi\rangle=\left(\langle\Psi|\hat{J}(e_{X, \Delta X}, T_1)_{I}|\Psi\rangle\right) |\Psi\rangle +O(\hbar)\\\\
\hat{h}(e_{X, \Delta X}, T_1)^{\bar{I}}_{\bar{J}}|\Psi\rangle=\left(\langle\Psi|\hat{h}(e_{X, \Delta X}, T_1)^{\bar{I}}_{\bar{J}}|\Psi\rangle\right) |\Psi\rangle +O(\hbar)\\\\
\hat{\nu}_{\phi^0}(X, T_{1})|\Psi\rangle=\left(\langle\Psi|\hat{\nu}_{\phi^0}(X, T_{1})|\Psi\rangle\right)|\Psi\rangle +O(\hbar)\rule{20pt}{0pt}\\
\end{split}
\end{equation} 
Additionally, we also expect the expectation values to appear continuous in terms of the spatial coordinates for a semi-classical state. Therefore, for any two $\bar e^*_{n,i}(\bar\Gamma)$ and $\bar e^*_{m,j}(\bar\Gamma)$ that share a common node and form a smooth path, we will impose:
\begin{equation}
\begin{split}
\langle\Psi|\hat{J}(e_{X_n, \Delta X_i}, T_1)_{I}|\Psi\rangle= \langle\Psi|\hat{J}(e_{X_m, \Delta X_j}, T_1)_{I}|\Psi\rangle +O(d)\\\\
\langle\Psi|\hat{h}(e_{X_n, \Delta X_i}, T_1)^{\bar{I}}_{\bar{J}}|\Psi\rangle=\langle\Psi|\hat{h}(e_{X_m, \Delta X_j}, T_1)^{\bar{I}}_{\bar{J}}|\Psi\rangle+O(d)\\\\
\langle\Psi|\hat{\nu}_{\phi^0}(X_n, T_{1})|\Psi\rangle=\langle\Psi|\hat{\nu}_{\phi^0}(X_m, T_{1})|\Psi\rangle +O(d)\rule{20pt}{0pt}\\
\end{split}
\end{equation} 
In this thesis, we will assume the existence of the states that satisfy $(7.1)$ and $(7.2)$ without explicitly constructing them. Because of the appearance of $\hbar$ or $l_p^2$ in every result of $(6.13)$, this assumption is rather mild.
Note that the coherence conditions are defined with respect to the observer who carries the clock $\phi^0$. The conditions say that the quantum spacetime $|\Psi \rangle$ has a sharply defined spatial geometry and extrinsic curvature on the spatial slice at clock time $T_1$. Therefore, the quantum spacetime $|\Psi\rangle$ is expected to be semi-classical around that moment. 

\section{ Emergent Gravitational Fields}

To make contacts with classical general relativity, the expectation values in $(7.1)$ need to be interpreted in terms of classical fields. The matter coordinates provide a natural way to make such a translation from the discrete structure of knot states into the continuous configuration of classical fields.

For an explicit example, we will first pick a simple spatial matter coordinate system. Recall that the lattice torus $\mathcal{T}_{torus}$ can be constructed by identifying the opposite boundary faces of a lattice rectangular prism $\bar I_{\mathbb{Z}}^3 \subset \mathbb{R}^3$, which consists of the vertices $\{\bar V_n=X_n\}$ and links $\{\bar l_i\}$. Here the bars indicate the embedding in $\mathbb{R}^3$. Such construction naturally gives a coordinate map $ v^*_n\to \bar V_n=X_n$ that satisfies the spatial coordinate conditions (fig.3). Also, the approximated spatial coordinate space is naturally a rectangle region $\bar I^3\supset\bar I_{\mathbb{Z}}^3 $ inside of $\mathbb{R}^3$. In this case, the set $\{ v^*_{n_b}\}$ where the map is discontinuous is the set that is mapped into the boundary of $\bar I^3 $.

Once a matter spatial coordinates are chosen, our model identifies every $ e^*_{n.i}$ satisfying $\bar e^*_{n.i}(\bar\Gamma)\cap \{\bar v^*_{n_b}(\bar\Gamma)\}  =\emptyset$ with an embedded path in $ {\mathbb{R}}^3$ under the guidance of the matter coordinate values. Specifically under our choice of coordinates, each such $ e^*_{n.i}$ is identified with the oriented path $\bar{e}_{X_n, \Delta X_{n,i}}$ that goes from the vertex $X_n$ to the vertex $X_n+\Delta X_{n,i} $, which overlaps completely with the link $\bar l_{i'}$ connecting the two vertices. Subsequently, the model choose a cubical cell decomposition dual to $\bar I_{\mathbb{Z}}^3$, dividing $\bar I^3 $ into a set of cubical cells $\{\bar c_{X_n }\}$. Each cell $\bar c_{X_n }$ uniquely contains a vertex $X_n$, and the boundaries of all the cells consist of a set of smooth faces $\{\bar s_i\}$ whose each element $\bar s_{i'}$ intersects transversely with an unique link $\bar l_{i'}$ among $\{\bar l_{i}\}$. Suppose $\bar l_{i'}$ links $X_n$ and $X_n+\Delta X_{n,i} $, and denote $\bar{S}_{X_n, \Delta X_{n,i}}\subset \bar I^3$ to be the oriented surface overlapping completely with $\bar s_{i'}$ that is dual to $\bar l_{i'}$ and has the same orientation as
 $\bar{e}_{X_n, \Delta X_{n,i}}$. The picture motivates an interpretation of the expectation values in $(7.1)$ as being given by smearing a smooth gravitational fields defined in $\bar I^3$ over the corresponding elements from $\{\bar{e}_{X, \Delta X}\}$ and $\{\bar{S}_{X, \Delta X}\}$. Explicitly, we pick a fitting algorithm that maps the expectation values $\{\langle \hat{J}(e_{X, \Delta X}, T)_{I}\rangle, \langle\hat{h}(e_{X, \Delta X}, T)^{\bar{I}}_{\bar{J}}\rangle\}$ and $\{\langle\hat{\nu}_{\phi}(X, T)\rangle\}$ to the values of the smooth fields $\{ E^{a}_{I}( X, T ), A^{J}_{b}( X, T)\}$ and $\{{\nu}_{\phi}(X, T)\}$ defined in $\bar I^{3}$. The model requires the fitting algorithm to obey the following rules:\footnote{ Note that such an algorithm is guaranteed to exist, since we are fitting the smooth fields with infinite degrees of freedom to the finitely many data points given by the expectation values of the local observables.}
\begin{equation}
\begin{split}
\nu_{\phi^0}(X, T) \equiv \langle\hat{\nu}_{\phi^0}(X, T)\rangle\rule{70pt}{0pt}\\\\
  \int_{\bar{S}_{X, \Delta X}}E^{a}_{I}(T)ds_{a}  \equiv \langle \widehat{J}(e_{X, \Delta X}, T )_I \rangle\rule{35pt}{0pt}\\\\
\mathcal{P}\exp[ \int_{\bar{e}_{X, \Delta X}} A^{J}_{b}(T)(\tau_{J})de^{b}]^{\bar{K}}_{\bar{L}} \equiv \langle \hat{h}(e_{X, \Delta X}, T)\rangle ^{\bar{K}}_{\bar{L}}\rule{20pt}{0pt}
\end{split}
\end{equation}
Due to the lattice torus topology, the continuous condition $(7.2)$ would imply boundary conditions on the emergent fields in the coordinate space $\bar I^{3}$. Denote the three pairs of boundary faces of $\bar I^{3}$ as $({\partial I}_x^+,{\partial I}_x^-)$, $({\partial I}_y^+,{\partial I}_y^-)$, and $({\partial I}_z^+,{\partial I}_z^-)$, the continuous condition $(7.2)$ on the lattice torus implies:
\begin{equation}
\begin{split}
\nu_{\phi^0}(X, T_1)|_{{\partial I}_{x,y,z}^+}=\nu_{\phi^0}(X, T_1)|_{{\partial I}_{x,y,z}^-}+O(d)\\\\
  E^{a}_{I}(X,T_1)|_{{\partial I}_{x,y,z}^+}=E^{a}_{I}(X,T_1)|_{{\partial I}_{x,y,z}^-}+O(d)\\\\
 A^{J}_{b}(X,T_1)|_{{\partial I}_{x,y,z}^+}= A^{J}_{b}(X,T_1)|_{{\partial I}_{x,y,z}^-}+O(d)\\\\
\end{split}
\end{equation}
The choices of $\bar I_{\mathbb{Z}}^3 \subset \mathbb{R}^3$, the cell decomposition $\{\bar s_i\}$, and the fitting algorithm described above are restricted but non-unique. However, any chosen set gives a valid correspondence between $|\Psi\rangle$ and the smooth fields. 

It is important to check that the emergent gravitational fields transform correctly under changes of spatial coordinates and frames.  First, we consider two observers using the same clock and current frames, but two different spatial coordinates given by $\hat{\phi}(v_n^*)$ and $\hat{\phi}'(v_n^*)$.\footnote{ It would be interesting to consider the case where the two spatial coordinate operators are based on different preferred dynamical nodes. In such a case the two observers has different notions of physical spatial points. Here we would focus on the simpler case with their agreement on $\{v_n^*\}$.} Suppose that at $T_1$, the two spatial coordinate systems $v^*_n \to X_n \in \bar I_{\mathbb{Z}}^3$ and $v^*_n \to X'_n \in \bar {I'}_{\mathbb{Z}}^3$ satisfy the spatial coordinate conditions, and are approximately continuous except at $\{ v^*_{n_b}\}$ and $\{ v^*_{{n_b}'}\}$. For the coordinate gaps, we arbitrarily set $d\geq d'$. Any $e^*_{n.i}$ satisfying $\bar e^*_{n.i}(\bar\Gamma)\cap\{\bar v^*_{n_b}(\bar\Gamma)\}=\emptyset$ and $\bar e^*_{n.i}(\bar\Gamma)\cap\{\bar v^*_{{n_b}'}(\bar\Gamma)\}=\emptyset $ would be mapped into an embedded path in each of the two coordinate spaces $\bar I^3$ and $\bar {I'}^3$. The two embedded paths in the two coordinate spaces are respectively $\bar e_{X_n,\Delta X_{n,i}}$ and $\bar e_{X'_n,\Delta X'_{n,i}}$. Also, the cell decomposition of $\bar I^3$ dual to $ \bar I_{\mathbb{Z}}^3$, and of $\bar {I'}^3$ dual $\bar {I'}_{\mathbb{Z}}^3$ induce surfaces $\{\bar S_{X_n, \Delta X_{n,i}}\}$ and $\{\bar S_{X'_n, \Delta X'_{n,i}}\}$ in the two coordinate spaces. By construction, the transformation $(\bar e_{X_n,\Delta X_{n,i}},\bar S_{X_n, \Delta X_{n,i}})\to(\bar e_{X'_n,\Delta X'_{n,i}},\bar S_{X'_n, \Delta X'_{n,i}} )$ is contravariant under the coordinate transformation $X \to X'$, up to an error of order $d$. On the other hand, we have $\langle \hat{J}( e_{X_n,\Delta X_{n,i}}, T_1 )_I \rangle=\langle \hat{J}( e_{X'_n,\Delta X'_{n,i}}, T_1 )_I \rangle +O(\hbar)$ and $\langle \hat{h}(e_{X_n,\Delta X_{n,i}}, T_1)^{\bar{K}}_{\bar{L}}\rangle=\langle \hat{h}( e_{X'_n,\Delta X'_{n,i}}, T_1)^{\bar{K}}_{\bar{L}}\rangle+O(\hbar)$. Thus from $(8.1)$, we conclude that the emergent gravitational fields transform as they do in the classical theory under a change of matter spatial coordinates, up to an error which is suppressed when both $\hbar$ and $d$ are small.

 Next, we consider two observers using the same spatial coordinates but two different frames, each satisfying the conditions $(6.8)$. At the clock time $T_1$, the transformation matrices between the two frames are given by
\begin{equation}
\begin{split}
\langle\Psi|\hat{\bar{V}}^{I}_{i}\hat{V}^{'i}_{J'}(X,T_1)|\Psi\rangle \equiv  \mathcal{R}^I_{J'}(X,T_1) + O(\hbar)\\
\langle\Psi|\hat{\bar{V}}^{'I'}_{i}\hat{V}_{J}^{i}(X,T_1)|\Psi\rangle \equiv\mathcal{R}_J^{\dagger I'}(X,T_1) + O(\hbar)\\
\langle\Psi|\hat{\bar{U}}^{\bar{I}}_{\bar{i}} \hat{U}^{'\bar{i}}_{\bar{J'}}(X,T_1)|\Psi\rangle \equiv  \mathcal{U}^{\bar{I}}_{\bar{J'}}(X,T_1) + O(\hbar)\\
\langle\Psi|\hat{\bar{U}}^{'\bar{I'}}_{\bar{i}} \hat{U}^{\bar{i}}_{\bar{J}}(X,T_1)|\Psi\rangle \equiv\mathcal{U}^{\dagger\bar{I'}}_{\bar{J}}(X,T_1) + O(\hbar)\\
\end{split}
\end{equation}
It follows from $(6.8)$ and $(8.3)$ that $\mathcal{R}^{-1}=\mathcal{R}^{\dagger}$ and  $\mathcal{U}^{-1}=\mathcal{U}^{\dagger}$.
Therefore the clock time $T_1$, the observables' expectation values for the two observers are related by $SU(2)$ transformations
\begin{equation}
\begin{split}
\mathcal{R}^I_{I'}(X,T_1)  \langle\hat{J}(S_{X, \Delta X}, T_1 )_I\rangle=\langle\hat{J}(S_{X, \Delta X}, T_1 )_{I'}\rangle+ O(\hbar)+O(d^3)\\\\
 \mathcal{U}^{\dagger\bar{K'}}_{\bar{K}}(X+\Delta X,T_1)\langle\hat{h}(e_{X, \Delta X}, T_1)^{\bar{K}}_{\bar{L}}\rangle\mathcal{U}^{\bar{L}}_{\bar{L'}}(X,T_1)=\langle\hat{h}(e_{X, \Delta X}, T_1)^{\bar{K'}}_{\bar{L'}}+ O(\hbar)+O(d^2)\\
\end{split}
\end{equation}
Again, combining $(8.4)$ with $(8.1)$, we conclude that the emergent gravitational fields transform as they do in the classical theory under a change of matter frames, up to an error which is suppressed when both $\hbar$ and $d$ are small.

Altogether, we have the transformation between the two emergent gravitational fields, given by two different sets of matter coordinates and frames, of the form
\begin{equation}
\begin{split}
\frac{E_{I'}^{a'}}{\sqrt {\det E'}}(X',T_1)
=\frac{ \partial X^{a'}}{\partial X^{a}}\mathcal{R}^I_{I'} \frac{E_{I}^{a}}{\sqrt{\det E}}(X,T_1) + O(\hbar)+ O(d)\\\\
( A_{a'}(X', T_1))^{\bar{K}'}_{\bar{L}'}
=\frac{ \partial X^{a}}{\partial X^{a'}}(\mathcal{U}^{\dagger}A_{a}{\mathcal{U}}(X,T_1)+ i \mathcal{U}^{\dagger}\partial_{a}{\mathcal{U}}(X,T_1))^{\bar{K}'}_{\bar{L}'}+ O(\hbar)+ O(d)\\\\
\end{split}
\end{equation}
which has the correct semi-classical limit.

The emergent gravitational fields describe the large scale limit of the physical quantum geometry we have been searching for in section $2.2.5$. Corresponding to each set of $\{e_{n,i}\}$ and $\{v_n\}$, we construct a set of dynamical surfaces $\{S_{n,i}\}$ and regions $\{R_{n}\}$ satisfying the following conditions. For any $\bar\Gamma$, the embedded surface $\bar S_{n,i}(\bar\Gamma)$ transversely intersects once with \emph{only} $\bar e_{n,i}(\bar\Gamma)$ among $\{\bar e_{m,j}(\bar\Gamma)\}$, in the same orientation. The embedded region $\bar R_{n}(\bar\Gamma)$ contains \emph{only} $\bar v_n(\bar\Gamma)$ among $\{\bar v_m(\bar\Gamma)\}$. Using $(5.1)$ and $(5.4)$, one obtains the local area and volume operators  $\{ \hat A(S_{X,\Delta X},T)\}$ and $\{\hat V(R_X,T)\}$. Moreover, the expectation values of the localized geometric observables $\{ \hat A(S_{X,\Delta X},T_1)\}$ and $\{\hat V(R_X,T_1)\}$ measures the areas and volumes of $\{S^*_{n,i}\}$ and $\{R^*_{n}\}$ corresponding to $\{e^*_{n,i}\}$ and  $\{v^*_n\}$. The sets $\{S^*_{n,i}\}$ and $\{R^*_{n}\}$ represent the sets of \emph{physical surfaces and regions} for the state $|\Psi\rangle$ at $T_1$ under the matter spatial coordinates. Further, the coordinate map $ v^*_n\to \bar V_n=X_n$ naturally identify the physical surfaces and regions with $\{\bar S_{X_n, \Delta X_{n,i}}\} $ and $\{\bar c_{X_n }\}$ in the coordinate space $\bar I^3$. Finally, the quantum geometry given by the action of $\{ \hat A(S_{X,\Delta X},T_1)\}$ and $\{\hat V(R_X,T_1)\}$ on $|\Psi\rangle$ is coherent, and has a large scale limit conforming with the emergent gravitational fields 
\begin{equation}
\begin{split}
\\
\int_{\bigcup_{n,i} \bar S_{X_n, \Delta X_{n,i}} }\sqrt{\delta^{IJ}{E}^{a}_{I}{E}^{b}_{J}(T_1)d\bar{S}_{a}d\bar{S}_{b}}= \sum_{n,i}\langle\hat A(S_{X_n,\Delta X_{n,i}},T_1) \rangle + O(\hbar)\\\\
  \int_{\bigcup_{n} \bar c_{X_n }}\sqrt{\frac{1}{3!}\epsilon^{IJK}\epsilon_{abc}{E}^{a}_{I}{E}^{b}_{J}{E}^{c}_{K}(T_1)} dX^3= \sum_{n} \langle\hat V(R_{X_n},T_1)\rangle + O(\hbar)\\
\end{split}
\end{equation}
where the left hand sides are exactly the classical expressions of the area and volume in terms of the emergent densitized triad fields.

\section{Emergent Constraints and Diffeomorphism Algebra}
Now we move on to explore the consequences of the symmetry of $|\Psi\rangle$, or equivalently the symmetry of $\hat{\mathbb P}$. In particular our goal is to see, in large scales and up to the matter back reactions, what equations are imposed by the symmetry on the emergent gravitational fields.

The right invariance of $\hat{\mathbb P}$  implies that the physical state $|\Psi\rangle$ satisfies 
\begin{equation}
\langle\Psi|\hat{\mathbb P}\hat{\exp}\left(i\hat{H}(\epsilon N_{p})\right)\hat{\Pi}_{T_1}|\Psi\rangle=\langle\Psi|\hat{\mathbb P}\hat{\Pi}_{T_1}|\Psi\rangle
\end{equation}
where $N_{p}$ and the real number $\epsilon$ are arbitrary. Denoting $\hat{O}^{\centerdot}\equiv \hat{O}+ \hat{O}^{\dagger}$ and taking $\epsilon$ to be small, we have
\begin{equation}
\begin{split}
\langle\Psi|\hat{\mathbb P}\hat{H}(N_p)\hat{\Pi}_{T_1}|\Psi\rangle=\langle\Psi|\hat{\mathbb P}\hat{H}^{\centerdot}_{g}(N_p)\hat{\Pi}_{T_1}|\Psi\rangle+\langle\Psi|\hat{\mathbb P}\hat{H}^{\centerdot}_{m}(N_p)\hat{\Pi}_{T_1}|\Psi\rangle=0
\end{split}
\end{equation}
Specifically, we can set the lapse function to be $ N_{p}= N_{p^*}^{\mathcal N}$ corresponding to an arbitrary ${\mathcal N}(X)$. Then it follows from $(6.12)$ that
\begin{equation}
\begin{split}
\langle\Psi|\hat{\mathbb P}\hat{\mathcal{H}}^{\centerdot}_g( N_{p^*}^{\mathcal N})\hat{\Pi}_{T_1}|\Psi\rangle + \langle\Psi|\hat{\mathbb P}\hat{H}^{\centerdot}_{m}( N_{p^*}^{\mathcal N})\hat{\Pi}_{T_1}|\Psi\rangle\rule{30pt}{0pt}\\
= \langle\Psi|\hat{\mathbb P}\hat{\mathcal{H}}^{\centerdot}_g(\mathcal N)\hat{\Pi}_{T_1}|\Psi\rangle+\langle\Psi|\hat{\mathbb P}\hat{H}^{\centerdot}_{m}( N_{p^*}^{\mathcal N})\hat{\Pi}_{T_1}|\Psi\rangle+ O(\hbar)\\
 \equiv \langle\Psi|\hat{\mathbb P}\hat{\mathcal{H}}^{\centerdot}_g(\mathcal N)\hat{\Pi}_{T_1}|\Psi\rangle+O(\hbar)+\epsilon_m\rule{95pt}{0pt}\\ = 0\rule{252pt}{0pt}\\
\end{split}
\end{equation}
for any $\mathcal{N}(X)$, where we denote the contribution from the matter sector as $\epsilon_m$
. Recall that $\hat{\mathcal{H}}_g(\mathcal N)$ is an ordered series of spatially local operators. To indicate this we write $\hat{\mathcal{H}}_g(\mathcal N)\equiv \mathcal{H}_{g}( \mathcal{N},\hat{J}, \hat{h})$, where $\hat{J}$ and $\hat{h}$ denote  the collection of all the spatially local operators. Moreover,  the condition $(6.3)$ for the clock gives
\begin{equation}
\begin{split}
 \langle\Psi|\hat{\mathbb P}\hat{\mathcal{H}}^{\centerdot}_g(\mathcal N)\hat{\Pi}_{T_1}|\Psi\rangle+\epsilon_m\rule{80pt}{0pt}\\
\equiv\langle\Psi|\hat{\mathbb P} \mathcal{H}^{\centerdot}_{g} (\mathcal{N}, \hat{J}, \hat{h})\hat{\Pi}_{T_1}|\Psi\rangle+\epsilon_m\rule{60pt}{0pt}\\
=\langle\Psi|  \mathcal{H}^{\centerdot}_{g} (\mathcal{N},\hat{J}(T_1 ), \hat{h}( T_1))|\Psi\rangle+O(\hbar)+\epsilon_m\rule{5pt}{0pt}\\
= 0\rule{202pt}{0pt}
\end{split}
\end{equation}
The coherence conditions $(7.1)$  says that the local observables in $(9.4)$ can be replaced by their expectation values, with errors of orders of $\hbar$: 
\begin{equation}
\begin{split}
\langle\Psi|  \mathcal{H}^{\centerdot}_{g} (\mathcal{N},\hat{J}(T_1 ), \hat{h}( T_1))|\Psi\rangle+O(\hbar)+\epsilon_m
=\langle\Psi|  \mathcal{H}^{\centerdot}_{g} (\mathcal{N},\langle\hat{J}(T_1 )\rangle, \langle\hat{h}( T_1)\rangle)|\Psi\rangle +O(\hbar) +\epsilon_m
= 0\\\\
\end{split}
\end{equation}
 Under the correspondence $(8.1)$, we now rewrite $(9.5)$ in terms of the emergent gravitational fields.
\begin{equation}
\begin{split}
\langle\Psi|  \mathcal{H}^{\centerdot}_{g} (\mathcal{N},\langle\hat{J}(T_1 )\rangle, \langle\hat{h}(T_1)\rangle)|\Psi\rangle+O(\hbar)+\epsilon_m
=H_g(\mathcal N)\big|_{E_{I}^{a}(T_1), A_{b}^{J}(T_1)}+O(d^4) +O(\hbar)+\epsilon_{m}
= 0\\\\
\end{split}
\end{equation}
where $H_g(\mathcal N)$ is exactly the classical Hamiltonian constraint given by $(2.4)$, written in terms of the emergent gravitational fields and matter coordinates
\begin{equation}
\begin{split}
H_g (\mathcal N)\big|_{E_{I}^{a}(T_1), A_{b}^{J}(T_1)}=\int_{\bar I^3} d^3 X \mathcal N(X) \frac{E^a_I E^b_J}{\sqrt{\det E}} \left[ {\epsilon^{IJ}}_K F^K_{ab} +2(1-\gamma^2) K^I_{[a} K^J_{b]}\right](X, T_1)
\\
\nonumber
\end{split}
\end{equation}
It follows that up to the corrections coming from the discretization of space, quantum fluctuations, and matter back reactions, the emergent gravitational fields of $|\Psi\rangle$ satisfy gravitational Hamiltonian constraint.

We now move on to show that Gauss and momentum constraints also emerge correctly in the semi-classical limit. The invariance of $\hat{\mathbb P}$ again implies
\begin{equation}
\begin{split}
\langle\Psi|\hat{\mathbb P}\hat{\exp}\left(i\hat{H}(\epsilon N_p)\right)\hat{\exp}\left(\hat{H}(i\epsilon' M_p)\right)\hat{\exp}\left(\hat{H}(-i\epsilon N_p)\right)\hat{\Pi}_{T_1}|\Psi\rangle=\langle\psi|\hat{\mathbb P}\hat{\Pi}_{T_1}|\Psi\rangle
\end{split}
\end{equation}
 Setting $\epsilon$ and $\epsilon'$ to be small, we have
\begin{equation}
\langle\Psi|\hat{\mathbb P}[\hat{H}(M_p),\hat{H}(N_p)]\hat{\Pi}_{T_1}|\Psi\rangle= 0 
\end{equation}
for any $M_p$ and $N_p$. Setting the lapse functions to be $M^{ \mathcal M}_{p^*}$ and $N^{\mathcal N}_{p^*}$ with arbitrary $\mathcal{M}(X)$ and $\mathcal{N}(X)$, we have 
\begin{equation}
\begin{split}
\langle\Psi|\hat{\mathbb P}\frac{i}{\hbar}[\hat{H}(M^{ \mathcal M}_{p^*}),\hat{H}(N^{\mathcal N}_{p^*})]\hat{\Pi}_{T_1}|\Psi\rangle\rule{60pt}{0pt}\\
\equiv\langle\Psi|\hat{\mathbb P}\frac{i}{\hbar}[\hat{H}^{\centerdot}_g(M^{ \mathcal M}_{p^*}),\hat{H}^{\centerdot}_g(N^{\mathcal N}_{p^*})]\hat{\Pi}_{T_1}|\Psi\rangle+\epsilon'_m\rule{22pt}{0pt}\\
=\langle\Psi|\hat{\mathbb P}\frac{i}{\hbar}[\hat{\mathcal{H}}^{\centerdot}_g(\mathcal{M}),\hat{\mathcal{H}}^{\centerdot}_g(\mathcal{N})]\hat{\Pi}_{T_1}|\Psi\rangle+O(\hbar)+\epsilon'_m\\
=0\rule{214pt}{0pt}\\\\
\end{split}
\end{equation}
Here we have singled out the gravity-gravity term of the commutator and denoted the terms involving $\hat{H}_m$ as $\epsilon'_m$.
The commutator $[\hat{\mathcal{H}}^{\centerdot}_g(\mathcal{M}),\hat{\mathcal{H}}^{\centerdot}_g(\mathcal{N})]$ is then carried out using $(6.13)$: 
\begin{equation}
\begin{split}
\langle\Psi|\hat{\mathbb P}\frac{i}{\hbar}[\hat{\mathcal{H}}^{\centerdot}_g(\mathcal{M}),\hat{\mathcal{H}}^{\centerdot}_g(\mathcal{N})]\hat{\Pi}_{T_1}|\Psi\rangle+O(\hbar)+\epsilon'_m\rule{180pt}{0pt}\\
\equiv\langle\Psi|\hat{\mathbb P} \mathcal{C}_g(\mathcal{M},\mathcal{N},\hat{J}, \hat{h} )\hat{\Pi}_{T_1}|\Psi\rangle+\langle\Psi|\hat{\mathbb P} \mathcal{C}_{g,\alpha}(\mathcal{M},\mathcal{N},\hat{J}, \hat{h}, \hat{\alpha})\hat{\Pi}_{T_1}|\Psi\rangle+O(\hbar)+\epsilon'_m\rule{50pt}{0pt}\\
=\langle\Psi|\mathcal{C}_g(\mathcal{M},\mathcal{N}, \hat{J}( T_1), \hat{h}(T_1))|\Psi\rangle+\langle\Psi| \mathcal{C}_{g,\alpha}(\mathcal{M},\mathcal{N},\hat{J}(T_1), \hat{h}(T_1), \hat{\alpha}(T_1) )|\Psi\rangle+O(\hbar)+\epsilon'_m\\
=0\rule{420pt}{0pt}
\end{split}
\end{equation}
where we separate the result into two terms according to the presence of $\hat{\alpha}$. Among the two terms involving $\hat{\mathcal{C}}_g(\mathcal M)\equiv \mathcal{C}_g(\mathcal{M},\mathcal{N},\hat{J}, \hat{h} )$ and $\hat{\mathcal{C}}_{g,\alpha}(\mathcal{M},\mathcal{N})\equiv \mathcal{C}_{g,\alpha}(\mathcal{M},\mathcal{N},\hat{J}, \hat{h}, \hat{\alpha})$, only the latter contains $\hat{\alpha}$.
Similar to the previous case, the semi-classical limit of $(9.10)$ in terms of the emergent gravitational fields is obtained as
\begin{equation}
\begin{split}
\langle\Psi| \mathcal{C}_g(\mathcal{M},\mathcal{N},\langle\hat{J}(T_1)\rangle, \langle\hat{h}(T_1)\rangle)|\Psi\rangle+\langle\Psi| \mathcal{C}_{g,\alpha}(\mathcal{M},\mathcal{N},\langle\hat{J}(T_1)\rangle,\langle\hat{h}(T_1)\rangle, \langle\hat{\alpha}(T_1)\rangle )|\Psi\rangle+O(\hbar)+\epsilon'_m\\
=\{H_g(\mathcal{M}), H_g(\mathcal{N})\}\big|_{E_{I}^{a}(T_1), A_{b}^{J}(T_1)}+O(d^4)+ O(\hbar)+\epsilon'_m\rule{195pt}{0pt}\\
=0\rule{450pt}{0pt}
\end{split}
\end{equation}
Note that the expectation of $\hat{\alpha}$ for $|\Psi\rangle$ is of order of one, since its spectrum contains only $+1$ and $-1$. Given that, the term $\mathcal{C}_{g,\alpha}$ only contributes as $O(d^4)$. Thus the leading order contribution only involves the expectation values that can be translated into the emergent gravitational fields.
Since $\mathcal M$ and $\mathcal N$ are arbitrary, it follows from $(2.6)$ that the emergent gravitational fields also satisfy gravitational Gauss and momentum constraints, up to the corrections due to the discretization of space, quantum fluctuations, and matter back reaction. 

The above calculations suggest that the operator $\hat{\mathcal{H}}^{\centerdot}_{g} (\mathcal{N})$ is the quantum counterpart of $H_g(\mathcal{N})$. According to the classical case, $\hat{\mathcal{H}}^{\centerdot}_g$ is then expected to generate the translations on only the gravitational fields, in the direction perpendicular to the spatial slices specified by the clock. Moreover, the calculations also suggest that $\hat{\mathcal{C}}_g(\mathcal{M},\mathcal{N})$ is the quantum counterpart of $\{H_g(\mathcal{M}), H_g(\mathcal{N})\}$, which generates combinations of spatial diffeomorphisms and local $SU(2)$ transformations on only the gravitational fields, with respect to the matter spatial coordinates and frames. As a first step to validate this interpretation, we list the results of the semi-classical limits of the commutators
\begin{equation}
\begin{split}
\langle\Psi|\hat{\mathbb P}\frac{i}{\hbar}\left[\hat{\mathcal{H}}^{\centerdot}_{g} (\mathcal{M}),\hat{\mathcal{H}}^{\centerdot}_{g} (\mathcal{N})\right]\hat{\Pi}_{T_1}|\Psi\rangle\rule{185pt}{0pt}\\
=\left\{H_g(\mathcal {M}), H_g(\mathcal {N})\right\}\big|_{E_{I}^{a}(T_1), A_{b}^{J}(T_1)}+O(\hbar)+O(d^4)\rule{105pt}{0pt}
\\\\
\langle\Psi|\hat{\mathbb P}\frac{i}{\hbar}\left[\hat{\mathcal{H}}^{\centerdot}_{g} (\mathcal{M}),\hat{\mathcal{C}}_{g}(\mathcal{N},\mathcal{N}')\right]\hat{\Pi}_{T_1}|\Psi\rangle\rule{175pt}{0pt}\\
=\left\{H_g(\mathcal {M}), \{H_g(\mathcal{N}), H_g(\mathcal{N'})\}\right\}\big|_{E_{I}^{a}(T_1), A_{b}^{J}(T_1)}+O(\hbar)+O(d^4)\rule{53pt}{0pt}
\\\\
\langle\Psi| \hat{\mathbb P}\frac{i}{\hbar}\left[\hat{\mathcal{C}}_{g}(\mathcal{M},\mathcal{N}),\hat{\mathcal{C}}_{g}(\mathcal{M'},\mathcal{N'})\right]\hat{\Pi}_{T_1}|\Psi\rangle\rule{157pt}{0pt}\\
=\left\{\{H_g(\mathcal{M}), H_g(\mathcal{N})\},\{H_g(\mathcal{M'}), H_g(\mathcal{N'})\}\right\}\big|_{E_{I}^{a}(T_1), A_{b}^{J}(T_1)}+O(\hbar)+O(d^4)
\\\\
\end{split}
\end{equation}
More generally, it can be shown that
\begin{equation}
\begin{split}
\langle\Psi|\hat{\mathbb P}\left(\frac{i}{\hbar}\right)^{n-1}\left[\hat{\Phi}_n,....\left[\hat{\Phi}_3,\left[\hat{\Phi}_2, \hat{\Phi}_1\right]\right]...\right]\hat{\Pi}_{T_1}|\Psi\rangle\rule{85pt}{0pt}\\
=\left\{{\Phi}_n,....\left\{{\Phi}_3,\left\{{\Phi}_2,{\Phi}_1\right\}\right\}...\right\}\big|_{E_{I}^{a}(T_1), A_{b}^{J}(T_1)}+O(\hbar)+O(d^4)\rule{55pt}{0pt}
\end{split}
\end{equation}
where each pair $(\hat{\Phi}_i, \Phi_i)$ can be set to be either the pair $(\hat{\mathcal H}^{\centerdot}_g(\mathcal{M}_i),H_g(\mathcal{M}_i))$, or the pair $(\hat{\mathcal{C}}_{g}(\mathcal{M}_i,\mathcal{N}_i),\{H_g(\mathcal{M}_i), H_g(\mathcal{N}_i)\}) $.
In this sense the four-dimensional diffeomorphism algebra for pure gravitational fields emerges in the semi-classical limit for $|\Psi\rangle$, up to the corrections.

The quantum spacetime $|\Psi\rangle$, including the gravitational and matter sectors, is  invariant under the actions of four-dimensional diffeomorphisms acting on both sectors. On the other hand, the generators identified in $(9.12)$ are expected to generate the four-dimensional diffeomorphisms of the gravitational sector relative to the matter sector, and act nontrivially on the quantum spacetime $|\Psi\rangle$. It is these relative transformations that give the physics we observe.

 \section{ Emergent Dynamics }
The Gauss constraint, momentum constraint and (modified) Hamiltonian constraint operators are simply zero operators in the physical Hilbert space $\mathbb H$. In this sense, the fundamental diffeomorphism symmetry becomes silent after providing the structure of the physical Hilbert space. Just as in classical general relativity, the physics in $\mathbb H$ should to be instead given by relations between the dynamical fields. In the model, the relation between gravitational field and the matter coordinate fields are expressed by the local observables $\hat O(X,T)$. The values $\{\langle\Psi|\hat O(X,T)|\Psi\rangle\}$ with different $(X,T)$ are expected to be related through certain quantum transformations. Moreover, semi-classical limits of such transformations have clear classical counterparts that have to be agreed with; specifically, the transformations from $\langle\Psi|\hat O(X,T)|\Psi\rangle$ to $\langle\Psi|\hat O(X,T+\Delta T)|\Psi\rangle$ has to give the clock-time dynamics of $O(X,T)$ in general relativity. In the last chapter, we found that the operators $\hat{\mathcal{H}}^{\centerdot}_{g} (\mathcal{N})$ and $\hat{\mathcal{C}}_g(\mathcal{M},\mathcal{N})$ reproduce the algebra of relative diffeomorphisms between the gravitational and matter fields, and are candidates for generators of such transformations in the quantum theory. In this chapter, we explicitly evaluate the relative transformations of the emergent gravitational fields with respect to the clock time. We will show that these transformations are indeed generated by $\hat{\mathcal{H}}^{\centerdot}_{g} (\mathcal{N})$ and $\hat{\mathcal{C}}_g(\mathcal{M},\mathcal{N})$, and they recover the dynamics of general relativity in appropriate semi-classical limits. 

Due to their self-adjointness, $\hat{\mathcal H}^{\centerdot}_g(\mathcal{M})$ and $\hat{\mathcal{C}}_{g}(\mathcal{M},\mathcal{N})$ generate the following unitary operators in $\mathfrak{K}$:
\begin{equation}
\begin{split}
\hat{\exp}\left(i\rule{1pt}{0pt}\frac{\epsilon_1}{\hbar}\hat{\mathcal H}^{\centerdot}_g(\mathcal{M})\right)\equiv \hat{U}_{\mathcal{H}_g}(\mathcal{M},\epsilon_1)\rule{2pt}{0pt},\rule{5pt}{0pt}\hat{\exp}\left(i\rule{1pt}{0pt}\frac{\epsilon_2}{\hbar}\rule{1pt}{0pt}\hat{\mathcal{C}}_{g}(\mathcal{M},\mathcal{N})\right) \equiv\hat{U}_{\mathcal{C}_g}(\mathcal{M},\mathcal{N},\epsilon_2)
\nonumber
\end{split}
\end{equation}
When $\epsilon_1$ is small, we have the infinitesimal transformations of the emergent gravitational fields by $ \hat{U}_{\mathcal{H}_g}(\mathcal{M},\epsilon_1)$ as:
\begin{equation}
\begin{split}
\delta_{\epsilon_1} \int_{\bar{S}_{X, \Delta X}}E^{a}_{I}(T_1)ds_{a}\rule{228pt}{0pt}\\
=\epsilon_1\langle\Psi|\hat{\mathbb P}\frac{i}{\hbar}[\hat{J}(e_{X, \Delta X} )_I,\mathcal{\hat{H}}^{\centerdot}_{g} (\mathcal{N})]\hat{\Pi}_{T_1}|\Psi\rangle+O({\epsilon_1}^2)+O(\hbar)+O(d^4)\rule{23pt}{0pt}\\\
\\
\delta_{\epsilon_1} \mathcal{P}\exp[ \int_{\bar{e}_{X, \Delta X}} A^{J}_{b}(T_1)(\tau_{J})de^{b} ]^{\bar{K}}_{\bar{L}}\rule{170pt}{0pt}\\
=\epsilon_1\langle\Psi|\hat{\mathbb P}\frac{i}{\hbar}[ \hat{h}(e_{X, \Delta X})^{\bar{K}}_{\bar{L}},\mathcal{\hat{H}}^{\centerdot}_{g} (\mathcal{N})]\hat{\Pi}_{T_1}|\Psi\rangle+O({\epsilon_1}^2)+O(\hbar)+O(d^4)\rule{25pt}{0pt}\\\
\\
\end{split}
\end{equation}
Similarly, the infinitesimal transformations by $\hat{U}_{\mathcal{C}_g}(\mathcal{M},\mathcal{N},\epsilon_2)$ with small $\epsilon_2$ are given by:
\begin{equation}
\begin{split}
\delta_{\epsilon_2} \int_{\bar{S}_{X, \Delta X}}E^{a}_{I}(T_1)ds_{a}\rule{228pt}{0pt}\\ 
=\epsilon_2\langle\Psi|\hat{\mathbb P}\frac{i}{\hbar}[\hat{J}(e_{X, \Delta X} )_I,\mathcal{\hat{C}}_g(\mathcal{M},\mathcal{M'})]\hat{\Pi}_{T_1}|\Psi\rangle +O({\epsilon_2}^2)+O(\hbar)+O(d^4)\rule{3pt}{0pt}\\\
\\
\delta_{\epsilon_2} \mathcal{P}\exp[ \int_{\bar{e}_{X, \Delta X}} A^{J}_{b}(T_1)(\tau_{J})de^{b} ]^{\bar{K}}_{\bar{L}}\rule{170pt}{0pt}\\
=\epsilon_2\langle\Psi|\hat{\mathbb P}\frac{i}{\hbar}[ \hat{h}(e_{X, \Delta X})^{\bar{K}}_{\bar{L}},\mathcal{\hat{C}}_g(\mathcal{M},\mathcal{M'})]\hat{\Pi}_{T_1}|\Psi\rangle+O({\epsilon_2}^2)+O(\hbar)+O(d^4)
\\\\
\end{split}
\end{equation}
Equations $(10.1)$ are of special interest in deriving the dynamics in terms of the clock time. These equations are expected to give the variations of the gravitational fields at the clock time $T_1$, when the fields are translated perpendicularly to clock time slices. The perpendicular direction of the time slices might or might not agree with the clock time coordinate direction at the time $T_1$. When the two agree, the clock time derivative of the local gravitational observables should be described by $(10.1)$ with a specific choice of $\mathcal N$, and possibly a term generated by the Gauss constraint. In this case the matter coordinates simply correspond to the gauge where the shift function $V^a$ is set to be zero. When the two directions differ, the equations of evolution should further include a term that is generated by the momentum constraint. In short, aside from the contribution $(10.1)$, the presence of the terms generated by momentum and the Gauss constraints should be determined by the dynamics of the spatial coordinate and frame fields. 

To calculate the clock-time derivative of a gravitational observable $\hat O$, one needs to choose an appropriate value for $\mathcal N$ in $(10.1)$. Ignoring matter back reactions, the commutator $\frac{i}{\hbar}[ \hat O(\bar v^*_n (\bar \Gamma)), \hat{ H}^{\centerdot}_g(\bar N)]$ approximates $\frac{i}{\hbar}[ \hat O(\bar v^*_n (\bar \Gamma)), \hat{ H}(\bar N)]$, which corresponds to  $\{O(\text x), H(\bar N)\}$ in the classical theory. Recall that when all fields are on-shell, $H(\bar N=1)$ generates the diffeomorphisms given by unit normal flows orthogonal to spatial slices, transforming $O(\text x)\equiv O(\text x,0)$ into $ O(\text x,t)$. Therefore, when all the constraints are satisfied, $\{ O(\text x), H(\bar N=1)\}=\partial_t O(\text x,t)|_{t=0}$ is simply the speed of $O$ with respect to the proper time carried by the flow. Also, in chapter $5$ we have defined the speed of the clock ${\nu}_{\phi^0}(\text x)\equiv\{\phi^0(\text x), H(\bar N=1)\}=\partial_t\phi^0(\text x,t)|_{t=0}$ (on shell). Combining the two different speeds, Leibniz rule instructs us to obtain the derivative of $O$ with respect to $\phi^0$ at the proper time $t=0$, by using ${\nu}^{-1}_{\phi^0}(\text x)\{ O(\text x), H(\bar N=1)\}=\{ O(\text x), H(\bar N={\nu}^{-1}_{\phi^0})\}$. Therefore, we expect the appropriate lapse function to be $\mathcal{N}(X)={\nu}^{-1}_{\phi^0}(X,T_1)$ .

To independently verify these statements about the dynamics, we now carry out a calculation directly based on the definition of the observables. Since there is a set of local observables corresponding to each clock time, the dynamics can be evaluated by comparing the expectation values of the observables at a sequence of different clock times. In particular, the comparison between two adjacent clock times leads to the equation of motion for $\langle O(X,T)\rangle$ as
\begin{equation}
\frac{d}{dT}\bigg|_{T_{1}}\langle\hat{O}(X,T)\rangle=\lim_{\Delta T \to 0} \frac{1}{\Delta T} \left[\langle\Psi|\hat{\mathbb{P}}\hat{O}(X)\hat{\Pi}_{T_{1}+\Delta T}|\Psi\rangle-\langle\Psi|\hat{\mathbb{P}}\hat{O}(X)\hat{\Pi}_{T_{1}}|\Psi\rangle\right]
\end{equation}
To evaluate this quantity, we start from the invariance of the local observables. Being physical, a local observable $\hat O(X,T)=\hat{\mathbb P}\hat O(X)\hat{\Pi}_{T}$ commutes with $\hat H(N_p)$ for arbitrary $N_p$. Setting $p=p^*$ and $N_p=N^{\mathcal N}_{p^{*}}$ we have
\begin{equation}
\begin{split}
0=\langle\Psi| \frac{i}{\hbar} \left[\hat{H}\left( N^{\mathcal N}_{p^{*}}\right),\hat{\mathbb P}\hat O(X)\hat{\Pi}_{T_1}\right]|\Psi\rangle\rule{175pt}{0pt}\\
\rule{20pt}{0pt}=\langle\Psi|\hat{\mathbb P}\hat O(X) \frac{i}{\hbar} \left[\hat{H}\left( N^{\mathcal N}_{p^{*}}\right),\hat{\Pi}_{T_1}\right]|\Psi\rangle +\langle\Psi| \hat{\mathbb P}\frac{i}{\hbar} \left[\hat{H}\left( N^{\mathcal N}_{p^{*}}\right),\hat O(X)\right]\hat{\Pi}_{T_1}|\Psi\rangle \\
\end{split}
\end{equation}
The second term is familiar from the previous chapter, while the first term can be calculated using the notation defined in $(3.7)$ and $(6.4)$: 
\begin{equation}
\begin{split}
\langle\Psi|\hat{\mathbb P}\hat O(X) \frac{i}{\hbar} \left[\hat{H}\left( N^{\mathcal N}_{p^{*}}\right),\hat{\Pi}_{T_1}\right]|\Psi\rangle\rule{280pt}{0pt}\\\\
=\langle\Psi|\hat{\mathbb P}\hat O(X)\sum_{[\bar\Gamma]} |s_{[\bar\Gamma]}\rangle\langle s_{[\bar\Gamma]}|\frac{i}{\hbar} \left[\hat{H}\left( N^{\mathcal N}_{p^{*}}\right),\hat{\Pi}_{T_1}\right]|\Psi\rangle \rule{220pt}{0pt}\\
=\sum_{[\bar\Gamma]}\langle\Psi|\hat{\mathbb P}\hat O(X) |s_{[\bar\Gamma]}\rangle\langle S_{\bar\Gamma}|\frac{i}{\hbar} \left[\hat{H}_{(\bar\Gamma)}\left( N^{\mathcal N}_{p^{*}}\right),\hat{\Pi}_{T_1(\bar\Gamma)}\right]\hat{\mathbb P}_{diff}|\Psi\rangle\rule{180pt}{0pt} \\
\\
=\sum_{[\bar\Gamma]}\langle\Psi|\hat{\mathbb P}\hat O(X) |s_{[\bar\Gamma]}\rangle\rule{360pt}{0pt}\\
\times\left[\langle S_{\bar\Gamma}|\frac{i}{\hbar} \sum _m\left[\hat{H}_{(\bar\Gamma)}( N^{\mathcal N}_{p^{*}}), \hat{\nu}_{\phi^0} (\bar{v}^*_{m}(\bar\Gamma))\right] \hat{\nu}^{-1}_{\phi^0}(\bar{v}^*_{m}(\bar\Gamma)) \prod_{n} \hat{\nu}_{\phi^0}(\bar{v}^*_{n}(\bar\Gamma)) \hat\delta(\hat\phi^0(\bar{v}^*_{n}(\bar\Gamma))-T_1)\hat{\mathbb P}_{diff}|\Psi\rangle\right.\\\left.
-\langle S_{\bar\Gamma}|\frac{i}{\hbar}\sum _m \left[\hat{H}_{(\bar\Gamma)}(N^{\mathcal N}_{p^{*}}), \hat\phi^0(\bar{v}^*_{m}(\bar\Gamma))\right]\frac{\partial}{\partial T_m}\bigg|_{T1}\prod_{n} \hat{\nu}_{\phi^0}(\bar{v}^*_{n}(\bar\Gamma))\hat\delta(\hat\phi^0(\bar{v}^*_{n}(\bar\Gamma))-T_{n})\hat{\mathbb P}_{diff}|\Psi\rangle\right]\rule{20pt}{0pt}\\+O(\hbar)\rule{418pt}{0pt}\\
=\langle\Psi|\hat{\mathbb P}\hat O(X)\sum _m\mathcal N(X_m)\hat\Phi_{int}(X_m) \hat{\Pi}_{T_1}|\Psi\rangle
+\langle\Psi|\hat{\mathbb P}\hat O(X)\sum _m\mathcal N(X_m) \hat{\nu}_{\phi^0}(X_m)\frac{\partial}{\partial T_m}\bigg|_{T1}\hat{\Pi}_{\{T_n\}}|\Psi\rangle\rule{3pt}{0pt}\\
+O(d)+O(\hbar)\rule{383pt}{0pt}\\\\
\end{split}
\end{equation}
where $\hat{\Pi}_{\{T_n\}}$ gives the deformed spatial slices with nonconstant clock field values around $T_1$, and is defined by
\begin{equation}
\begin{split}
\langle s_{[\bar\Gamma]}|\hat{\Pi}_{\{T_n\}}\equiv \langle S_{\bar\Gamma}|sym\left\{\prod_{n} \hat{\nu}_{\phi^0}(\bar{v}^*_{n}(\bar\Gamma))\hat\delta(\hat\phi^0(\bar{v}^*_{n}(\bar\Gamma))-T_{n})\right\}\hat{\mathbb P}_{diff}\\
\end{split}
\end{equation}
for any $\langle s_{[\bar\Gamma]}| \in \mathfrak K $. Since $\hat{\Pi}_{\{T_n=T_1\}}= \hat{\Pi}_{T_1}$, the coherence condition $(7.1)$ implies
\begin{equation}
\begin{split}
\hat{\mathbb P}\hat O(X)\frac{\partial}{\partial T_m}\bigg|_{T1}\hat{\Pi}_{\{T_n\}}|\Psi\rangle\rule{200pt}{0pt}\\
 =\frac{\partial}{\partial T_m}\bigg|_{T1}\hat{\mathbb P} O(X,\{T_n\})\hat{\Pi}_{T_1}|\Psi\rangle
+\frac{\partial}{\partial T_m}\bigg|_{T1} \hat{\mathbb P}O(X, T_1)\hat{\Pi}_{\{T_n\}}|\Psi\rangle
+ O(\hbar)
\end{split}
\end{equation}
where $\hat O(X,T)$ can be any local observables involving only gravitational and the clock's momentum field, and $O(X,\{T_n\})$ is a smooth function with $O(X, \{T_n=T_1\})=O(X,T_1)=\langle\hat O(X,T_1)\rangle$. As proposed near the begining of this chapter,  the lapse function will now be set to be $\mathcal N(X)={\nu}^{-1}_{\phi^0}(X,T_1)$. Applying the coherence conditions to $(10.5)$, we derive
\begin{equation}
\begin{split}
\langle\Psi|\hat{\mathbb P}\hat O(X) \frac{i}{\hbar} \left[\hat{H}\left( N ^{{\nu}^{-1}_{(T_1)}}_{p^{*}}\right),\hat{\Pi}_{T_1}\right]|\Psi\rangle\rule{297pt}{0pt}\\
=O(X,T_1)\langle\Psi|\hat{\mathbb P}\sum _m\mathcal{\nu}^{-1}_{\phi^0}(X_m,T_1)\hat\Phi_{int}(X_m) \hat{\Pi}_{T_1}|\Psi\rangle
+O(X,T_1)\langle\Psi|\hat{\mathbb P}\sum _m\frac{\partial}{\partial T_m}\bigg|_{T1}\hat{\Pi}_{\{T_n\}}|\Psi\rangle\rule{70pt}{0pt}\\
+O(X,T_1) \sum _m{\nu}^{-1}_{\phi^0}(X_m,T_1)\frac{\partial}{\partial T_m}\bigg|_{T1}{\nu}_{\phi^0}(X_m,\{T_n\} )\langle\Psi|\hat{\mathbb P}\hat{\Pi}_{T_1}|\Psi\rangle\rule{190pt}{0pt}\\
+\sum _m\frac{\partial}{\partial T_m}\bigg{|}_{T1} O(X,\{T_n\})\langle\Psi|\hat{\mathbb P}\hat{\Pi}_{T_1}|\Psi\rangle
+O(d)+O(\hbar)\rule{235pt}{0pt}\\
= O(X,T_1)\left[\sum _m \mathcal{\nu}^{-1}_{\phi^0}(X_m,T_1)\left(\Phi_{int}(X_m,T_1)  
+\frac{\partial}{\partial T_m}\bigg|_{T1}{\nu}_{\phi^0}(X_m,\{T_n\} )\right)\right]
+\frac{d}{d T}\bigg{|}_{T1} O(X, T)\rule{63pt}{0pt}\\
+O(d)+O(\hbar)\rule{425pt}{0pt}\\
\end{split}
\end{equation}
where we have used the ADM clock condition $(6.1)$. The first term above involves the dynamics of the clock, and the second term is the desired clock time derivative of the local observable.

Further, combining $(10.8)$ and $(10.4)$, we obtain
\begin{equation}
\begin{split}
0=\langle\Psi| \frac{i}{\hbar} \left[\hat{H}\left(   N^{{\nu}^{-1}_{(T_1)}}_{p^{*}}\right),\hat{\mathbb P}\hat O(X)\hat{\Pi}_{T_1}\right]|\Psi\rangle\rule{255pt}{0pt}\\
= O(X,T_1)\left[\sum _m \mathcal{\nu}^{-1}_{\phi^0}(X_m,T_1)\left(\Phi_{int}(X_m,T_1)  
+ \frac{\partial}{\partial T_m}\bigg|_{T1}{\nu}_{\phi^0}(X_m,\{T_n\} )\right)\right]
+\frac{d}{d T}\bigg{|}_{T1} O(X, T)\\
+\langle\Psi| \hat{\mathbb P}\frac{i}{\hbar} \left[\hat{H}\left(   N^{{\nu}^{-1}_{(T_1)}}_{p^{*}}\right),\hat O(X)\right]\hat{\Pi}_{T_1}|\Psi\rangle +O(d)+O(\hbar)\rule{175pt}{0pt}\\
\end{split}
\end{equation}
Specifically, when $\hat O(X,T)=\hat{\mathbb P}\hat{\Pi}_{T}$, equation $(10.9)$ gives the clock dynamics
\begin{equation}
\begin{split}
0= \left[\sum _m \mathcal{\nu}^{-1}_{\phi^0}(X_m,T_1)\left(\Phi_{int}(X_m,T_1)  
+ \frac{\partial}{\partial T_m}\bigg|_{T1}{\nu}_{\phi^0}(X_m,\{T_n\} )\right)\right]
 +O(d)+O(\hbar)\\\\
\end{split}
\end{equation}
Finally, $(10.9)$ and $(10.10)$ leads to the relative Heisenberg equation with respect to the clock time
\begin{equation}
\begin{split}
\frac{d}{d T}\bigg{|}_{T1} O(X, T)\rule{330pt}{0pt}\\
= \langle\Psi| \hat{\mathbb P}\frac{i}{\hbar} \left[\hat O(X),\hat{H}\left(  N^{{\nu}^{-1}_{(T_1)}}_{p^{*}}\right)\right]\hat{\Pi}_{T_1}|\Psi\rangle +O(d)+O(\hbar)\rule{150pt}{0pt}\\
= \langle\Psi| \hat{\mathbb{P}}\frac{i}{\hbar} \left[ \hat{O}(X), \hat{H}^{\centerdot}_g \left( N^{{\nu}^{-1}_{(T_1)}}_{p^{*}}\right) \right] \hat{\Pi}_{T_{1}}|\Psi\rangle+ \langle\Psi| \hat{\mathbb{P}}\frac{i}{\hbar} \left[ \hat{O}(X), \hat{H}^{\centerdot}_m\left( N^{{\nu}^{-1}_{(T_1)}}_{p^{*}}\right) \right] \hat{\Pi}_{T_{1}}|\Psi\rangle\rule{20pt}{0pt}\\
+ O(\hbar)+O(d)\rule{340pt}{0pt}\\
\end{split}
\end{equation}

The second term in $(10.7)$ involving $\hat{H}_m$ requires some attention. Referring to $(5.1)$ and $(5.3)$, we recall that a spatially localized operator is a product of gravitational operators and the matter operators that provide the coordinates and frames. Accordingly the commutator between the spatially localized operator and $\hat{H}_m(N_{p^*})$ can be sorted  into two parts. The first part comes from commuting the gravitational operators with $\hat{H}_m(N_{p^*})$, and gives the contribution of matter back reaction, which will be denoted as $\epsilon_m$. The second part comes from commuting the matter operators with $\hat{H}_m(N_{p^*})$, and gives the dynamics of the matter coordinates and frames. From the previous discussion, this part should contain terms generated by momentum and Gauss constraints. 

Recall that the emergent gravitational fields transform as $(8.2)$ between two spatial matter coordinates and frames. Consequently, we only need to evaluate the dynamics in one specific spatial coordinate and frame. Here we pick our specific coordinates and frames to be fixed at each $v_{\bar{n}}$, which means
\begin{equation}
\begin{split}
 \langle\Psi| \hat{\mathbb{P}}\frac{i}{\hbar} \left[ \hat{\phi}(v^*_{n}), \hat{H}^{\centerdot}_m\left( N^{{\nu}^{-1}_{(T_1)}}_{p^{*}}\right)  \right] \hat{\Pi}_{T_{1}}|\Psi\rangle=0\\
\\
 \langle\Psi| \hat{\mathbb{P}}\frac{i}{\hbar} \hat{O}_{i}(v^*_{{n}})\left[ \hat{J}(v^*_{{n}})^{i}_{I},  \hat{H}^{\centerdot}_m\left(  N^{{\nu}^{-1}_{(T_1)}}_{p^{*}}\right) \right] \hat{\Pi}_{T_{1}}|\Psi\rangle=0\\
\langle\Psi| \hat{\mathbb{P}}\frac{i}{\hbar} \hat{O}^{i}(v^*_{{n}})\left[ \hat{\bar J}(v^*_{{n}})_{i}^{I},  \hat{H}^{\centerdot}_m\left(  N^{{\nu}^{-1}_{(T_1)}}_{p^{*}}\right) \right] \hat{\Pi}_{T_{1}}|\Psi\rangle=0\\
\\
 \langle\Psi| \hat{\mathbb{P}}\frac{i}{\hbar} \hat{O}_{\bar{i}}(v^*_{{n}})\left[ \hat{{U}}(v^*_{{n}})_{\bar{I}}^{\bar{i}},  \hat{H}^{\centerdot}_m\left(  N^{{\nu}^{-1}_{(T_1)}}_{p^{*}}\right) \right] \hat{\Pi}_{T_{1}}|\Psi\rangle=0\\
 \langle\Psi| \hat{\mathbb{P}}\frac{i}{\hbar} \hat{O}^{\bar{i}}(v^*_{{n}})\left[ \hat{\bar{U}}(v^*_{{n}})^{\bar{I}}_{\bar{i}},  \hat{H}^{\centerdot}_m\left(  N^{{\nu}^{-1}_{(T_1)}}_{p^{*}}\right) \right] \hat{\Pi}_{T_{1}}|\Psi\rangle=0\\
\\
\end{split}
\end{equation}
The dynamics of these specific matter spatial coordinates and frames have zero contribution to $(10.11)$, up to an error of order of $\hbar$. In this setting, the clock time derivatives of the gravitational local observables can be evaluated using only $\hat{\mathcal{H}}_g$ and the matter back reactions:
\begin{equation}
\begin{split}
\frac{d}{dT}\bigg|_{T_{1}} \langle \hat{J}(e_{X, \Delta X}, T)_{I} \rangle\rule{212pt}{0pt} \\
=\frac{d}{dT}\bigg|_{T_{1}}\int_{\bar{S}_{X, \Delta X}}E^{a}_{I}(T)ds_{a}\rule{205pt}{0pt}\\
= \langle\Psi| \hat{\mathbb{P}}\frac{i}{\hbar} \left[ \hat{J}(e_{X, \Delta X})_{I}, \hat{H}^{\centerdot}_g\left( N^{{\nu}^{-1}_{(T_1)}}_{p^{*}}\right)\right] \hat{\Pi}_{T_{1}}|\Psi\rangle +\epsilon_m+ O(\hbar)+O(d)\rule{10pt}{0pt}\\
= \langle\Psi| \hat{\mathbb{P}}\frac{i}{\hbar} \left[  \hat{J}(e_{X, \Delta X})_{I} , \hat{\mathcal H}^{\centerdot}_{g}\left({{\nu}_{\phi^0}}^{-1}(T_1)\right) \right] \hat{\Pi}_{T_{1}}|\Psi\rangle+\epsilon_m+ O(\hbar)+O(d)
\\\\
\frac{d}{dT}\bigg|_{T_{1}} \langle\hat{h}(e_{X, \Delta X}, T)^{\bar{I}}_{\bar{J}}\rangle\rule{215pt}{0pt}\\
=\frac{d}{dT}\bigg|_{T_{1}} \mathcal{P}\exp[ \int_{\bar{e}_{X, \Delta X}} A^{J}_{b}(T)(\tau_{J})de^{b} ]^{\bar{I}}_{\bar{J}}\rule{154pt}{0pt}\\ 
= \langle\Psi| \hat{\mathbb{P}}\frac{i}{\hbar} \left[ \hat{h}(e_{X, \Delta X})^{\bar{I}}_{\bar{J}}, \hat{H}^{\centerdot}_g\left( N^{{\nu}^{-1}_{(T_1)}}_{p^{*}}\right)\right] \hat{\Pi}_{T_{1}}|\Psi\rangle+\epsilon_m+ O(\hbar)+O(d)\rule{17pt}{0pt}\\
= \langle\Psi| \hat{\mathbb{P}}\frac{i}{\hbar} \left[  \hat{h}(e_{X, \Delta X})^{\bar{I}}_{\bar{J}},\hat{\mathcal H}^{\centerdot}_{g}\left( {{\nu}_{\phi^0}}^{-1}(T_1)\right)\right] \hat{\Pi}_{T_{1}}|\Psi\rangle+\epsilon_m+ O(\hbar)+O(d))\rule{4pt}{0pt}
\\\\
\end{split}
\end{equation}
By setting $\epsilon_1=\Delta T$ and $\mathcal{N}= {{\nu}_{\phi^0}}^{-1}(T_1)$ in $(10.1)$, we see that $(10.1)$ agrees perfectly with $(10.13)$. This means that $\hat{\mathcal{H}}_g$ indeed generates the diffeomorphisms perpendicular to the equal-clock-time spatial slices, when acting on $|\Psi\rangle$.

Next, we evaluate the semi-classical limit of $(10.13)$.
The commutators in the equations can be carried out again using $(6.13)$ to obtain
\begin{equation}
\begin{split}
\frac{d}{dT}\bigl|_{T_{1}}\int_{\bar{S}_{X, \Delta X}}E^{a}_{I}(T)ds_{a}\rule{318pt}{0pt}\\
= \langle\Psi| \hat{\mathbb{P}}\frac{i}{\hbar} \left[ \hat{J}(e_{X, \Delta X})_{I}, \hat{\mathcal H}^{\centerdot}_{g}({{\nu}_{\phi^0}}^{-1}(T_1))  \right] \hat{\Pi}_{T_{1}}|\Psi\rangle+\epsilon_m+ O(\hbar)+O(d)\rule{118pt}{0pt}\\
\equiv \langle\Psi|\hat{\mathbb{P}}\Phi_{J}^{X, \Delta X}(\hat{J},\hat{h})_{I}\hat{\Pi}_{T_{1}}|\Psi\rangle
+ \langle\Psi|\hat{\mathbb{P}}\Phi_{J,\alpha}^{X, \Delta X}(\hat{J},\hat{h}, \hat{\alpha})_{I}\hat{\Pi}_{T_{1}}|\Psi\rangle+\epsilon_m+ O(\hbar)+O(d)\rule{52pt}{0pt}\\
 =\langle\Psi|\Phi_{J}^{X, \Delta X}(\hat{J}(T_1),\hat{h}( T_1))_{I}|\Psi\rangle
+ \langle\Psi|\Phi_{J,\alpha}^{X, \Delta X}(\hat{J}(T_1),\hat{h}(T_1), \hat{\alpha}(T_1))_{I}|\Psi\rangle+\epsilon_m+ O(\hbar)+O(d)\\
\\\\
\frac{d}{dT}\bigl|_{T_{1}} \mathcal{P}\exp[ \int_{\bar{e}_{X, \Delta X}} A^{J}_{b}(T)(\tau_{J})de^{b} ]^{\bar{K}}_{\bar{L}}\rule{266pt}{0pt}\\ 
= \langle\Psi| \hat{\mathbb{P}}\frac{i}{\hbar} \left[ \hat{h}(e_{X, \Delta X})^{\bar{I}}_{\bar{J}}, \hat{\mathcal H}^{\centerdot}_{g}( {{\nu}_{\phi^0}}^{-1}(T_1))\right] \hat{\Pi}_{T_{1}}|\Psi\rangle+\epsilon_m+ O(\hbar)+O(d)\rule{119pt}{0pt}\\
\equiv \langle\Psi|\hat{\mathbb{P}}\Phi_{h}^{X, \Delta X}(\hat{J},\hat{h})^{\bar{K}}_{\bar{L}}\hat{\Pi}_{T_{1}}|\Psi\rangle+\epsilon_m+ O(\hbar)+O(d)\rule{206pt}{0pt}\\
 =\langle\Psi|\Phi_{h}^{X, \Delta X}(\hat{J}(T_1),\hat{h}(T_1))^{\bar{K}}_{\bar{L}}|\Psi\rangle
+\epsilon_m+ O(\hbar)+O(d)\rule{190pt}{0pt}\\
\end{split}
\end{equation}
Note that the equation for $E$ fields contains a term $\Phi_{J,\alpha}$ which involves the operator $\hat{\alpha}$.
Finally, using the coherence conditions $(7.1)$, we derive the equations of motion for the emergent gravitational fields
\begin{equation}
\begin{split}
\frac{d}{dT}\bigg|_{T_{1}}\int_{\bar{S}_{X, \Delta X}}E^{a}_{I}(T)ds_{a}\rule{270pt}{0pt}\\\\
 =\langle\Psi|\Phi_{J}^{X, \Delta X}(\langle\hat{j}( T_1)\rangle,\langle\hat{h}(T_1)\rangle)_{I}|\Psi\rangle
+ \langle\Psi|\Phi_{J,\alpha}^{X, \Delta X}(\langle\hat{J}(T_1)\rangle,\langle\hat{h}(T_1)\rangle, \langle\hat{\alpha}(T_1)\rangle)_{I}|\Psi\rangle\rule{20pt}{0pt}\\+\epsilon_m+ O(\hbar)+O(d)\rule{295pt}{0pt}\\
=\left\{  \int_{\bar{S}_{X, \Delta X}} E^{a}_{I}ds_{a},H_g({{\nu}_{\phi^0}}^{-1}(T_1)) \right\}\bigg|_{E(T_1), A(T_1)} +\epsilon_m+ O(\hbar)+O(d^3)\rule{72pt}{0pt}
\\\\
\frac{d}{dT}\bigg|_{T_{1}} \mathcal{P}\exp[ \int_{\bar{e}_{X, \Delta X}} A^{J}_{b}(T)(\tau_{J})de^{b}]^{\bar{K}}_{\bar{L}}\rule{227pt}{0pt}\\ \\
 =\langle\Psi|\Phi_{h}^{X, \Delta X}(\langle\hat{J}( T_1)\rangle,\langle\hat{h}(T_1)\rangle)^{\bar{K}}_{\bar{L}}|\Psi\rangle+\epsilon_m+ O(\hbar)+O(d)\rule{135pt}{0pt}\\
=\left\{ \mathcal{P}\exp[ \int_{\bar{e}_{X, \Delta X}} A^{J}_{b}(T)(\tau_{J})de^{b} ]^{\bar{K}}_{\bar{L}}, H_g({{\nu}_{\phi^0}}^{-1}(T_1))\right\}\bigg|_{E(T_1), A(T_1)} +\epsilon_m+ O(\hbar)+O(d^2)\\\\
\end{split}
\end{equation}
where $H_g(\mathcal N)$ is exactly the classical gravitational Hamiltonian constraint. Note that the term of $\Phi_{J,\alpha}$ contributes only to the order of $d^3$, and is suppressed similarly to the previous case. Since $(10.15)$ holds for any ${\bar{S}_{X, \Delta X}}$ and ${\bar{e}_{X, \Delta X}}$, it gives the equations of motion for the emergent fields up to errors of $O(d)$
\begin{equation}
\begin{split}
\frac{d}{dT}\bigg|_{T_{1}}E^{a}_{I}(X,T)
=\left\{E^{a}_{I}(X,T), \left[H_g(\mathcal N) +M_g(V)+G_g(\Lambda)\right]\right\}\bigg|_{E(T_1), A(T_1),\mathcal N={{\nu}_{\phi^0}}^{-1}(T_1) ,V=0, \Lambda=0}\\ +\epsilon_m+ O(\hbar)+O(d)\rule{230pt}{0pt}
\\\\
\frac{d}{dT}\bigg|_{T_{1}}  A^{J}_{b}(X,T)
=\left\{ A^{J}_{b}(X,T), \left[H_g(\mathcal N) +M_g(V)+G_g(\Lambda)\right] \right\}\bigg|_{E(T_1), A(T_1),\mathcal N={{\nu}_{\phi^0}}^{-1}(T_1),V=0,\Lambda=0}\\ +\epsilon_m+ O(\hbar)+O(d)\rule{230pt}{0pt}\\\\
\end{split}
\end{equation}
Referring to $(2.4)$ we conclude that $(10.16)$ is the gauge-fixed (with $N={{\nu}_{\phi^0}}^{-1}$, $V=0$ and $\Lambda=0$) classical equations of motion in Hamiltonian form, up to the corrections. Once again the correction terms are given by matter back reaction, quantum fluctuations, and the discretization of space. Up to these errors, with the matter coordinates and frames satisfying $(10.8)$, the equations of motion agree with classical general relativity.

The dynamics of emergent gravitational fields using other matter coordinates and frames follows immediately, from applying the clock time dependent transformation $(8.5)$ to $(10.16)$. Since $(8.5)$ agrees with the classical transformations up to the errors, it will simply re-express $(10.16)$ into another gauge in which $V^a$ and $\Lambda^I$ are nonzero.

\section{Corrections to the Classical Limit}

Before finishing our exploration, let us look into the correction terms that appear in the constraint equations $(9.6)$ and $(9.11)$, the emergent algebra $(9.13)$, and the equations of motion $(10.16)$. For the purpose of recovering general relativity, we assume that the matter back reaction $\epsilon_m$ is small and focus on other corrections.

The corrections of order $\hbar$ result from two different kinds of quantum effects. The first is the regularization of the inverse triad factor $(\det E)^{-1/2}$ appearing in ${H}_g$. In the standard operator $\hat{H}_g(\bar N)$ defined in $(2.29)$-$(2.31)$, this factor is regularized by the commutator between the total volume and holonomy operators, and thus is always finite by construction. This is true even when the state being acted on has zero total spatial volume. On the other hand, the classical counterpart $(\det E)^{-1/2}$ clearly diverges when the $E$ fields vanishes. This contrast signifies the departure of the quantum Hamiltonian constraint from the classical Hamiltonian constraint when the scale of space becomes comparable to $l_p$. The gravitational Hamiltonian constraint operator $\hat{H}_g( N_p)$ for our model, defined in $(3.3)$-$(3.5)$, preserves this feature and therefore also contains the inverse triad corrections at the Plank scale. The second source of corrections is the uncertainty principle. All the fields are quantum mechanical in the model, so there are corrections coming from quantum fluctuations in both gravitational and matter sectors. In our context of describing local gravitational fields, the uncertainty in the matter fields results in fuzziness of the coordinates and frames. As a result, the conditions on the coordinates and frames listed in chapter $6$ all come with errors of order $\hbar$. Further, in obtaining the semi-classical limit, the replacement of the local gravitational and clock momentum observable operators with their expectation values also introduces errors of order $\hbar\l_p^2$ and order $\hbar$. Summarizing all the corrections denoted by $O(\hbar)$, we see that they are negligible when the coherent state $|\Psi\rangle$ is observed at a large scale, when the competing classical terms are much greater.

The corrections of order $d$ result from the discretization of space. Remarkably, these terms are of zeroth order of $\hbar$ and could still dominate in large scales when the quantum effects are insignificant. Recall that $d$ is the upper bound of the spatial coordinate gap, and it is finite since there is only finitely many physical points $\{v^*_{ n}\}$ in the discretized space. This is in contrast with the regularization of a theory in a continuous space, in which coordinate gaps are merely regularization parameters to be set infinitesimal. The effect of the finite value of $d$ comes in two parts. First, in the correspondence $(8.1)$, we see that the expectation values of the local observables determine the emergent fields only up to errors of  order $d$. These errors account for the fact that one can only specify a smooth field up to corrections of order $d$, by making finitely many measurements on coordinate sites separated by gaps of typical value $d$. Indeed, this seems nothing more than a fact of reality, that we can only estimate continuous fields using finite data points. However, instead of the insufficiency of discrete data in describing a continuous reality, the model attributes the errors to our intention of obtaining a continuous picture from the intrinsically discrete reality. Second, the use of holonomies instead of the $A$ fields as configuration variables introduces corrections in all orders of $|A|d$. In the regularized classical Hamiltonian constraint, the $A$ fields appear through the linear terms in holonomies divided by the coordinate lengths of their paths. The nonlinear terms in the holonomies contribute to errors, and vanish in the limit of the paths shrinking to zero length. In the quantum theory, it no longer makes sense to take such limits, since the knot space is topological. In our model, we introduces matter coordinates such that we can define emergent A fields from the holonomy observables, through the correspondence $(8.1)$. Similar to the classical case, when written in terms of the emergent fields, nonlinear terms of holonomy observables contribute to corrections in the semi-classical limit of the model. Moreover, the coordinate lengths of the embedded paths in the coordinate space are given by the finite coordinate gap $d$ that can not be taken to zero as a limit. Therefore, the discretization of space in our model also leads to finite values of  holonomy corrections, which contain all powers of $|A|d$. The holonomy corrections are clearly suppressed by small $d$ value when $A$ fields are nonsingular and bounded. However, the corrections would become important when $A$ fields become singular near, for example, an initial singularity or a  black hole. Overall, we see that all the corrections denoted by $O(d)$ are negligible when the matter spatial coordinate gaps are small and the emergent gravitational fields are nonsingular.

Lastly, all the corrections in the model are well-formulated, and can be explicitly calculated. Moreover, the model is ready for the inclusion of emergent matter fields defined in a way analogous to $(8.10)$, and also their full coupling with the emergent gravitational fields. The explicit expressions of the corrections will be in terms of the emergent fields and their quantum fluctuations. Therefore, the model serves as a promising testing ground for the emergence of loop quantum cosmology from loop quantum gravity.

\section{Conclusions}
Let us briefly review the results of this thesis. The thesis is based on the current stage of loop quantum gravity, which results from a specific canonical quantization of general relativity, and possesses a rigorously constructed knot space that solves the Gauss and momentum constraints. The space encodes the Planck-scale quantum geometry of space in a background independent way that respect the diffeomorphism symmetry in general relativity. Due to its background independence and discrete structure, the theory's semi-classical limit has been difficult and unclear. However, the success of loop quantum cosmology provides evidence of the possible correct limits. In parallel with loop quantum cosmology, the model in this thesis proposes a concrete method to overcome the difficulties and obtain the semi-classical limit of the full theory.

The model we have constructed shares the same knot space with loop quantum gravity coupled to matter fields, although it uses the modified graph-preserving Hamiltonian constraint operator in constructing the physical Hilbert space. Strictly respecting background independence, the model utilizes its matter sector to provide the coordinates and frames to describe the local gravitational observables.  In obtaining the physical Hilbert space, the model assumes the validity of the group averaging procedure described in chapter 4. The semi-classical limits given by the coherent state $|\Psi\rangle$ agree with full general relativity, when the criteria discussed in chapter 11 are met so that the corrections are insignificant. Also, the corrections of order $\hbar$ and $d$ have clear interpretations, as discussed in the preceding chapter.

Finally, the model provides a set-up to explicitly calculate the correction terms for the dynamics of emergent gravitational fields. It is of great interest to see how these corrections behave near the initial singularity, and near the singularity of a black hole. In loop quantum cosmology, the quantum and holonomy corrections have been extensively studied. Among the remarkable results of these models is that the corrections replace the initial singularity with a well-behaved bouncing of the scale factor, preceded by a contracting universe \cite{lqc1}\cite{lqc2} and followed by a built-in slow-roll inflationary phase \cite{test}\cite{test1}\cite{test2}. These predictions are testable and give distinguishable signals in the spectrum of cosmic microwave background radiation \cite{test}\cite{test1}\cite{test2}. It  is then important to ask whether loop quantum cosmology emerges from a certain symmetrical semi-classical limit of loop quantum gravity, and if so,  what additional details the full theory would provide regarding the predicted signals. Explicitly evaluating the correction terms in our model could provide answers to these questions. Hopefully, it would show that the observable predictions of loop quantum cosmology are indeed tests of loop quantum gravity, as a fundamental theory of gravity. (A concrete, affirmative result for emergence of the bouncing has been obtained and will appear in \cite{lin}.)

\end{document}